\documentclass[12pt]{article}

\usepackage{cite,amsmath,epsfig}

\newcommand{\sll}{{\tilde{l}}}
\newcommand{\slR}{{\tilde{l}_{\rm R}}}
\newcommand{\slL}{{\tilde{l}_{\rm L}}}
\newcommand{\msl}[1]{m_{\tilde{l_{#1}}}}
\newcommand{\smu}{{\tilde{\mu}}}
\newcommand{\smuR}{{\tilde{\mu}_{\rm R}}}
\newcommand{\smuL}{{\tilde{\mu}_{\rm L}}}
\newcommand{\se}{{\tilde{e}}}
\newcommand{\seR}{{\tilde{e}_{\rm R}}}
\newcommand{\seL}{{\tilde{e}_{\rm L}}}
\newcommand{\st}{{\tilde{\tau}}}
\newcommand{\stR}{{\tilde{\tau}_1}}

\newcommand{\mse}[1]{m_{\tilde{e}_{#1}}}

\newcommand{\msmu}[1]{m_{\tilde{\mu}_{#1}}}

\newcommand{\MsQ}{\mathswitch M_{\tilde{\rm Q}}}
\newcommand{\cha}{\tilde{\chi}}
\newcommand{\neu}{\tilde{\chi}^0}
\newcommand{\mcha}[1]{m_{\tilde{\chi}^\pm_{#1}}}
\newcommand{\mneu}[1]{m_{\tilde{\chi}^0_{#1}}}
\newcommand{\sbt}{\mathswitch {s_\beta}}
\newcommand{\cbt}{\mathswitch {c_\beta}}

\newcommand{\h}[1][1]{\frac{#1}{2}}
\newcommand{\hh}[1][1]{\tfrac{#1}{2}}
\newcommand{\dZe}{\delta Z_e}
\newcommand{\dMW}{\delta \MW^2}
\newcommand{\dMZ}{\delta \MZ^2}
\newcommand{\dsw}{\delta \sw}

\newcommand{\PL}{\mathswitchr L}
\newcommand{\PR}{\mathswitchr R}
\newcommand{\PSL}{\mathswitchr {SL}}
\newcommand{\PSR}{\mathswitchr {SR}}
\newcommand{\dZsl}{\mathswitch \delta Z^\sll}

\newcommand{\dzL}{\mathswitch \delta \tilde{Z}^\PL}
\newcommand{\dzR}{\mathswitch \delta \tilde{Z}^\PR}

\newcommand{\dzn}{\mathswitch \delta \tilde{Z}^0}

\newcommand{\seCL}{\Sigma^{\pm\PL}}
\newcommand{\seCR}{\Sigma^{\pm\PR}}
\newcommand{\seCSL}{\Sigma^{\pm\PSL}}

\newcommand{\seN}{\Sigma^0}
\newcommand{\seNL}{\Sigma^{0\PL}}

\newcommand{\seNSL}{\Sigma^{0\PSL}}

\def\mathswitch#1{\relax\ifmmode#1\else$#1$\fi}
\def\mathswitchr#1{\relax\ifmmode{\mathrm{#1}}\else$\mathrm{#1}$\fi}
\newcommand{\PW}{\mathswitchr W}
\newcommand{\PZ}{\mathswitchr Z}
\newcommand{\PH}{\mathswitchr H}
\newcommand{\Pe}{\mathswitchr e}
\newcommand{\Pb}{\mathswitchr b}
\newcommand{\Pt}{\mathswitchr t}
\newcommand{\PA}{\mathswitchr A}
\newcommand{\MW}{\mathswitch {M_\PW}}
\newcommand{\MZ}{\mathswitch {M_\PZ}}
\newcommand{\MH}{\mathswitch {M_\PH}}
\newcommand{\MA}{\mathswitch {M_\PA}}
\newcommand{\me}{\mathswitch {m_\Pe}}
\newcommand{\mb}{\mathswitch {m_\Pb}}
\newcommand{\mt}{\mathswitch {m_\Pt}}
\newcommand{\mf}{m_f}
\newcommand{\scrs}{{}}
\newcommand{\sw}{\mathswitch {s_{\scrs\PW}}}
\newcommand{\cw}{\mathswitch {c_{\scrs\PW}}}

\newcommand{\gev}{\,\, \mathrm{GeV}}
\newcommand{\mev}{\,\, \mathrm{MeV}}
\newcommand{\re}{\Re e \,}
\newcommand{\im}{\Im m \,}
\newcommand{\dd}{\partial}

\newcommand{\SLASH}[2]{\makebox[#2ex][l]{$#1$}/}

\newcommand{\dslash}{\SLASH{\dd}{.15}}

\newcommand{\pslash}{\SLASH{p}{.2}}

\newcommand{\Eslash}{\SLASH{E}{.5}\,}
\newcommand{\RR}{{\rm R}}
\newcommand{\LL}{{\rm L}}
\newcommand{\eR}{e_{\rm R}}
\newcommand{\eL}{e_{\rm L}}
\newcommand{\wL}{\mathswitch \omega_\LL}
\newcommand{\wR}{\mathswitch \omega_\RR}
\newcommand{\anc}{\rule{0mm}{0mm}}
\newcommand{\lesim}{\,\raisebox{-.1ex}{$_{\textstyle <}\atop^{\textstyle\sim}$}\,}
\newcommand{\gesim}{\,\raisebox{-.3ex}{$_{\textstyle >}\atop^{\textstyle\sim}$}\,}
\newcommand{\knickpfeil}{\;\raisebox{1.12ex}{$\lfloor$} \!\!\! \to}
\newcommand{\limit}[1]{\stackrel{#1}{-\!\!\!\longrightarrow}}
\newcommand{\drbar}{{\mathswitch {\overline{\rm DR}}} }
\newcommand{\OO}{{\mathcal O}}
\newcommand{\mwe}{M_{\rm ew}}

\newcommand{\mycaption}[1]{\caption{\sl #1}}

\begin{document}
\thispagestyle{empty}

\def\thefootnote{\fnsymbol{footnote}}

\begin{flushright}
DESY 03--111\\
FERMILAB--Pub--03/314--T\\
\end{flushright}

\vspace{1cm}

\begin{center}

{\Large\sc {\bf Slepton Production \\[1ex]
at \boldmath{$e^+e^-$} and \boldmath{$e^-e^-$} Linear Colliders}}
\\[3.5em]
{\large
{\sc
A.~Freitas$^{1}$%
,
A.~von Manteuffel$^{2}$
and
P.~M.~Zerwas$^{2}$
}
}

\vspace*{1cm}

{\sl
$^1$ Fermi National Accelerator Laboratory, Batavia, IL 60510-500, USA

\vspace*{0.4cm}

$^2$ DESY Theorie, Notkestr. 85, D--22603 Hamburg, Germany
}

\end{center}

\vspace*{2.5cm}

\begin{abstract}

High-precision analyses are presented for the production of scalar sleptons,
selectrons and smuons in supersymmetric theories, at future $e^+e^-$ and
$e^-e^-$ linear colliders. Threshold production can be exploited for
measurements of the selectron and smuon masses, an essential ingredient for the
reconstruction of the fundamental supersymmetric theory at high scales. The
production of selectrons in the continuum will allow us to determine the Yukawa
couplings in the selectron sector, scrutinizing the identity of the Yukawa and
gauge couplings, which is a basic consequence of supersymmetry. The theoretical
predictions are elaborated at the one-loop level in the continuum, while at
threshold non-zero width effects and Sommerfeld rescattering
corrections are included. The phenomenological analyses are performed for
$e^+e^-$ and $e^-e^-$ linear colliders with energy up to about 1 TeV 
and with high
integrated luminosity up to 1 ab$^{-1}$ to cover 
the individual slepton channels
separately with high precision.

\end{abstract}

\def\thefootnote{\arabic{footnote}}
\setcounter{page}{0}
\setcounter{footnote}{0}

\newpage


\section{Introduction}

Supersymmetry \cite{susy4d,mssm} provides us with a stable bridge
\cite{Witten:1981kv} between the electroweak scale of $\sim 10^2 \gev$ where
laboratory experiments in particle physics are performed, and the Grand
Unification / Planck scale of $\sim 10^{16}$ / $10^{19} \gev$ where all
phenomena observed at low energies are expected to be rooted in a fundamental
theory including gravity. Bridging more than fourteen orders of magnitude
requires a base of high precision experiments from which the extrapolation to
the Planck scale can be carried out in a solid way. Such a program has already
been pursued very successfully for the three gauge couplings which appear to
unify at the high scale \cite{gaugeuni}. A parallel program should be carried
out in supersymmetric theories for the other fundamental parameters
\cite{blair:00p}, including the parameters of soft supersymmetry breaking,
which may be transferred from a hidden sector near the Planck scale by
gravitational interactions to our visible world.

A solid base for these extrapolations can be built by experiments at
high-energy $e^+e^-$ and $e^-e^-$ colliders \cite{lc1,tesla,nlc,clic}
which, if operated with high luminosity,
will enable us to map out a comprehensive and precise picture of the
supersymmetric sector at the electroweak scale. After the chargino and
neutralino sectors \cite{ckmz,neuhigh} have been explored earlier, we will concentrate in this
analysis on the charged scalar lepton sector of the first and second
generation, in which mixing phenomena are expected to be strongly suppressed%
\footnote{For
a summary of earlier work on this subject see Refs.~\cite{tesla,af_ichep}.}.
[The third
generation and the neutral sector will be summarized in two later addenda
while the colored sector will be analyzed in a separate report.] We have
elaborated the processes
\begin{align}
               e^+e^- \to \smu^+_i \smu^-_i 
	       \phantom{\se^+_i \se^-_i}
	       & [i = \LL, \RR]
\intertext{and}
\begin{aligned}
               e^+e^- &\to \se^+_i \se^-_j \\
               e^-e^- &\to \se^-_i \se^-_j
\end{aligned} \phantom{\smu^+_i \smu^-_i}
& [i,j = \LL,\RR]
\end{align}
at the level of one-loop accuracy. At the thresholds we have calculated
the production cross-sections for off-shell particles including the non-zero
width effects and the Coulombic Sommerfeld rescattering corrections, while
in the continuum the supersymmetric one-loop corrections have been
calculated for on-shell slepton production.

The threshold production of smuons, mediated by s-channel photon and $Z$-boson
exchanges, proceeds through P-waves, giving rise to the moderately steep
$\beta^3$ behavior of the cross-sections in the velocity $\beta = (1-4
\msmu{}^2/s)^{1/2}$ of the smuons. The accuracy that can be reached in
measurements of the masses $\msmu{L,R}$ through threshold scans, is
nevertheless competitive with the accuracy achieved in the continuum by
reconstructing the particles through decay products in the final states.
Non-diagonal and diagonal pairs of selectrons however can be 
excited in S-waves in $e^+e^-$ and $e^-e^-$ collisions, mediated by
t-channel neutralino exchange, and they give rise to the linear $\beta$
dependence of the cross-sections near the thresholds \cite{ee}.
This steep onset of the
excitation curves allows us to measure the selectron masses with unrivaled
precision.

Selectron production in the continuum is strongly affected by the
electron-selectron-gaugino Yukawa couplings and, as result, they can be
determined very precisely by measuring the cross-sections for the production
processes. In this way the identity of the Yukawa couplings ($\hat{g}$) with
the gauge couplings ($g$), $\hat{g}=g$, a basic consequence of supersymmetry,
can be thoroughly investigated with high precision,
as explored first in Ref.~\cite{susyid}.

The threshold analyses in this report adopt techniques outlined earlier in
Ref.~\cite{thr1}. The one-loop calculations in the continuum are
performed in the dimensional reduction scheme (DRED) for regularization and with
on-shell renormalization of masses and couplings. This program must be carried
out consistently for the slepton sector and the neutralino sector. In the loop
corrections all sectors of the electroweak supersymmetric model contribute,
which in general do not decouple for large supersymmetry breaking masses
\cite{eYuk,superoblique}.
A remarkable feature is the appearance of anomalous threshold singularities
\cite{anom}, which show up as discontinuities in the cross-section as a
function of the center-of-mass energy. They are induced by specific mass
patterns of the particles in the loops \cite{Liu:xy}, which are generally
expected to be realized in supersymmetric models but  which are atypical for
Standard Model calculations.

The phenomenological analyses are performed in the Minimal Supersymmetric
Standard Model (MSSM), based on the parameters of the mSUGRA
Snowmass Point SPS1a \cite{sps}. They include effects of initial-state 
beamstrahlung radiation as well as the decays of the sleptons.
All contributions are taken into account that lead to the same final state.
They have been elaborated at the level typical for phenomenological simulations
of processes at an $e^+e^-$ linear collider in the TeV range. Besides Standard
Model backgrounds the most important background channels inside SUSY are taken
into account explicitly. The dominant standard backgrounds, in particular from
$W^+W^-$, $ZZ$ and $Z\gamma$ production, are eliminated {\it a priori}
by proper cuts adopted from previous experimental studies \cite{martynco}.
At the level of precision required here, it is also necessary to
include sub-dominant contributions from off-shell production of gauge bosons 
and SUSY particles.

The final picture is quite exciting: Selectron masses can be determined
at an accuracy of 50 MeV, {\it i.e.} in the per-mille range, while the masses
of the less frequently produced smuons are still accessible at the per-cent
level. The same level of accuracy can also be realized in measurements
of the Yukawa couplings of the selectron sector, thus allowing for a
high-precision comparison with the corresponding gauge couplings. {\it In
summa}: A high-resolution picture of the charged slepton sector in the
first and second generation can be drawn by experiments at prospective
$e^+e^-$ and $e^-e^-$ linear colliders.

The report is organized as follows. In Section \ref{basics} we
summarize the main features of slepton production and decay 
in $e^+e^-$ and $e^-e^-$ collisions
at the Born level. Section \ref{threshold}
presents the predictions for smuon and
selectron production at threshold in detail, leading us to the aforementioned
accuracies expected from selectron and smuon mass measurements in threshold
scans. In Section \ref{continuum} slepton pair production 
in the continuum is
described, and exploited finally for measurements of the Yukawa couplings
in the selectron sector. 
Partial results had been presented earlier in Ref.~\cite{susy02}, while
additional technical details can be found in Ref.~\cite{thesis}. 
Spectrum and properties of supersymmetric particles in
the reference point SPS1a, relevant for the present study, are summarized in the
Appendix for the sake of completeness and the reader's convenience.


\section{\hspace{-2mm}Basics of Smuon and Selectron Production and Decay}
\label{basics}

\subsection{Notation and Conventions}
\label{nota}

In this report we restrict ourselves to the Minimal Supersymmetric Standard
Model (MSSM) as a well-defined framework. Since the muon and electron masses
are very small, the mixing among L- and R-smuon and -selectron states, partners
of the left- and right-chiral leptons, can be neglected and the mass
eigen-states correspond to the L,R eigen-states.

In the other sectors of the MSSM,
mixing needs to be taken into account. The MSSM requires two Higgs doublets
$H_{\rm u}$ and $H_{\rm d}$, which both acquire non-zero vacuum
expectation values $v_{\rm u}$ and $v_{\rm d}$.
The fields mix to form the Goldstone and the physical degrees of freedom with the
mixing angle
\begin{equation}
\tan\beta \equiv v_{\rm u}/v_{\rm d}, \label{eq:tanbeta}
\end{equation}
given by the ratio of the vacuum expectation values.

The charged higgsinos $\widetilde{H}_{\rm u,d}^\pm$ and the winos
$\widetilde{W}^\pm$ mix to form two charginos $\cha^\pm_i$ $(i=1,2)$, while the
neutral higgsinos $\widetilde{H}_{\rm u,d}^0$ and the gauginos $\widetilde{B},
\widetilde{W}^0$ form four neutralino  mass eigen-states $\neu_i$
$(i=1,2,3,4)$.

Apart from the electroweak parameters,
the spectrum of the charginos and neutralinos is described by
three mass parameters, the Higgs/higgsino parameter $\mu$ in the superpotential
and the soft SU(2) and U(1) gaugino parameters, $M_2$ and $M_1$, respectively.
For the charginos the mass term reads
\begin{equation}
\mathcal{L}_{\rm m_{\cha^\pm}} = -\bigl( \widetilde{W}^-,
\widetilde{H}_{\rm d}^- \bigr) \,
X \begin{pmatrix} \widetilde{W}^+ \\ \widetilde{H}_{\rm u}^+
\end{pmatrix} + \mbox{h.c.}
\end{equation}
where $\widetilde{W}^\pm, \widetilde{H}_{\rm u,d}^\pm$ are the Weyl spinors of
the charged winos and higgsinos. The mass matrix
\begin{equation}
X = \begin{pmatrix} M_2 & \sqrt{2} \MW \sin\beta \\
                \sqrt{2} \MW \cos\beta & \mu \end{pmatrix}
\label{eq:Mcha}
\end{equation}
can be diagonalized by two unitary matrices $U$ and $V$ according to
\begin{align}
U^*XV^{-1} &= \begin{pmatrix} \mcha{1} & 0 \\ 0 & \mcha{2} \end{pmatrix}, &
\begin{pmatrix} \chi^-_1 \\ \chi^-_2  \end{pmatrix} &=
  U \begin{pmatrix} \widetilde{W}^- \\ \widetilde{H}_{\rm d}^- \end{pmatrix}, &
\begin{pmatrix} \chi^+_1 \\ \chi^+_2  \end{pmatrix} &=
  V \begin{pmatrix} \widetilde{W}^+ \\ \widetilde{H}_{\rm u}^+ \end{pmatrix},
\label{eq:chamix}
\end{align}
generating the mass eigen-states $\chi^\pm_i$.
In the chiral representation, the Dirac spinors $\cha^\pm_i$ of the charginos
are constructed from the Weyl spinors as follows,
\begin{equation}
\cha^-_i = \begin{pmatrix} \chi^-_i \\[1ex] \overline{\chi^+_i} \end{pmatrix}
\quad\mbox{ and }\quad
\cha^+_i = \begin{pmatrix} \chi^+_i \\[1ex] \overline{\chi^-_i} \end{pmatrix}.
\label{eq:chaspin}
\end{equation}

\noindent
The neutralino mass term in the current eigen-basis is given by
\begin{equation}
\mathcal{L}_{\rm m_{\cha^0}} = -\h {\psi^0}^\top \,
Y \, \psi^0 + \mbox{h.c.}, \qquad
\psi^0 = \bigl( \widetilde{B}, \widetilde{W}^0,
  \widetilde{H}_{\rm d}^0, \widetilde{H}_{\rm u}^0 \bigr)^{\!\top},
\label{eq:neuLagr}
\end{equation}
with the symmetric mass matrix
\begin{equation}
Y = \begin{pmatrix} M_1 & 0 & -\MZ\,\sw\,\cbt & \MZ\,\sw\,\sbt \\
                    0 & M_2 & \MZ\,\cw\,\cbt & -\MZ\,\cw\,\sbt \\
                    -\MZ\,\sw\,\cbt & \MZ\,\cw\,\cbt & 0 & -\mu \\
                    \MZ\,\sw\,\sbt & -\MZ\,\cw\,\sbt & -\mu & 0 \end{pmatrix},
\label{eq:Mneu}
\end{equation}
in which the abbreviations $\sbt = \sin\beta$ and $\cbt = \cos\beta$ have been
introduced; $\sw$ and $\cw$ are the sine and cosine of the electroweak
mixing angle. The transition to the mass eigen-basis is performed by the 
unitary mixing matrix $N$,
\begin{equation}
N^*YN^{-1} = \mbox{diag}
  \bigl(\mneu{1}^2, \mneu{2}^2, \mneu{3}^2, \mneu{4}^2 \bigr) 
\qquad\mbox{ with }\qquad
\chi^0_i = N_{ij} \psi^0_j.
\label{eq:neumix}
\end{equation}
The Majorana spinors $\neu_i$ of the physical neutralinos are composed of the
Weyl spinors as
\begin{equation}
\neu_i = \begin{pmatrix} \chi^0_i \\[1ex] \overline{\chi^0_i} \end{pmatrix}.
\label{eq:neuspin}
\end{equation}
Explicit analytical solutions for the mixing matrices can be found in
Ref.~\cite{ckmz}\footnote{Note that a convention for the chargino mass matrix
$X$ different from eq.~\eqref{eq:Mcha} is used in Ref.~\cite{ckmz}.}

\subsection{Production Mechanisms}

In supersymmetric theories with R-parity conservation scalar leptons are
produced in pairs. Since mixing can be neglected, the pairs are built of the
current eigen-states with chiral index L or R.

\renewcommand{\arraystretch}{1.3}
\begin{table}[tb]
\begin{center}
\begin{tabular}{|@{\hspace{1.8ex}}l@{\hspace{2ex}}|c|cl|}
\hline
Process & Exchange particles & Orbital wave & Threshold excitation \\
\hline \hline
$\eL^+\eR^- \,/\, \eR^+\eL^- \to \smuR^+ \smuR^- \,/\, \smuL^+ \smuL^-$ &
	$\gamma,Z\phantom{,\neu}$ & P-wave & $\propto \beta^3$ \\
\hline
$\eL^+\eR^- \to \seR^+ \seR^-$ & $\gamma,Z,\neu$
        & P-wave & $\propto \beta^3$ \\[-.5ex]
$\eR^+\eL^- \to \seR^+ \seR^-$ & $\gamma,Z\phantom{,\neu}$
        & P-wave & $\propto \beta^3$ \\
\hline
$\eL^+\eR^- \to \seL^+ \seL^-$ & $\gamma,Z\phantom{,\neu}$
        & P-wave & $\propto \beta^3$ \\[-.5ex]
$\eR^+\eL^- \to \seL^+ \seL^-$ & $\gamma,Z,\neu$
        & P-wave & $\propto \beta^3$ \\
\hline
$\eL^+\eL^- \to \seR^+ \seL^-$ & $\phantom{\gamma,Z,}\neu$
        & S-wave & $\propto \beta$   \\[-.5ex]
$\eR^+\eR^- \to \seL^+ \seR^-$ & $\phantom{\gamma,Z,}\neu$
        & S-wave & $\propto \beta$   \\
\hline \hline
$\eR^-\eR^- \to \seR^- \seR^-$ & $\phantom{\gamma,Z,}\neu$
        & S-wave & $\propto \beta$   \\
$\eL^-\eL^- \to \seL^- \seL^-$ & $\phantom{\gamma,Z,}\neu$
        & S-wave & $\propto \beta$   \\
\hline
$\eL^-\eR^- \to \seL^- \seR^-$ & $\phantom{\gamma,Z,}\neu$
        & P-wave & $\propto \beta^3$ \\
\hline
\end{tabular}
\end{center}
\vspace{-1em}
\mycaption{Classification of smuon and selectron production modes in terms of the
exchanged particles, the orbital angular momentum of the final state wave
function and the rise of the excitation curve near threshold. Specific beam
polarization states are required for the individual channels.}
\label{tab:process}
\end{table}
\begin{figure}[t]
\vspace{2em}
\begin{tabular}{ccc}
\epsfig{file=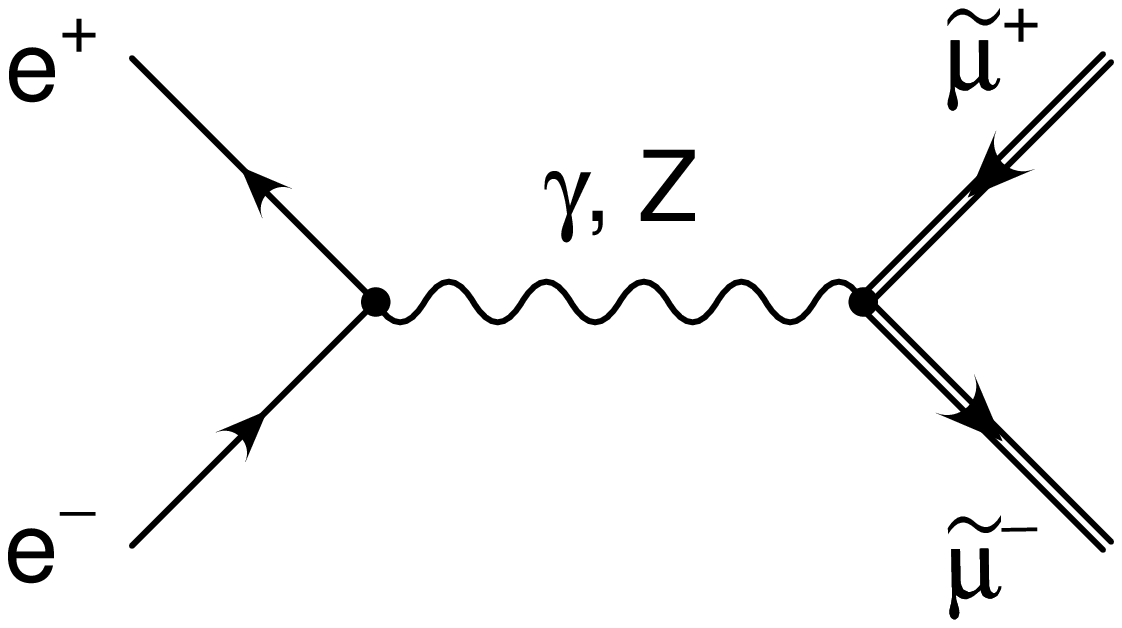,width=5cm} &
\epsfig{file=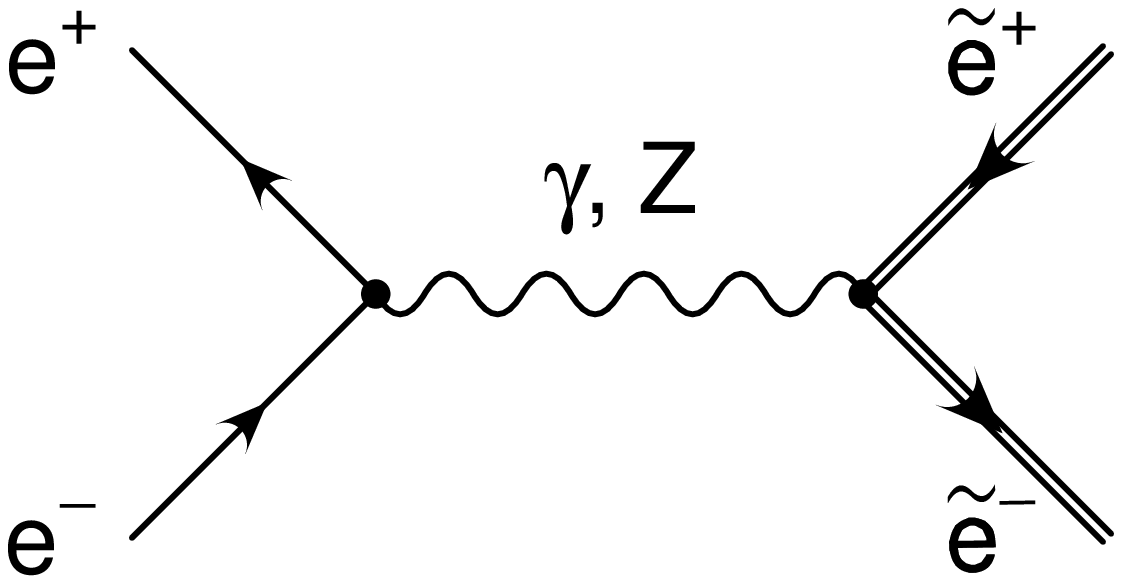,width=5cm} &
\epsfig{file=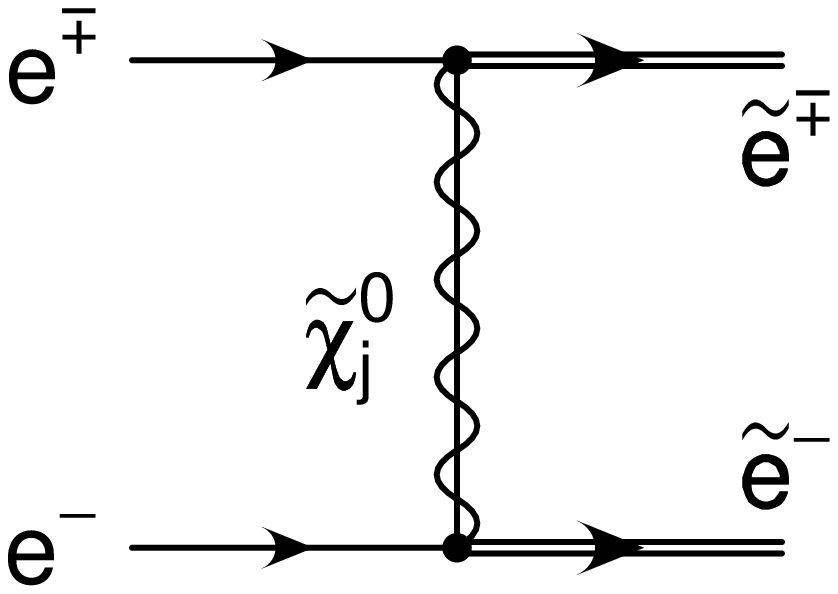,width=4.5cm} \\[-1em]
 (a) & (b) & (c)
\end{tabular}
\mycaption{Generic leading-order diagrams for the pair production 
of smuons and selectrons in $e^+e^-$ or $e^-e^-$ scattering.}
\label{fig:cdiag}
\end{figure}

\underline{{\bf Scalar smuons}} are produced in diagonal pairs 
via s-channel photon
and $Z$-boson exchanges in $e^+e^-$ collisions, see Tab.~\ref{tab:process} and
Fig.~\ref{fig:cdiag}~(a).

Since the intermediate state is a vector, the helicities of electron and
positron must be opposite to each other. By angular momentum conservation the
scalar smuons are therefore produced in P-wave states. This gives rise to the
characteristic $\beta^3$ behavior of the excitation curves close to threshold,
with $\beta$ denoting the velocity of the smuons in the
final state.

The cross-sections for the production of RR and LL smuon pairs by polarized
electron/positron beams may be written as
\begin{align}
\sigma[\eR^+ \, \eL^- \to \smu_i^+ \, \smu_i^-] &=
  \frac{2\pi \alpha^2}{3s} \beta^3
  \left[ 1 + g_i \, g_\LL \frac{s}{s-\MZ^2} \right]^2, \\
\sigma[\eL^+ \, \eR^- \to \smu_i^+ \, \smu_i^-] &=
  \frac{2\pi \alpha^2}{3s} \beta^3
  \left[ 1 + g_i \, g_\RR \frac{s}{s-\MZ^2} \right]^2,
\end{align}
with $i = \LL, \RR$ and the left and right-chiral Z couplings
\begin{equation}
g_{\rm L} = \frac{-1+2\sw^2}{2\sw\cw}, \qquad g_{\rm R} = \frac{\sw}{\cw}.
\end{equation}
As mentioned before, the polarization combinations with equal helicity of
electron and positron vanish.
The electromagnetic coupling $\alpha$ may conveniently be defined at the
energy scale $\sqrt{s}$, incorporating properly the running of the gauge
coupling.

The angular distribution of the smuons follows the familiar $\sin^2\theta$
rule so that the new particles are produced preferentially perpendicular
to the $e^+e^-$ beam axis.

\begin{figure}[tb]
\epsfig{file=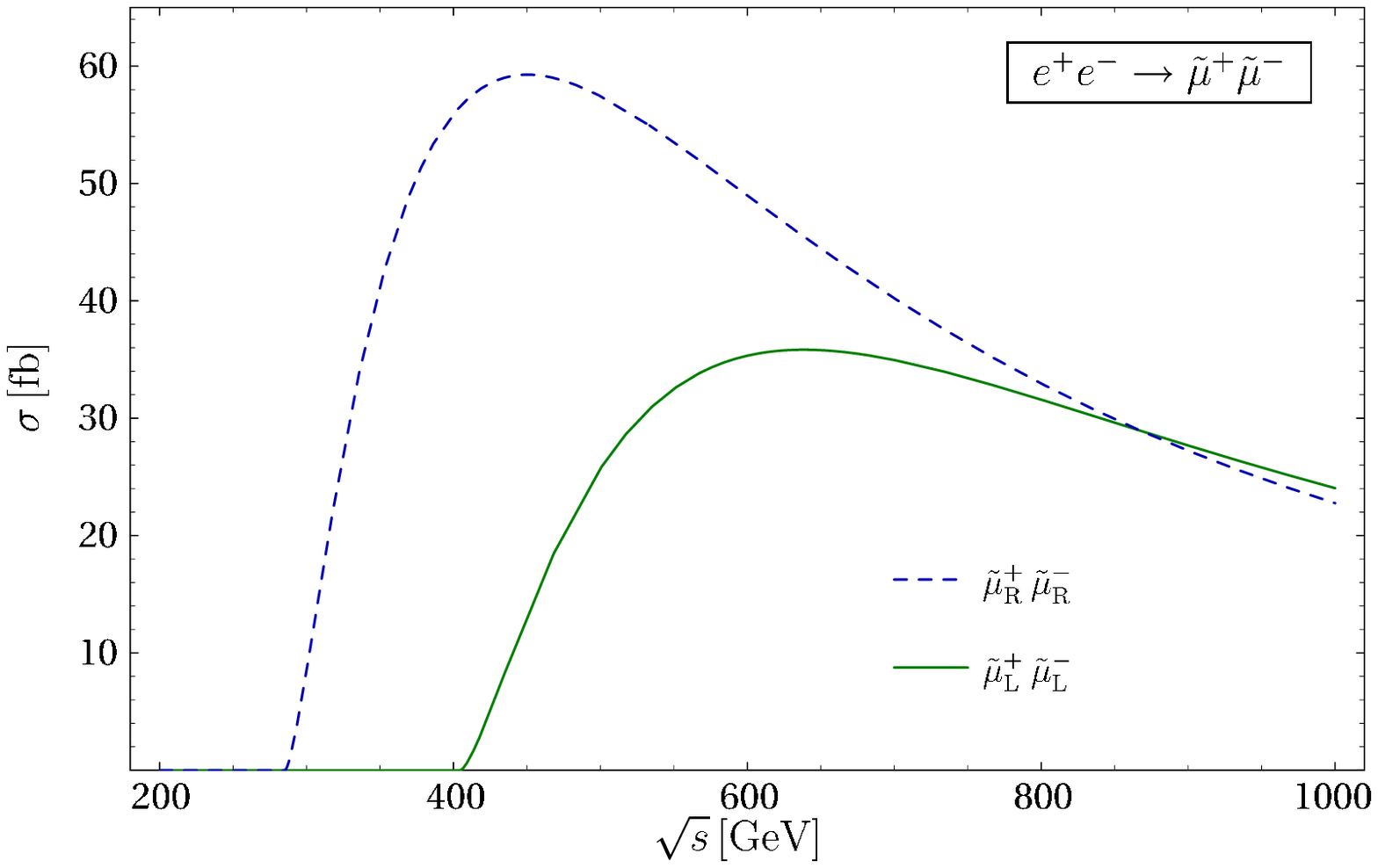,width=6in}
\vspace{-1ex}
\mycaption{Born cross-sections for right- and left-chiral smuon pair production in
unpolarized $e^+e^-$ annihilation. 
}
\label{fig:smuonXsec}
\end{figure}
The size of the cross-sections for smuons in the mSUGRA Snowmass point
SPS1a runs up to 35 fb  and 60 fb for left and right-chiral pairs,
cf.\ Fig.~\ref{fig:smuonXsec}. 
The cross-sections generally reach a maximum at
$s \simeq 10 \, \msmu{}^2$, 
while for asymptotically 
large energies they scale as $1/s$.

\underline{{\bf Scalar electrons}} can be produced, besides the standard 
photon and
$Z$-boson s-channel exchanges, via neutralino $\neu_j$ [$j=1\dots 4$] exchanges
in the t-channel, cf.\ Tab.~\ref{tab:process} and Fig.~\ref{fig:cdiag}~(b,c),
thereby generating, in addition to the diagonal, also non-diagonal L/R pairs.
Moreover, diagonal and non-diagonal selectron pairs can be generated by
t-channel neutralino exchanges in $e^-e^-$ collisions, see
Fig.~\ref{fig:cdiag}~(c).

In contrast to the vectorial s-channel amplitudes, the t-channel neutralino
exchanges allow for S-wave production at the thresholds. To reduce the total
angular momentum
to zero, electron and positron beams are required with equal helicities. If
the helicities are opposite, standard vectorial exchanges give rise to the
familiar P-wave states, cf.\ Tab.~\ref{tab:process}.

The S-wave production processes are particularly appealing for selectron mass
measurements in threshold scans due to the steep onset of the cross-sections
$\propto\beta$.
In $e^+e^-$ collisions only mixed selectron pairs, $\seL\seR$, can be
produced in an S-wave, while S-wave production of diagonal selectron pairs,
$\seR\seR$ and $\seL\seL$, is possible in the $e^-e^-$ mode.
Moreover, $e^-e^-$ collisions provide a nearly background-free environment for
selectron studies.

The Born formulae for selectron production by polarized beams read
\begin{align}
\sigma[e_{-i}^+ \, e_i^- \to \se_i^+ \, \se_i^-] &=
  \frac{2 \pi \alpha^2}{3s} \beta^3
  \biggl [ 1 + g_i^2 \frac{s}{s-\MZ^2} \biggr ]^2
\nonumber \\
&\;+ \frac{16 \pi \alpha^2}{s}
 \sum_{j=1}^4 \sum_{k=1}^4 |X_{ij}|^2 \, |X_{ik}|^2 \, h^{jk} \label{eq:seBorn}
 \\
&\;+ \frac{8 \pi \alpha^2}{s}
 \sum_{j=1}^4 |X_{ij}|^2 \left[ 1+ g_i^2 \, \frac{s}{s-\MZ^2} \right] f^j
 && [i = \LL/\RR, -i = \RR/\LL], \nonumber
\displaybreak[0] \\[1ex]
\sigma[e_i^+ \, e_{-i}^- \to \se_i^+ \, \se_i^-] &=
  \frac{2 \pi \alpha^2}{3s} \beta^3
  \biggl [ 1 + g_i \, g_{-i} \frac{s}{s-\MZ^2} \biggr ]^2
 && [i = \LL/\RR, -i = \RR/\LL],
\displaybreak[0] \\[2ex]
\sigma[\eL^+ \, \eL^- \to \seR^+ \, \seL^-] &= \frac{16 \pi \alpha^2}{s}
 \sum_{j=1}^4 \sum_{k=1}^4 X_{\LL j} \, X^*_{\RR j} \, X_{\RR k} \, X^*_{\LL k}
 \, H^{jk}, \\
\sigma[\eR^+ \, \eR^- \to \seL^+ \, \seR^-] &=
        \sigma[\eL^+ \, \eL^- \to \seR^+ \, \seL^-], \nonumber
\displaybreak[2] \\[1ex]
\sigma[e_i^- \, e_i^- \to \se_i^- \, \se_i^-] &= \frac{16 \pi \alpha^2}{s}
 \sum_{j=1}^4 \sum_{k=1}^4 X_{ij}^2 \, X^{*2}_{ik}
 \left[ G_+^{jk} + H^{jk} \right]
 && [i = \LL/\RR],\\
\sigma[\eL^- \, \eR^- \to \seL^- \, \seR^-] &= \frac{16 \pi \alpha^2}{s}
 \sum_{j=1}^4 \sum_{k=1}^4 X_{\LL j}^* \, X^*_{\RR j} \, X_{\LL k} \, X_{\RR k}
 \, h^{jk},
\end{align}
with
\begin{align}
f^j &= \Delta_j \beta - \frac{\Delta_j^2 - \beta^2}{2} \,
        \ln \frac{\Delta_j + \beta}{\Delta_j - \beta}, \\[1ex]
h^{jk} &= \left\{
\begin{array}{ll}
  \displaystyle -2 \beta + \Delta_j \ln\frac{\Delta_j + \beta}{\Delta_j - \beta}
    \quad & j = k \\
  \displaystyle \frac{f^k - f^j}{\Delta_j - \Delta_k}
    & j \neq k
\end{array} \right. , \displaybreak[0] \\[1em]
G^{jk}_\pm &= \frac{2}{s} \; \frac{\mneu{j}\mneu{k}}{\Delta_j \pm \Delta_k}
  \left[ \ln \frac{\Delta_k + \beta}{\Delta_k - \beta} \pm
         \ln \frac{\Delta_j + \beta}{\Delta_j - \beta} \right], \\[1ex]
H^{jk} &= \left\{
\begin{array}{ll}
  \displaystyle \frac{4\beta}{s} \; \frac{\mneu{j}^2}{\Delta_j^2 - \beta^2}
    \quad & j = k \\[2ex]
  G^{ij}_- & j \neq k
\end{array} \right. ,
\end{align}
where for the case of diagonal selectron pairs, $\seR \seR$ and $\seL \seL$,
\begin{equation}
\Delta_j = \frac{2}{s} (\mse{i}^2 - \mneu{j}^2) -1
\quad {\rm and} \quad
\beta = \sqrt{1 - 4 \mse{i}^2/s}, \label{eq:kindia}
\end{equation}
while for mixed pairs
\begin{equation}
\Delta_j = \frac{1}{s} (\mse{\LL}^2 + \mse{\RR}^2 - 2 \mneu{j}^2) -1
\quad {\rm and} \quad
\beta = \frac{1}{s} \sqrt{(s-\mse{\LL}^2-\mse{\RR}^2)^2 - 4 \mse{\LL}^2 \mse{\RR
}^2)^2}.
\label{eq:kinmix}
\end{equation}
The matrix 
\begin{equation}
X_{ij} = \bigl [(\cw + g_i \sw) N_{j1} + (\sw - g_i \cw)
N_{j2} \bigr ]/\sqrt{2} \label{eq:X}
\end{equation}
accounts for the neutralino mixing with $N$ being the
neutralino mixing matrix, see \eqref{eq:neumix}.

Since the higgsino components of the neutralino states couple with the
small electron mass to the electron-selectron system, the exchange
mechanism automatically projects on the gaugino components of the
neutralino wave functions. The exchange of relatively light 
neutralinos with dominant gaugino components in the t-channel
leads in general to large production cross-sections.
Only if the neutralinos mass is nearly zero, the contributions associated with
a Majorana mass insertion in the t-channel of amplitudes with zero total spin
is suppressed.
In the reference point SPS1a, however, the masses of the gaugino-like
neutralinos are sufficiently large, being of the order of the slepton masses, to 
generate large cross-sections in all selectron channels.

\begin{figure}[p]
(a)\\[-1em]
\epsfig{file=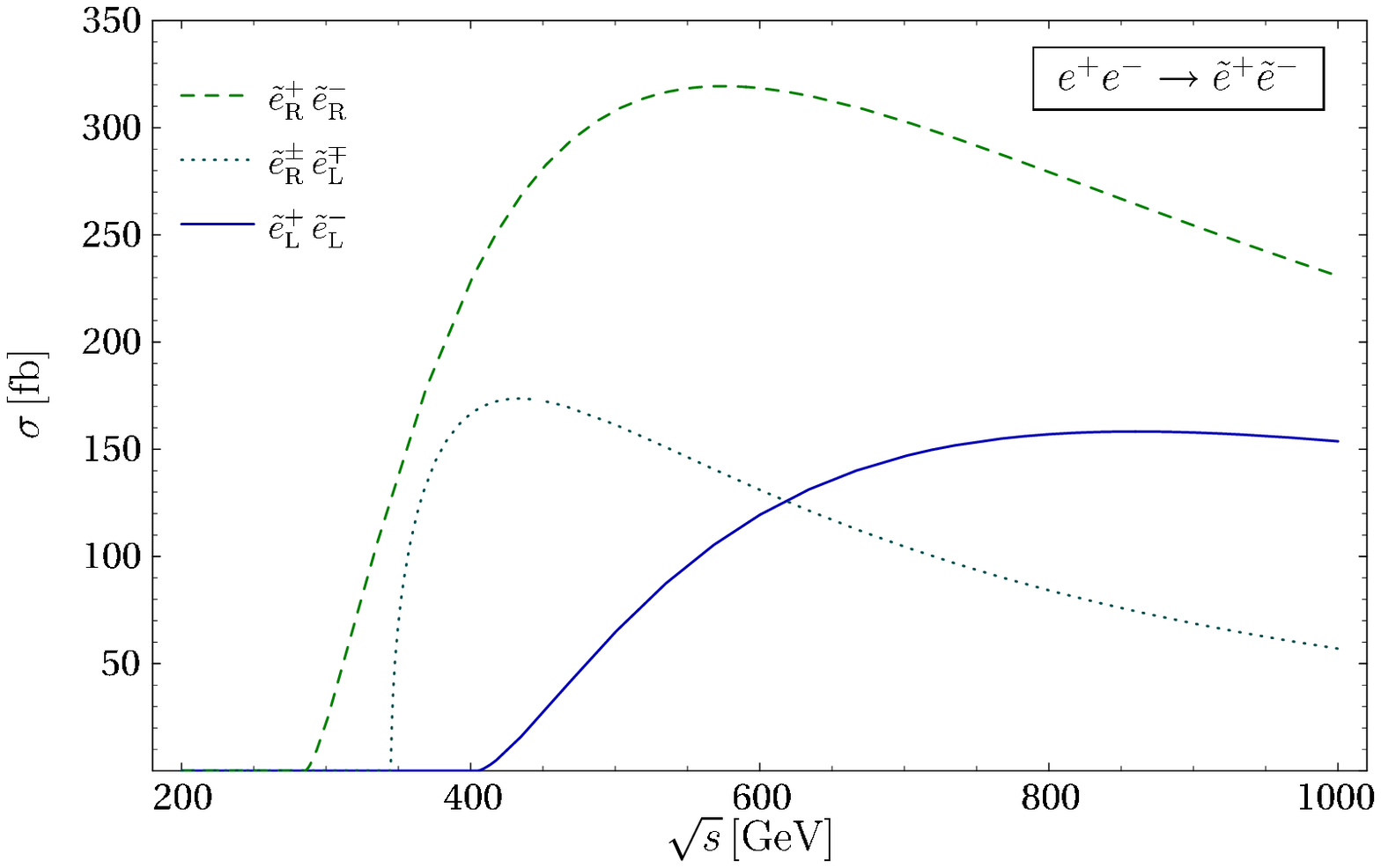,width=6in}\\[1em]
(b)\\[-1em]
\epsfig{file=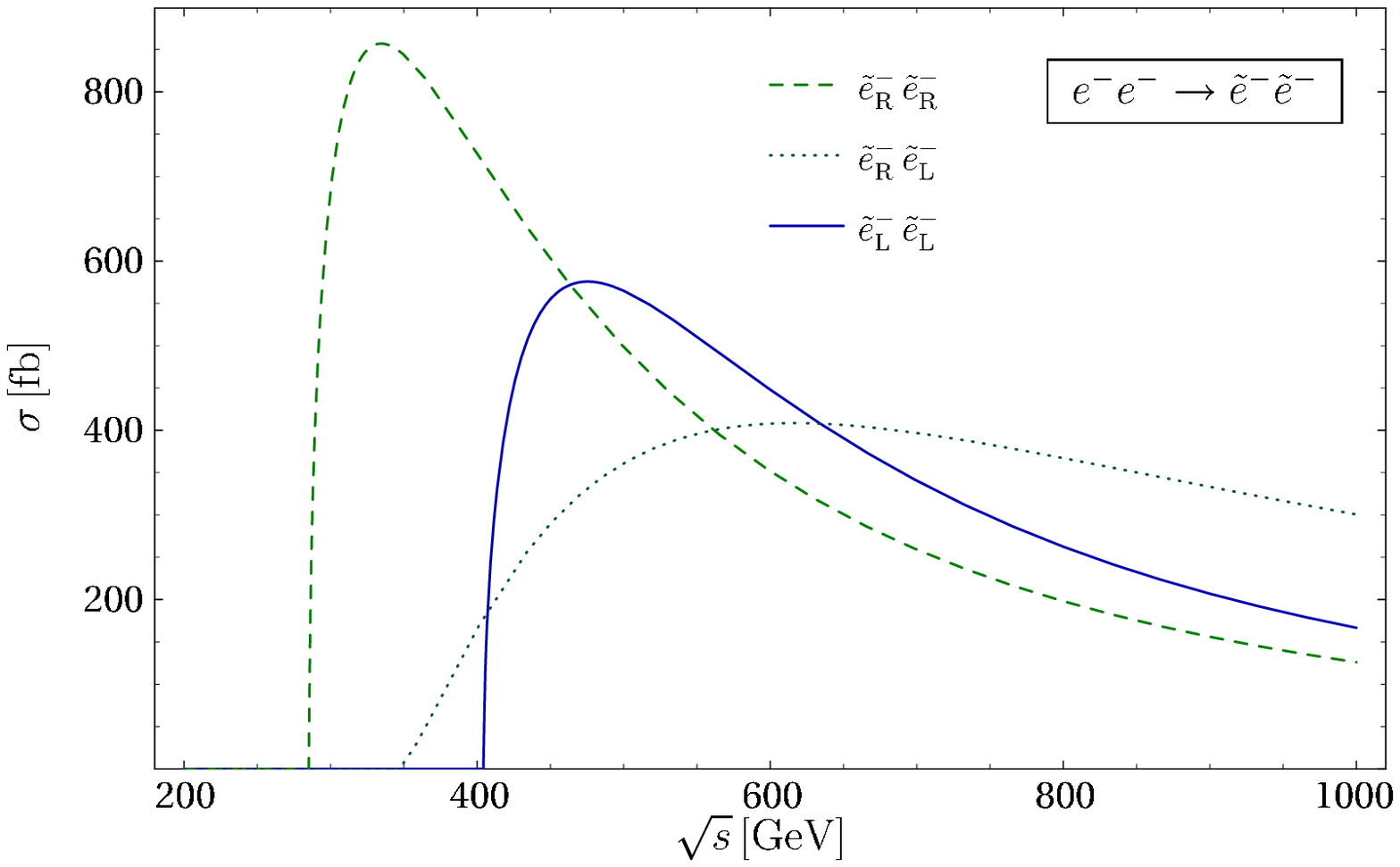,width=6in}
\mycaption{Born cross-sections for selectron pair production in unpolarized
$e^+e^-$ (a) and\newline $e^-e^-$ (b) collisions. 
}
\label{fig:selXsec}
\end{figure}
A typical set of selectron production cross-sections is shown in
Fig.~\ref{fig:selXsec}~(a/b) for $e^+e^-$ and $e^-e^-$ collisions,
respectively. As expected, the t-channel neutralino exchange enhances the
cross-sections considerably. The cross-sections exceed smuon production by
about an order of magnitude.

The angular sparticle distributions for the various production processes and
polarized beams read
\begin{align}
\frac{{\rm d}\sigma}{{\rm d}\Omega}[e_{-i}^+ \, e_i^- \to \se_i^+ \, \se_i^-] &=
  \frac{\alpha^2}{4s} \beta^3 \sin^2 \theta
  \biggl [ 1 + g_i^2 \frac{s}{s-\MZ^2} \biggr ]^2
\nonumber \\
+ \frac{4 \alpha^2}{s} \beta^3 &
 \sum_{j=1}^4 \sum_{k=1}^4 |X_{ij}|^2 \, |X_{ik}|^2 \, 
 \frac{\sin^2 \theta}{\bigl[\Delta_j - \beta\cos\theta\bigr]
 			\bigl[\Delta_k - \beta\cos\theta\bigr]}
 \\
+ \frac{2 \alpha^2}{s} \beta^3 &
 \sum_{j=1}^4 |X_{ij}|^2 \left[ 1+ g_i^2 \, \frac{s}{s-\MZ^2} \right]
 \frac{\sin^2 \theta}{\Delta_j - \beta\cos\theta}
 \qquad [i = \LL/\RR, -i = \RR/\LL], \nonumber
\displaybreak[0] \\[1ex]
\frac{{\rm d}\sigma}{{\rm d}\Omega}[e_i^+ \, e_{-i}^- \to \se_i^+ \, \se_i^-] &=
  \frac{\alpha^2}{4s} \beta^3 \sin^2 \theta
  \biggl [ 1 + g_i \, g_{-i} \frac{s}{s-\MZ^2} \biggr ]^2
 \qquad\qquad\: [i = \LL/\RR, -i = \RR/\LL],
\displaybreak[0] \\[2ex]
\frac{{\rm d}\sigma}{{\rm d}\Omega}[\eL^+ \, \eL^- \to \seR^+ \, \seL^-] &=
\frac{16 \alpha^2}{s} \beta
 \sum_{j=1}^4 \sum_{k=1}^4 X_{\LL j} \, X^*_{\RR j} \, X_{\RR k} \, X^*_{\LL k}
 \, \frac{\mneu{j}\mneu{k}/s}{\bigl[\Delta_j - \beta\cos\theta\bigr]
 			\bigl[\Delta_k - \beta\cos\theta\bigr]}, \\
\frac{{\rm d}\sigma}{{\rm d}\Omega}[\eR^+ \, \eR^- \to \seL^+ \, \seR^-] &=
        \frac{{\rm d}\sigma}{{\rm d}\Omega}[\eL^+ \, \eL^- \to \seR^+ \, \seL^-], \nonumber
\displaybreak[2] \\[1ex]
\frac{{\rm d}\sigma}{{\rm d}\Omega}[e_i^- \, e_i^- \to \se_i^- \, \se_i^-] &=
\frac{16 \alpha^2}{s} \beta
 \sum_{j=1}^4 \sum_{k=1}^4 X_{ij}^2 \, X^{*2}_{ik}
 \frac{4 \Delta_j \Delta_k \, \mneu{j}\mneu{k}/s}{\bigl[
 				\Delta_j^2 - \beta^2\cos^2\theta\bigr]
 			    \bigl[\Delta_k^2 - \beta^2\cos^2\theta\bigr]}
 \quad \raisebox{1ex}{[i = \LL/\RR],} \nonumber \\[-1.5em]
 \\
\frac{{\rm d}\sigma}{{\rm d}\Omega}[\eL^- \, \eR^- \to \seL^- \, \seR^-] &= 
\frac{4 \alpha^2}{s} \beta^3
 \sum_{j=1}^4 \sum_{k=1}^4 X_{\LL j}^* \, X^*_{\RR j} \, X_{\LL k} \, X_{\RR k}
 \, \frac{\sin^2 \theta}{\bigl[\Delta_j - \beta\cos\theta\bigr]
 			\bigl[\Delta_k - \beta\cos\theta\bigr]},
\end{align}
with $\theta$ being the angle between the incoming $e^-$ and the outgoing
$\se^-$ particles and $\Delta_j$ defined in
eqs.~\eqref{eq:kindia}, \eqref{eq:kinmix}, respectively.
Near the thresholds the angular sparticle distributions are
$\propto\sin^2\theta$ for P-waves while S-wave distributions are isotropic.
With rising center-of-mass energy, the t-channel neutralino exchange however
accumulates the selectrons in the forward and backward directions as the
exchange amplitudes peak near $\cos\theta \approx \pm 1$:
\begin{align}
\frac{{\rm d}\sigma}{{\rm d}\cos\theta}[e^+ \, e^- \to \se_i^+ \, \se_i^-]
 &\propto \sum_{j,k}
 \frac{1-\cos^2 \theta}{\bigl[\Delta_j - \beta
 \cos\theta\bigr] \bigl[\Delta_k - \beta
 \cos\theta\bigr]}
 \stackrel{s \gg \mse{i}^2}{\longrightarrow}
 \frac{1+\cos\theta}{1-\cos\theta} ,
\\
\frac{{\rm d}\sigma}{{\rm d}\cos\theta}[e^+ \, e^- \to \seR^\pm \, \seL^\mp]
 &\propto \sum_{j,k}
 \frac{1}{\bigl[\Delta_j - \beta
 \cos\theta\bigr] \bigl[\Delta_k - \beta
 \cos\theta\bigr]}
 \stackrel{s \gg \mse{i}^2}{\longrightarrow}
 \frac{1}{(1-\cos\theta)^2} . 
\end{align}

\subsection{Decay Mechanisms}

The R-sleptons $\smuR$ and $\seR$ are expected to decay
predominantly into the lightest neutralino 
if the latter has a dominant bino component:
$\slR^\pm \to l^\pm \neu_1$.

The dominant decay modes of the L-sleptons $\smuL$ and $\seL$ in the
SPS1a scenario are also
expected to be decays to the lightest neutralino. However, additional heavy
neutralino cascade decays and decays to charginos generate more complicated
final states \cite{spsval}. The tree-level decay widths for these two-particle
decays are given by
\begin{align}
\Gamma[\sll^-_i \to l^- \, \neu_j] &= \alpha \, |X_{ij}|^2 \; \msl{i}
 \Biggl ( 1-\frac{\mneu{j}^2}{\msl{i}^2} \Biggr )^{\!\!2}
\qquad [i = \LL/\RR, \; j = 1\dots4], \label{eq:neudec} \\
\Gamma[\sll^-_\LL \to \nu_l \, \cha^-_k] &= \frac{\alpha}{4} |U_{k1}|^2 \, \msl{\LL}
 \Biggl ( 1-\frac{\mcha{j}^2}{\msl{\LL}^2} \Biggr )^{\!\!2}
\qquad [k = 1,2], \label{eq:chadec}
\end{align}
where $X$ denotes the matrix defined in eq.~\eqref{eq:X}, 
while $U$ is the chargino mixing matrix defined in eq.~\eqref{eq:chamix}.

\renewcommand{\arraystretch}{1.2}
\begin{table}[tb]
\begin{center}
\begin{tabular}{|c||l|r@{\:}ll|}
\hline
Sparticle & Mass $m$ [GeV] & \multicolumn{3}{c|}{Decay modes} \\
 & Width $\Gamma$ [GeV] & & & \\
\hline \hline
$\slR = \seR/\smuR$ & $m = 142.72$ & $\slR^-$ & $\to l^- \, \neu_1$ & 100\% \\
        &       $\Gamma = 0.21$ & & & \\
\hline
$\slL = \seL/\smuL$ & $m = 202.32$ & $\slL^-$ & $\to l^- \, \neu_1$ & 48\% \\
        &       $\Gamma = 0.25$ && $\to l^- \, \neu_2$ & 19\% \\
        &                       && $\to \nu_l \, \cha^-_1$ & 33\% \\
\hline \hline
$\neu_1$ & $m = 96.18$ & \multicolumn{2}{c}{---} & \\
\hline
$\neu_2$ & $m = 176.62$ & $\neu_2$ & $\to \seR^\pm \, e^\mp$ & 6\% \\
        &       $\Gamma = 0.020$ && $\to \smuR^\pm \, \mu^\mp$ & 6\% \\
        &                       && $\to \stR^\pm \, \tau^\mp$ & 88\% \\
        &                       && $\to q \, \bar{q} \, \neu_1$ & 0.1\% \\
\hline
$\cha_1^\pm$ & $m = 176.06$ & $\cha^+_1$ & $\to \stR^+ \, \nu_\tau$ & 100\% \\
        &       $\Gamma = 0.014$ &&  & \\
\hline
\end{tabular}
\end{center}
\vspace{-1em}
\mycaption{Masses, widths and branching ratios of smuons, selectrons and the light
neutralino and chargino states for the reference points SPS1a \cite{sps,spsval}.}
\label{tab:sps1}
\end{table}
Masses, widths and branching ratios for the reference point SPS1a
\cite{sps,spsval} are collected
in Tab.~\ref{tab:sps1}. While R-sleptons decay almost exclusively
into light neutralinos plus leptons, the same decay modes are also dominant for
L-sleptons. 
Due to the fairly large value of $\tan\beta$ 
and, as a result, the significant stau mixing,  
charginos $\cha ^\pm _1$ and the $\neu_2$ neutralinos decay
primarily to $\tau$ final states, so that their experimental analysis is more
demanding. 
As significant rates are predicted for the decay modes of the sleptons into
$\neu_1$, we will focus the subsequent phenomenological analyses to these
exceedingly clear channels: the final states are oppositely charged leptons
plus missing energy, $e^+e^- \to l^+l^- + \Eslash$.
Decays to $\neu_2$ with subsequent $\tau$ decays can nevertheless be exploited
to discriminate between L- and R-sleptons.


\section{Slepton Production at Threshold and\\ Mass Measurements}
\label{threshold}

Smuon pairs are produced in $e^+e^-$ annihilation near threshold in P-waves as
a result of angular momentum conservation for spin-1 photon and $Z$-boson
s-channel exchanges. This leads to the $\beta^3$ behavior of the excitation
curve in the velocity of the produced particles. In contrast, t-channel
neutralino exchanges can give rise to a steep linear beta dependence of the
excitation curves for selectrons in $e^+e^-$ and $e^-e^-$ collisions,
characteristic for states with zero total angular momentum.

These rules are valid at the Born level but they are modified by the non-zero
widths of the produced resonances and by Sommerfeld rescattering effects
generated by Coulombic photon exchange between the slowly moving final-state
particles \cite{thr1}. While the non-zero widths smear out the onset of the
threshold excitation curves, Coulombic photon exchange enhances the
cross-section near threshold. For on-shell particle production, the Coulomb
correction factor is singular,
$\propto\beta^{-1}$, so that the excitation curves are enhanced
to $\beta^2$ for P-waves and they jump to non-zero values for S-waves.
For the production of unstable particles, this singular behavior is alleviated by
the off-shellness and finite width effects.

Moreover, studying off-shell production of sleptons, the calculation has
to be performed for the final states after the decays of the resonances.
Restricting ourselves to the simplest neutralino $\neu_1$ and $\neu_2$ decay
modes, the processes
\begin{align}
               e^+e^- &\to \mu^+ \mu^- \, \neu_1 \, \neu_1
\intertext{and}
               e^+e^- &\to e^+ e^- \, \neu_{1,2} \, \neu_{1,2} \\
               e^-e^- &\to e^- e^- \, \neu_{1,2} \, \neu_{1,2}
\end{align}
must be analyzed including all channels which give rise to these final states.

The decay of L-sleptons into the next-to-lightest neutralino $\neu_2$
with the subsequent decay $\neu_2 \to \tau^+ \tau^- \neu_1$ (cf.\
Tab.~\ref{tab:sps1}) can be used to distinguish them from R-sleptons, 
which predominantly decay into the lightest neutralino $\neu_1$, {\it i.e.}
$\slR^\pm \to l^\pm \, \neu_1$. This
is of particular importance since in most scenarios the R-sleptons are expected
to be lighter than the corresponding L-sleptons, so that the L-sleptons are
produced on top of a huge background of R-sleptons. For scenarios with
$\tan\beta \gesim 10$, the $\neu_2$ mainly decays into a $\tau$ pair and the
lightest neutralino, so that the production of an L-slepton is
signaled by the appearance of additional $\tau$ jets.

The small cross-section for L-smuon production together
with the branching ratio for $\smuL^\pm \to \mu^\pm \, \neu_2$ results in
expected event rates
that are too low to perform a measurement of the threshold excitation curve.
Therefore, only R-smuons, but selectrons of both L and R type
will be analyzed in detail.

\subsection{Off-shell Slepton Production}

\begin{figure}[tb]
(a) \underline{Double resonance diagram} \hspace{2ex}
(b) \underline{Single resonance diagrams}
\vspace{0.3cm} \\
\phantom{} \epsfig{file=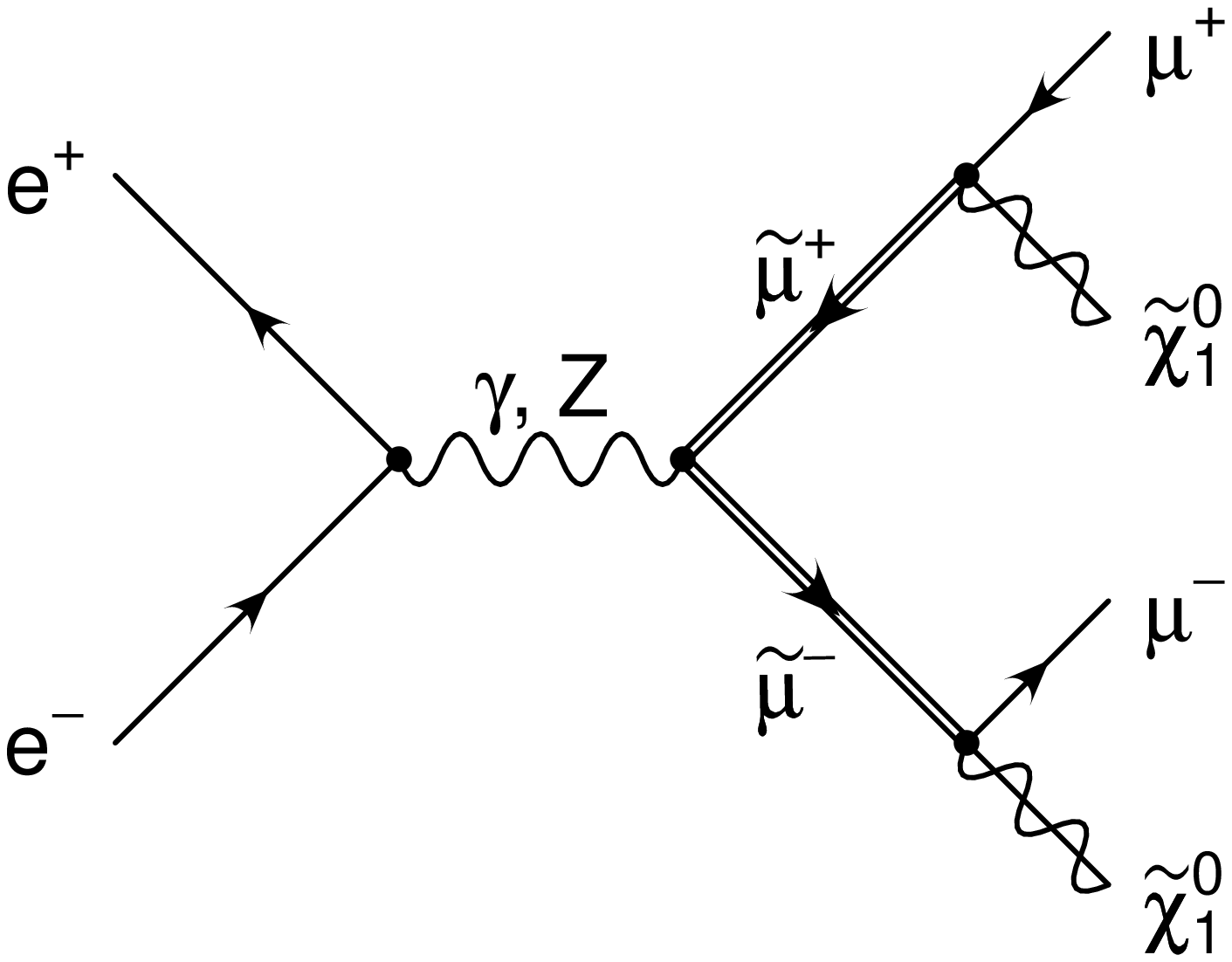,width=5cm, bb=113 240 540 544}
\hspace{5ex}
\epsfig{file=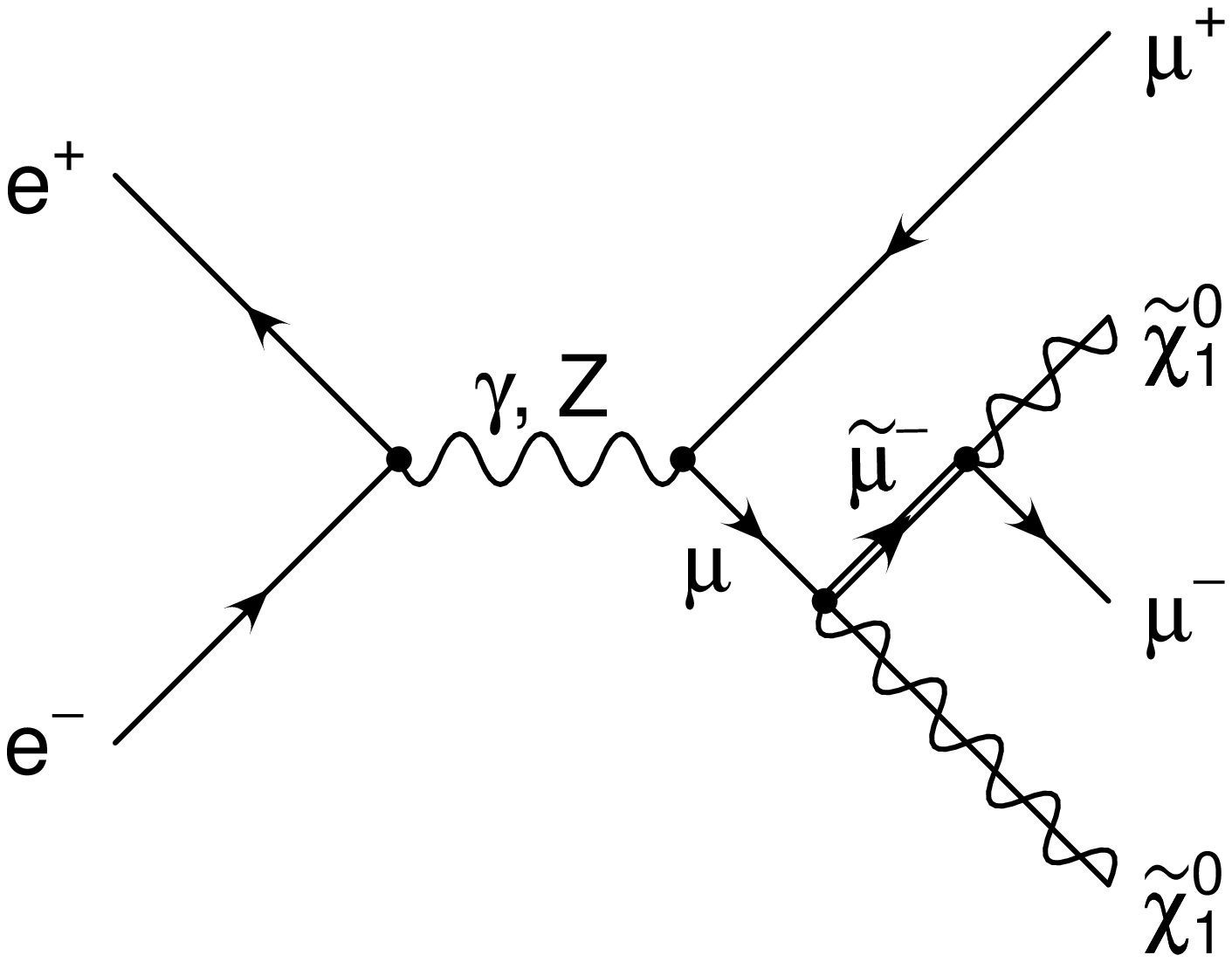,width=5cm, bb=113 240 540 544}
\epsfig{file=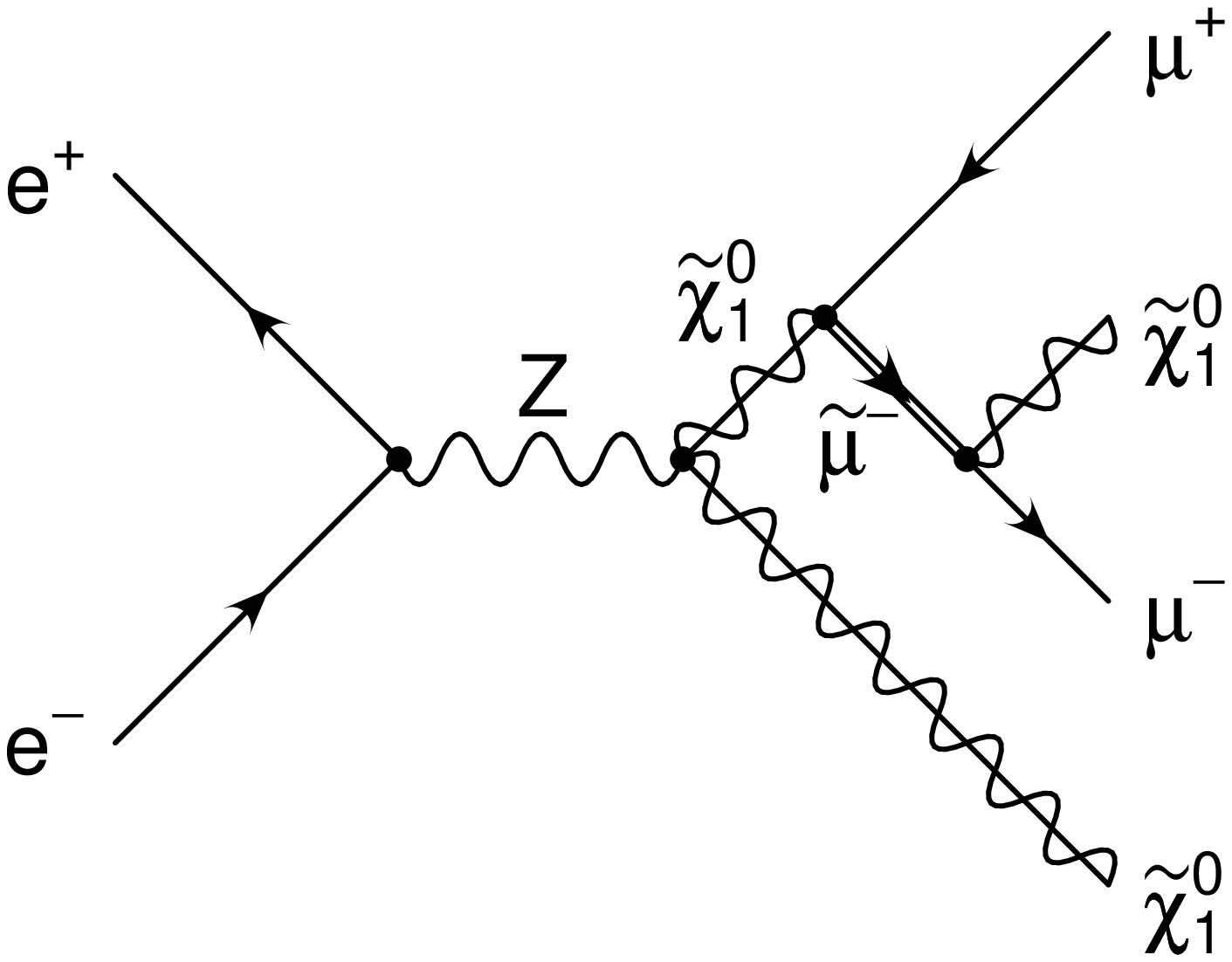,width=5cm, bb=113 240 540 544}
\mycaption{The doubly and singly resonant contributions
to the process $e^+e^- \to \mu^+\mu^- \neu_1 \neu_1$.}
\label{fig:signal}
\end{figure}
The leading contribution to the $\mu^+ \mu^- \, \neu_1 \, \neu_1$ final state,
and to electron final states correspondingly, is generated by the double
resonance diagram shown in Fig.~\ref{fig:signal}~(a). For invariant
$\mu\neu_1$ masses near the smuon mass, the smuon propagators must be replaced
by the Breit-Wigner form, which explicitly includes the non-zero width
$\Gamma_{\smu}$  of the resonance state. This is achieved by
substituting the complex parameter
\begin{equation}
m_{\smu}^2 \;\to\; M_{\smu}^2 = m_{\smu}^2 - im_{\smu}\Gamma_{\smu}
\end{equation}
for the smuon mass. To keep the amplitude gauge invariant, the double-resonance
diagram of Fig.~\ref{fig:signal}~(a) must be supplemented by the
single-resonance diagrams of Fig.~\ref{fig:signal}~(b).

\subsection{Coulombic Sommerfeld Correction}

\begin{figure}[tb]
\vspace{2ex}
\raisebox{-4mm}{\epsfig{file=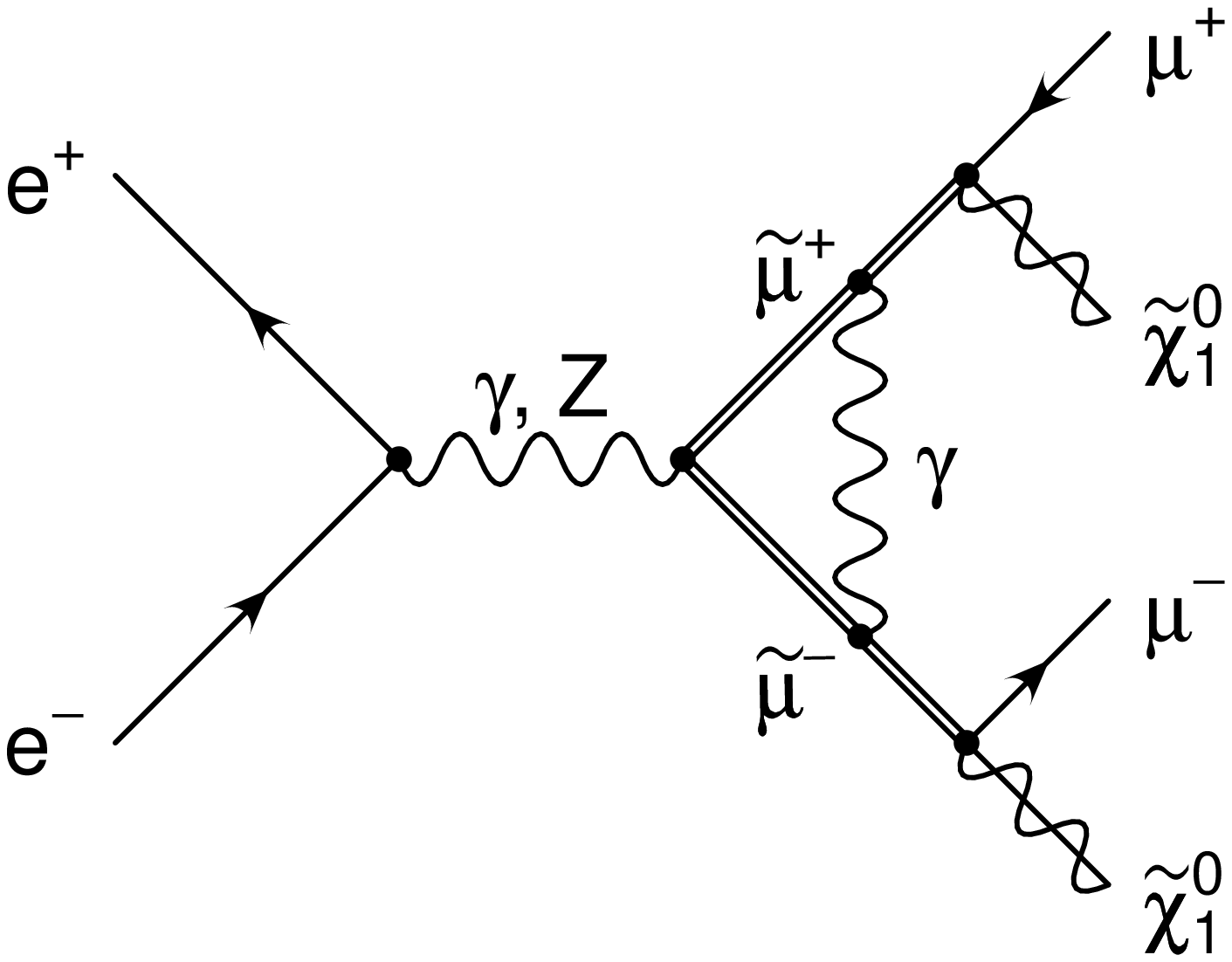,width=5cm, bb=113 240 540 544}} 
\hfill
\parbox{10cm}{\mycaption{Coulomb correction to smuon production,
$e^+e^- \to \mu^+\mu^- \neu_1 \neu_1$.}\label{fig:coulomb}}
\vspace{2ex}
\end{figure}
The Coulomb interaction due to photon exchange between slowly moving charged
particles in Fig.~\ref{fig:coulomb} gives rise to large corrections to the
excitation curve near threshold. For stable particles the cross-section is
modified universally by the singular coefficient $\sigma_{\rm Born} \to
(\alpha\pi/2\beta) \sigma_{\rm Born}$ at leading order. This Sommerfeld
correction \cite{sommerfeld} removes one power of the velocity $\beta$ off the
threshold suppression. For the production of off-shell particles the
singularity is screened \cite{FKoffs} and the remaining enhancement depends on
the orbital angular momentum $l$. 
The Coulomb correction is associated with the leading term in the $\beta$
expansion of the photon exchange diagram Fig.~\ref{fig:coulomb}.
Exploiting the fact that terms with the loop
momentum in the numerator generate terms proportional to $\beta$ in the 
diagrammatic analysis, it follows that
\begin{equation}
\sigma_{\rm Coul}^{\rm off-shell} = -\sigma_{\rm Born} \, \frac{\alpha s}{2\pi} 
\, C_0 \;\; \Re e \!
\left(\frac{2p_+p_- - 2 M_{\rm X}^2}{2p_+p_- - p_+^2 - p_-^2}\right)^l,
\end{equation}
for the complex pole masses $M^2_\pm = m_\pm^2 - i \, m_\pm \Gamma_{\!\pm}$ and
the momenta $p_\pm$ of the slepton $\sll^\pm$, $\sll=\se,\smu$.
The leading part in $\beta$ of the scalar triangle function $C_0$ can be
evaluated according to Ref.~\cite{c0coul}.
The last factor, taken to the $l$-th power, is of kinematical origin,
incorporating the effect of the angular momentum $l$ of the wave function.

\begin{figure}[tb]
\epsfig{file=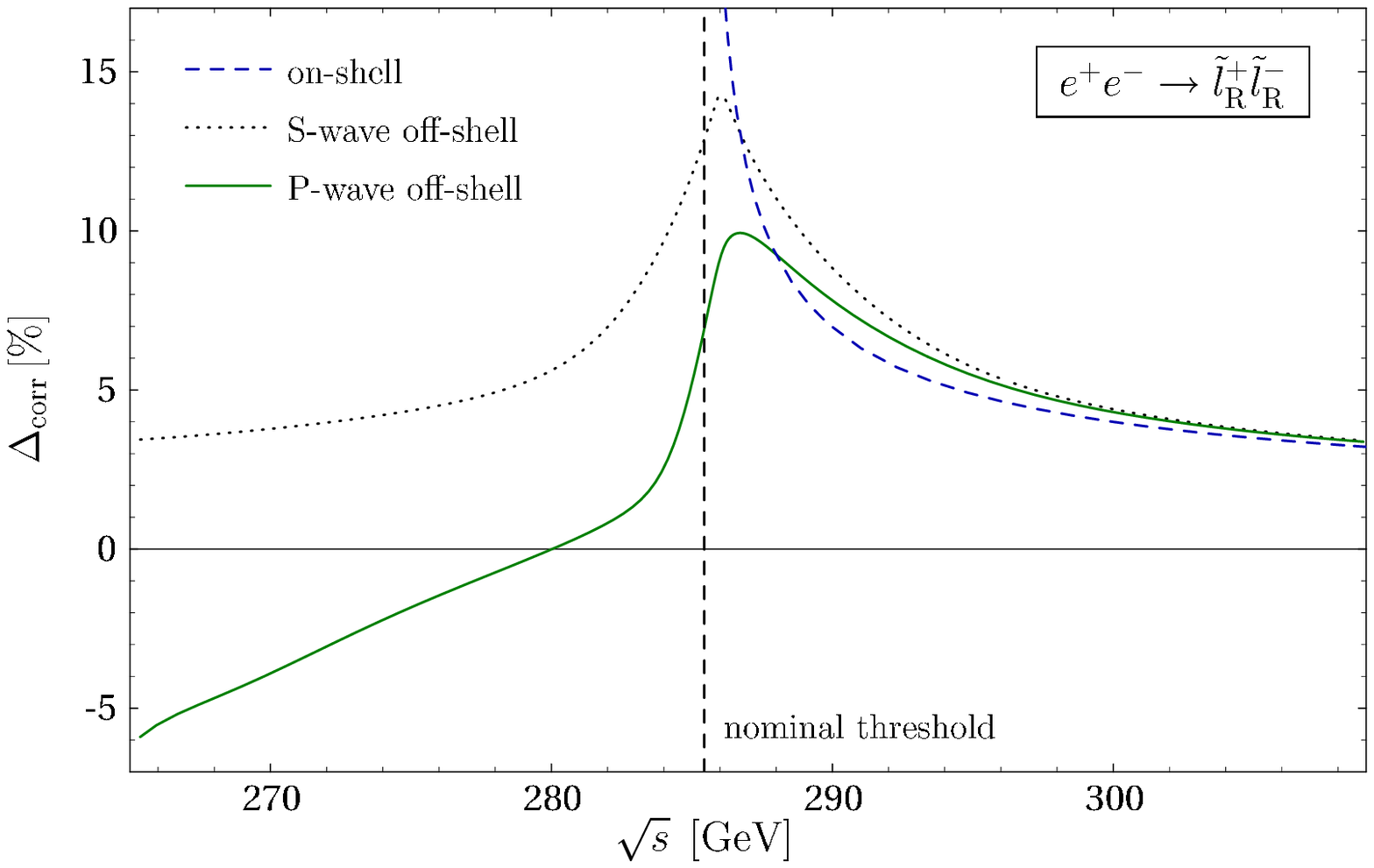,width=6in}
\mycaption{Correction factor $\Delta_{\rm corr}$ due to Coulomb rescattering
relative to the Born cross-section for on-shell and off-shell S- and P-wave
production.}
\label{fig:coulps}
\end{figure}

After carrying out the expansion for smuon and selectron P-wave production
and for selectron S-wave production, the leading order can be written in the
form
\begin{equation}
\sigma_{\rm Coul} =  \sigma_{\rm Born} \,
\frac{\alpha\pi}{2 \beta_p} 
\biggl [ 1\! -\! \frac{2}{\pi} \arctan
\frac{|{\beta_M}|^2 - \beta_p^2}{2 \beta_p \;
        \Im \! m \, \beta_M} \biggr ]
 \,\Re e
  \biggl [\frac{\beta_M^2 + \beta_p^2}{2 \beta_p^2}
  \biggr ]^l	\qquad [l = 0,1],
\end{equation}
with the generalized velocities
\begin{align}
\beta_M &= \frac{1}{s}\sqrt{(s-M_+^2-M_-^2)^2-4
M_+^2 M_-^2}, \\
\beta_p &= \frac{1}{s}\sqrt{(s-p_+^2-p_-^2)^2-4
p_+^2 p_-^2}.
\end{align}
The off-shellness damps the singularities as illustrated in
Fig.~\ref{fig:coulps} for the S- and P-waves.

\subsection{Final-state Analysis}
\label{fsana}

The final states in the general processes $e^+e^- \to l^+l^- + \Eslash$ and
$e^+e^- \to l^+l^- \, \tau^+\tau^- + \Eslash$ can be generated by a large
variety of background processes in addition to the signal slepton channels.
Within the SUSY sector itself, pair production of charginos and neutralinos
with subsequent (cascade) decays feed the final state $l^+l^- + \Eslash$. $ZZ$
and $Zh_0$/$ZH_0$ intermediate states with one particle decaying to lepton
pairs, the other to a pair of $\neu_1$ will also contribute to this class of
final states.
The final state with an additional tau pair is characteristic
for the production of a L-slepton together with a
R-slepton. SUSY backgrounds to this signature arise from neutralino production or
from stau $\st$ production with the decay chain $e^+e^- \to \st^+\st^-
\to \tau^+\tau^- \, \neu_2 \, \neu_1 \to \tau^+\tau^- \, e^+e^- \, \neu_1 \,
\neu_1$.

Moreover, pure Standard Model processes, like the production of gauge boson
pairs, $W^+W^-$, $ZZ$ and $Z\gamma^*$, leading to the final state
$l^+l^-\nu\bar{\nu}$, also have to be taken into account. They are generically
large and need to be reduced by appropriate cuts \cite{CDRMartyn}. The
background from resonant $Z$ production can easily be reduced by cutting on the
invariant mass of the lepton pair or the invisible recoil momentum around the
$Z$-pole.
Contributions from $WW$ pair production have a characteristic angular
distribution of the final state leptons. Because of the spin correlations and
the boost factor, the leptons tend to be aligned back to back and along the
beam direction. Therefore this background can be reduced effectively by
rejecting signatures with back-to-back leptons.
The explicit values for the cuts are summarized in Tab.~\ref{tab:WZ1}.
\renewcommand{\arraystretch}{1.3}
\begin{table}[tb]
\centering
\begin{tabular}{|p{5.7cm}|p{5cm}|l|}
\hline
Condition & Variable & Accepted range \\
\hline \hline
Reject leptons in forward/\newline backward region 
from Bhabha/\newline M\o ller scattering
& lepton polar angle $\theta_{
\rm l}$ &
  $|\cos \theta_{\rm l}| < 0.95$ \\
Reject soft leptons/jets from \newline radiative photon splitting and
$\gamma$-$\gamma$ background
 & lepton/jet energy $E_{\rm l}$ & $E_{\rm l} > 5$ GeV 
\\
Reject missing momentum in forward/backward region
from particles lost in the beam pipe &
  missing momentum polar \newline angle $\theta_{\vec{p}_{\rm miss}}$ &
  $|\cos \theta_{\vec{p}_{\rm miss}}| < 0.90$ \\
Angular separation of two \newline leptons & angle $\phi_{\rm l^+l^-}$ between
leptons & $|1- \cos \phi_{\rm l^+l^-}| > 0.002$ \\
Angular separation of two quark or tau jets & angle $\phi_{\rm jj}$ between
jets & $|1- \cos \phi_{\rm jj}| > 0.015$ \\
\hline
Cut on $Z$ decaying into \newline lepton pair &
  di-lepton invariant mass \newline $m_{\rm l^+l^-}$ &
  $|m_{\rm l^+l^-} - \MZ| > 10$ GeV \\
Cut on invisibly decaying $Z$ &
  invariant recoil mass $m_{\rm recoil}$ &
  $|m_{\rm recoil} - \MZ| > 15$ GeV \\
\hline
Reject back-to-back \newline leptons from $W$ pairs &
  angle $\phi_{\rm l^+l^-}$ between leptons &
  $|\cos \phi_{\rm l^+l^-}| < 0.7$ \\
\hline
\end{tabular}
\mycaption{General cuts to reduce large Standard Model backgrounds and
to account for the detector geometry and resolution.}
\label{tab:WZ1}
\end{table}
\renewcommand{\arraystretch}{1}

Triple gauge boson production, $W^+W^-Z$ and $W^+W^-\gamma^*$, contributes to
the final state $e^+e^- \to l^+l^- \, \tau^+\tau^- + \Eslash$. The total
cross-section for these processes is well below 1 fb \cite{3gaugebos} and can be
reduced further by applying cuts on the invariant di-lepton mass.

The dominant supersymmetric backgrounds involve decay cascades of neutralinos
and charginos that, for example following the decay chain
\begin{equation}
\begin{aligned}
e^+e^- \to \neu_1 \, &\neu_j\\[-.5ex]
&\knickpfeil l^+ \, l^- \, \neu_1,
\end{aligned}
\end{equation}
with $j>1$, generate $l^+l^- + \Eslash$ final states.
Since the lepton pair originates only from a single neutralino decay, these
backgrounds give rise to increased missing energy and lower lepton-pair
invariant mass compared to the signal, and they can effectively be reduced by cuts
on these two variables \cite{thr1}. Near threshold an alternative method for
reducing the backgrounds can be applied, based on the fact that the energy
of the leptons originating from a two-body decay is defined sharply in this
kinematical configuration. Thus by
selecting leptons with energies in a band $\Delta E \approx 10$ GeV around the
nominal threshold energy $E_{l,\rm thr} = (\msl{}^2 - \mneu{j}^2)/(2\msl{})$
greatly suppresses both SM and SUSY backgrounds.
This second cut choice is applied in the following examples.

The signal-to-background ratio can further be enhanced by using beam
polarization. The optimal polarization choices for the different
production processes are listed in Tab.~\ref{tab:process}. As evident from the
table, polarization of both the electron and positron beams can help to
discriminate between the slepton chiralities (see {\it e.g.} \cite{gudichep}).
In the following 80\% polarization for the electrons and 50\% polarization for
the positrons is assumed.
Without positron polarization the signal-to-background ratio would 
in general be reduced by a factor of 1.5.

For a realisitic description, it is necessary to include initial-state
radiation (ISR) and beamstrahlung effects. The leading logarithmic ISR
contributions are included using the structure-function method \cite{structf},
while effective beamstrahlung parametrizations, taken for the {\sc Tesla}
design for definiteness, are adopted from the program \textsl{Circe} \cite{circe}.

\begin{figure}[tb]
\hspace{-6mm}
\epsfig{file=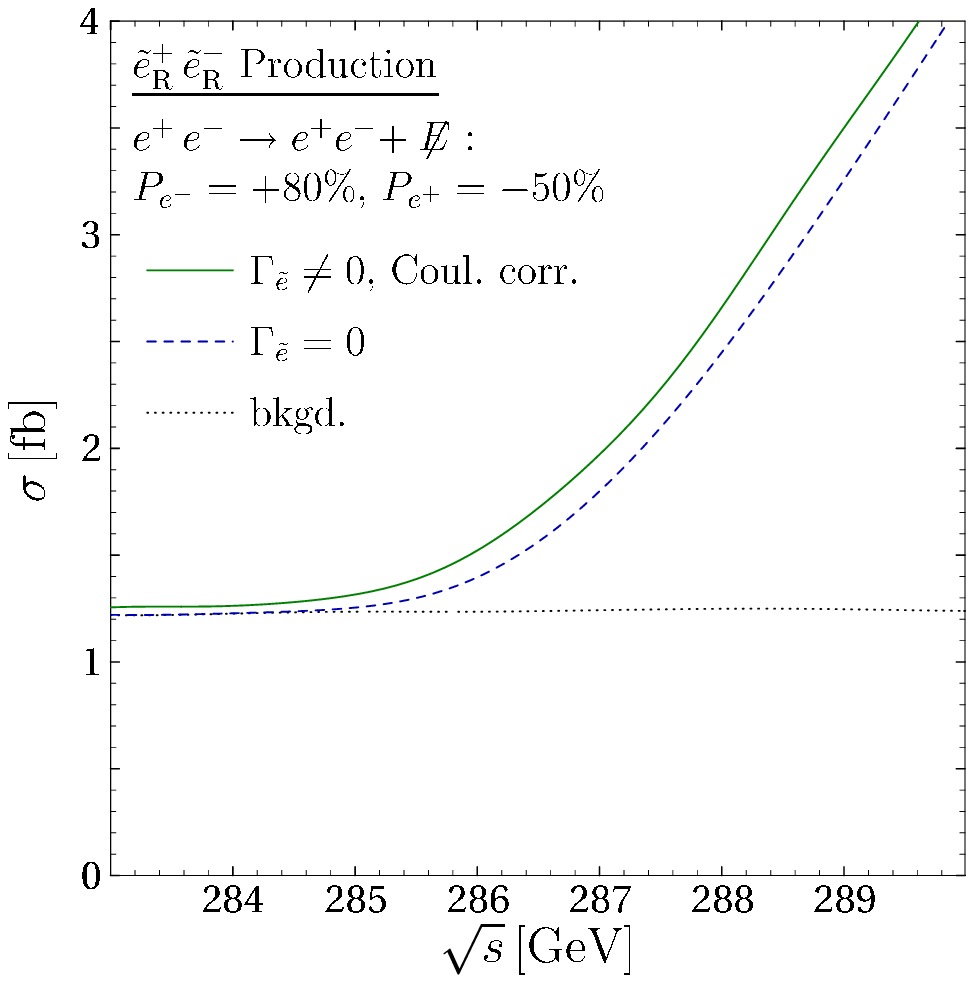,width=8.5cm}
\hspace{-3mm}
\epsfig{file=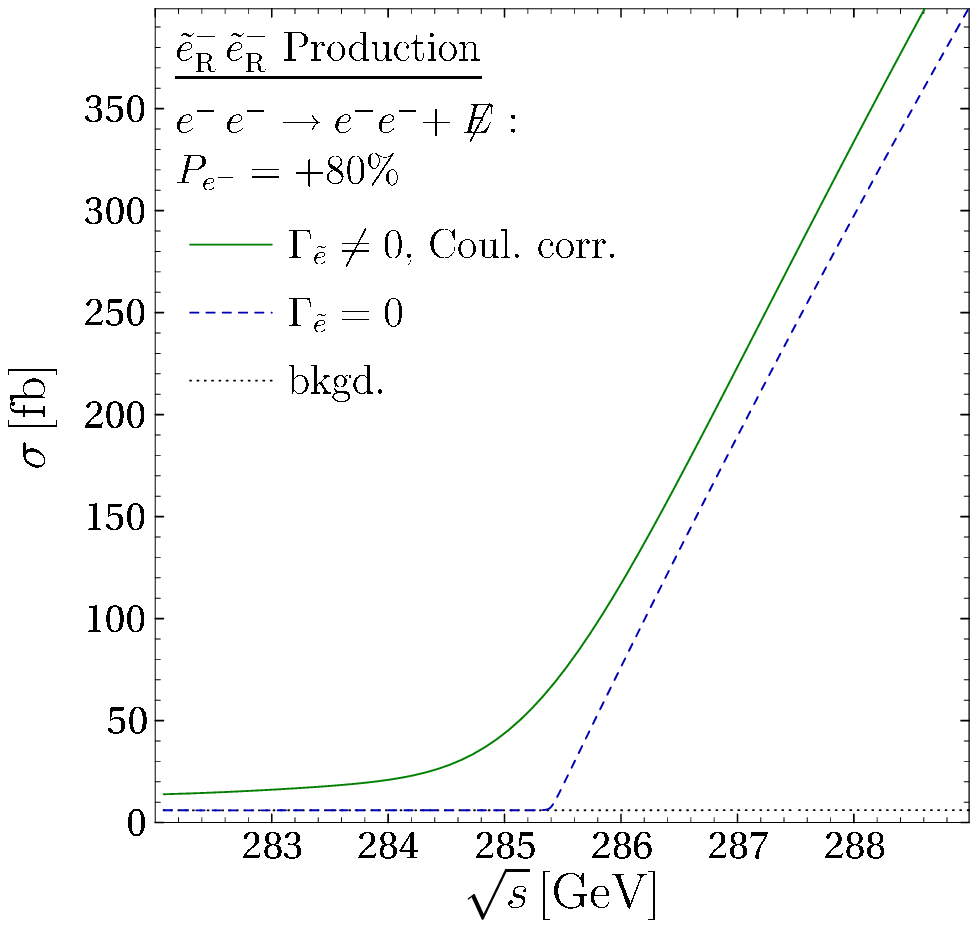,width=8.9cm}
\mycaption{The excitation curves for $\seR$ pair production over Standard Model and
supersymmetric
backgrounds for $e^+e^-$ annihilation (left) and $e^-e^-$ scattering (right).
The signal contribution with non-zero widths and Coulomb
rescattering is compared with the case of zero width and no higher order
corrections. The signal is enhanced with beam polarization as indicated, where
$(+)$ corresponds to right-handed and $(-)$ to left-handed polarization.
}
\label{fig:thr}
\end{figure}
Including the SUSY and SM backgrounds, the excitation curves, after the
beamstrahlung and ISR is switched on and the cuts defined before are applied, are
displayed in Fig.~\ref{fig:thr}
for two characteristic examples, $\seR^+\seR^-$ pair
production in $e^+e^-$ collisions as a P-wave process, 
and $\seR^-\seR^-$ pair production in $e^-e^-$
collisions as a typical S-wave process. Separately shown are the zero-width
Born prediction, the background contributions and the final prediction
including non-zero width and rescattering effects, with the backgrounds
added on.

\renewcommand{\arraystretch}{1.5}
\begin{table}[tb]
\centering{
\begin{tabular}{|l|ll|}
\hline
Process & \multicolumn{2}{l|}{Fitted values for slepton mass $m$ and width
$\Gamma$}\\
\hline \hline
$e^+e^- \to (\seR^+\seR^-) \to e^+e^- + \Eslash$ &
  $\mse{\RR} = 142.8^{+0.21}_{-0.19} \gev$
& $\Gamma_\seR = 150^{+300}_{-250} \mev$ \\
\hline
$e^-e^- \to (\seR^-\seR^-) \to e^-e^- + \Eslash$ &
  $\mse{\RR} = 142.70^{+0.048}_{-0.053} \gev$ 
&  $\Gamma_\seR = 200^{+50}_{-40} \mev$ \\
\hline\hline
$e^+e^- \to (\seR^\pm\seL^\mp) \to e^+ e^- \, \tau^+\tau^- + \Eslash$ &
  $\mse{\LL} = 202.2^{+0.37}_{-0.33} \gev$ 
&  $\Gamma_\seL = 240^{+20}_{-20} \mev$ \\
\hline
$e^-e^- \to (\seL^-\seL^-) \to e^- e^- \, \tau\tau\tau\tau + \Eslash$ &
  $\mse{\LL} = 202.1^{+0.62}_{-0.44} \gev$ 
&  $\Gamma_\seL = 240^{+500}_{-240} \mev$ \\
\hline \hline
$e^+e^- \to (\smuR^+\smuR^-) \to \mu^+\mu^- + \Eslash$ &
  $\msmu{\RR} = 142.8^{+0.42}_{-0.38} \gev$ 
&  $\Gamma_\smuR = 350^{+400}_{-400} \mev$ \\
\hline
\end{tabular}
}
\mycaption{Expected precision for the determination of slepton masses and widths
from threshold scans in $e^+e^-$ and $e^-e^-$ scattering. The reconstructed 
values are obtained from a four-parameter fit as outlined in the text.
}
\label{tab:massres}
\end{table}
\renewcommand{\arraystretch}{1}
The results expected from these simulations for the mass measurements are
presented in Tab.~\ref{tab:massres}. They are based on
data simulated at five equidistant points in a center-of-mass energy range of
10 GeV in the threshold regions for $\smuR$ pair production, and diagonal and
non-diagonal $\seR$ and $\seL$ production.  For the $e^+e^-$ mode a total
luminosity of 50 fb$^{-1}$ for each threshold scan is assumed, corresponding to
10 fb$^{-1}$ per scan point. In the $e^-e^-$ mode the anti-pinch effect leads
to a somewhat reduced machine luminosity. Therefore it is presumed that a total
of 5 fb$^{-1}$ is available for each scan measurement, corresponding to 1
fb$^{-1}$ per scan point. 
For the reconstruction of the mass a binned likelihood method is employed, using
four free parameters in the fit: the slepton mass and width, a constant
scale factor for the absolute normalization
of the excitation curve and a constant background
level\footnote{Since the remaining backgrounds after cuts are flat, they can
effectively be approximated by a constant.}. 
The last two parameters render the mass fit independent on details of other SUSY
sectors, in particular the masses and mixings of the heavier neutralinos that
are not accessible in the slepton decays.

Evidently, S-wave $\seR$ production in $e^-e^-$ collisions provides us with
mass measurements of 50 MeV, {\it i.e.} a relative error of less than 1
per-mille. This will presumably be the highest accuracy that can ever be
reached for sfermion mass measurements in the supersymmetric particle sector.


\section{Slepton Production in the Continuum and\\ Determination
         of Yukawa Couplings} 
\label{continuum}

The motivation for high-precision analyses of smuon and selectron
production in the continuum is twofold, different though for the
two species.

{\bf 1.)} Smuon pair production in the continuum serves as a rich source of
particles which after subsequent decays to muons and neutralinos allows us to
determine the smuon and neutralino masses. The smuon mass measurement in the
continuum is competitive with the accuracy expected from threshold scans. The
R-smuon decay to the lightest neutralino will be the gold-plated process
(besides the analogous R-selectron decay) for measuring the mass of the lightest
neutralino, which is a key
particle in cosmology. Smuon pair production leads to clean
$\mu^+\mu^-+\Eslash$ final states that can easily be identified and controlled
experimentally. The measurement of the cross-sections therefore provides a
valuable instrument for testing supersymmetry dynamics at the quantum level.

{\bf 2.)} Since the precision of threshold scans in mass measurements 
of selectrons
cannot be rivaled, the central target of selectron pair production in the
continuum, besides the neutralino mass measurement
\cite{Tsukamoto:1995gt,martynco},
is the analysis of the
selectron-electron-neutralino Yukawa couplings in the SU(2) and U(1) sectors
\cite{eYuk}.
They are predicted to be equal to the corresponding gauge couplings in
supersymmetric theories, even if the supersymmetry breaking is included by soft
terms in the Lagrangian, leaving the system theoretically self-consistent.
The relevant mechanism involves the t-channel exchange of neutralinos.
Knowledge of the neutralino masses and mixing parameters is therefore required
before high-sensitivity tests can be carried out. Thus this method is
one of the components in a complex experimental program including, in
addition to the analysis of selectron pair production, also pair production of
charginos and neutralinos that in turn are (partly) mediated by neutral and charged
slepton t-channel exchanges. The synopsis of all these channels will
finally provide us with a comprehensive and detailed picture of the entire
Yukawa sector in a model-independent form.

We will separate in this report the description of the theoretical
techniques necessary for controlling the higher-order corrections,
from the application to smuon and selectron pair production and their
phenomenological evaluation, with emphasis on the analysis of Yukawa
couplings.

\subsection{Renormalization of the MSSM}

At the one-loop level, which we work out in this report for slepton
production, dimensional reduction (DRED) provides us with a valid
regularization scheme including chiral currents%
\footnote{For the production of smuons, involving only gauge couplings at
tree level, the calculation has been repeated independently in dimensional
regularization and perfect agreement has been found \cite{thesis}.}.
By reducing the kinematics
in the propagation of particles to $D<4$ dimensions but leaving the number of
field components unchanged, supersymmetry is preserved in the higher-order
amplitudes, and so is gauge invariance \cite{dred}.

Multiplicative renormalization of masses, couplings and fields can
therefore be performed without introducing additional ad-hoc counter terms
to restore the supersymmetry. The renormalization factors $Z$,
and equivalently the shifts of variables, that absorb
all the ultra-violet divergences, will be fixed by on-shell renormalization,
{\it i.e.} the on-shell definition of the physical particle masses, the on-shell
definition of the electromagnetic gauge couplings in the trilinear
lepton-lepton-photon
vertex, and normalization of the on-shell renormalized fields to unity.
As a consequence of supersymmetry and gauge symmetry,
the renormalization of all other quantities, Yukawa couplings, quartic
couplings etc., induces calculable additional shifts. 
This program can be carried out consistently in theories including soft
supersymmetry breaking terms%
\footnote{More information about the renormalization of the MSSM and
a general overview of different renormalization techniques can be
found in Ref.~\cite{majsusy}.}. 

\begin{figure}[tp]
$
\parbox[c]{2cm}{\psfig{figure=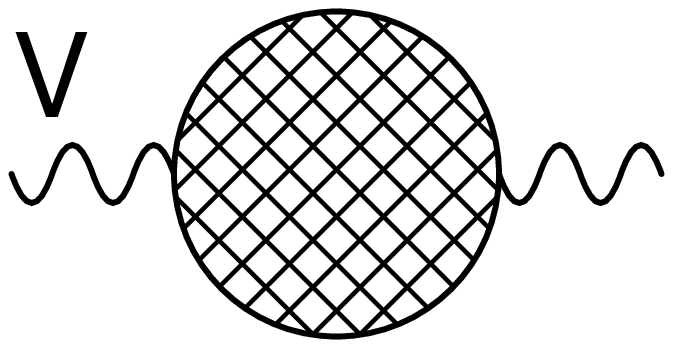, width=2cm}} =
\parbox[c]{2cm}{\psfig{figure=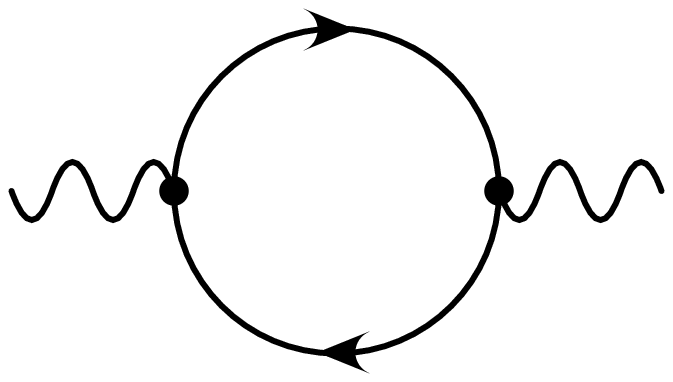, width=2cm}} +
\parbox[c]{2cm}{\psfig{figure=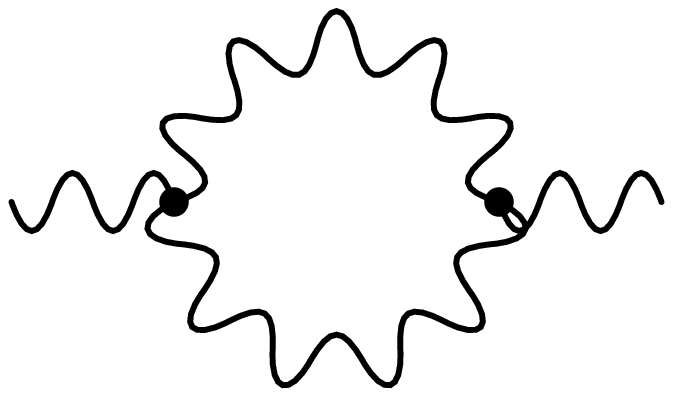, width=2cm}} +
\parbox[c]{2cm}{\psfig{figure=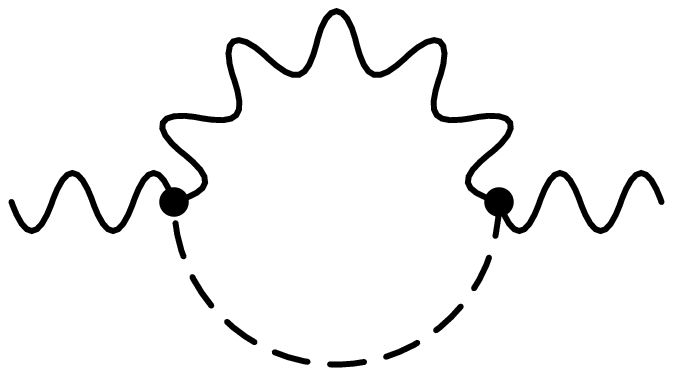, width=2cm}} +
\parbox[c]{2cm}{\psfig{figure=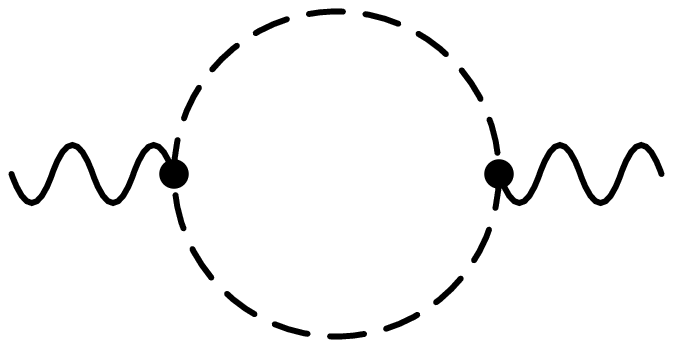, width=2cm}} +
\parbox[c]{2cm}{\psfig{figure=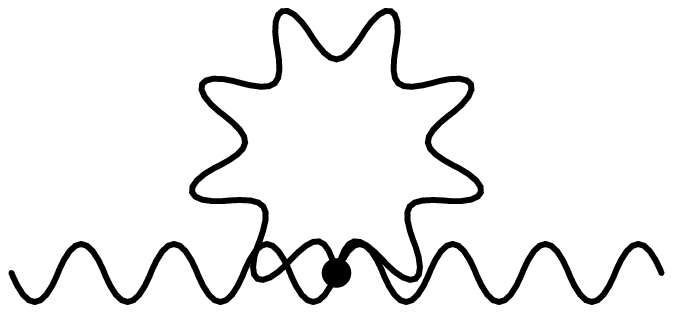, width=2cm}} +
\parbox[c]{2cm}{\psfig{figure=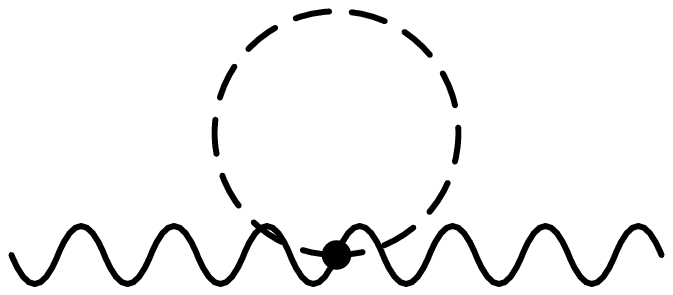, width=2cm}}
$\\[1em]
$\anc\hspace{2cm}\, +
\parbox[c]{2cm}{\psfig{figure=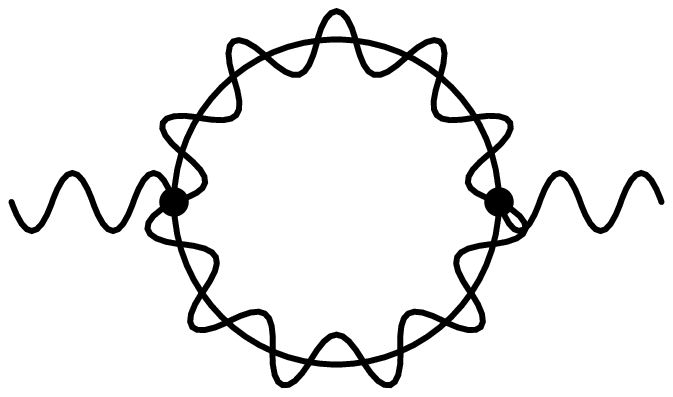, width=2cm}} +
\parbox[c]{2cm}{\psfig{figure=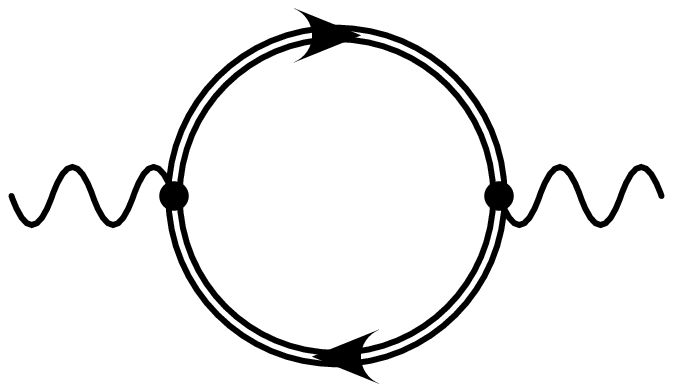, width=2cm}} +
\parbox[c]{2cm}{\psfig{figure=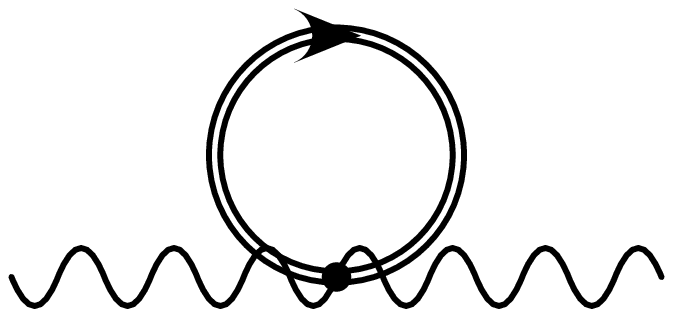, width=2cm}}
\phantom{+\hspace{2cm}+\hspace{2cm}+\hspace{2cm}\anc}
$\\[2em]
$
\parbox[c]{2cm}{\psfig{figure=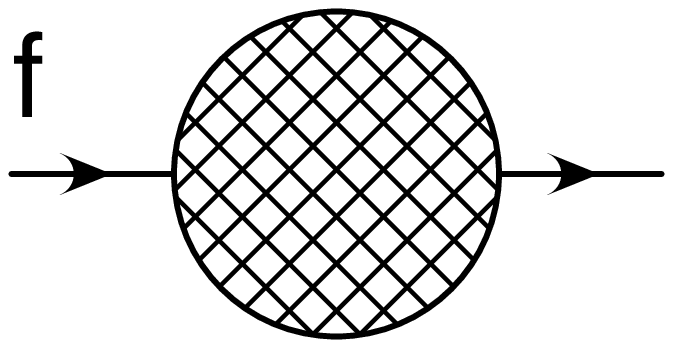, width=1.9cm}} =
\parbox[c]{2cm}{\psfig{figure=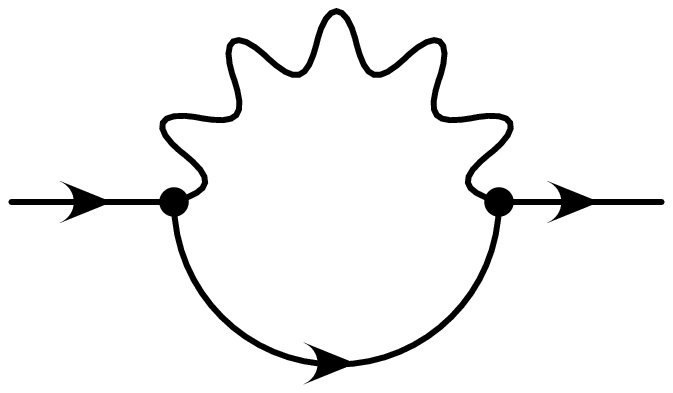, width=1.9cm}} +
\parbox[c]{2cm}{\psfig{figure=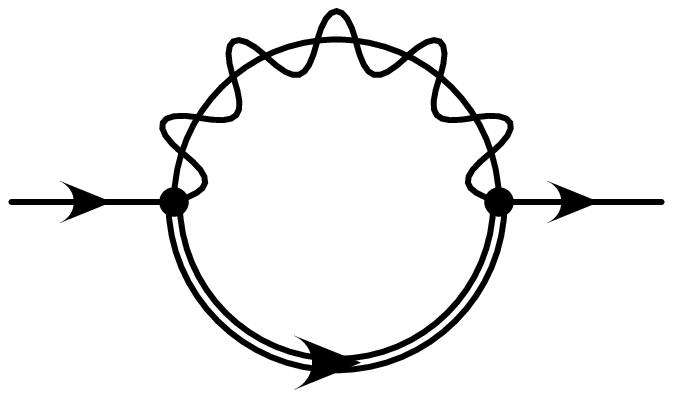, width=1.9cm}}
\phantom{+\hspace{1.9cm}+\hspace{2cm}+\hspace{2cm}+\hspace{2cm}\anc}
$\\[2em]
$
\parbox[c]{2cm}{\psfig{figure=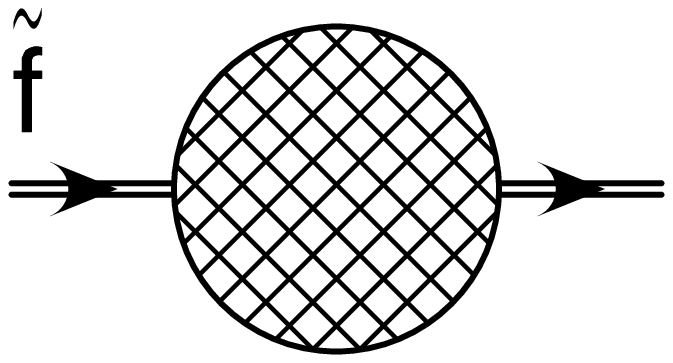, width=1.9cm}} =
\parbox[c]{2cm}{\psfig{figure=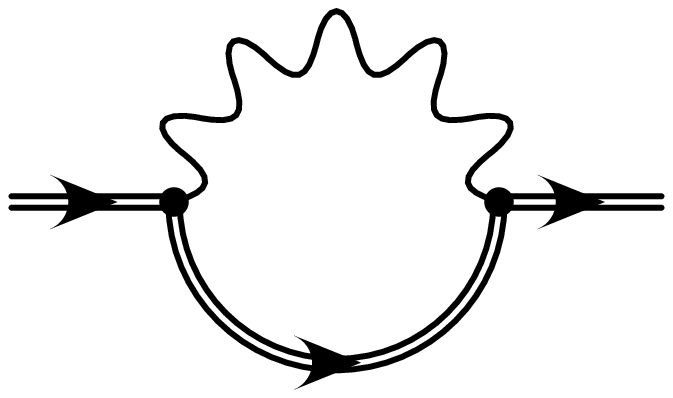, width=1.9cm}} +
\parbox[c]{2cm}{\psfig{figure=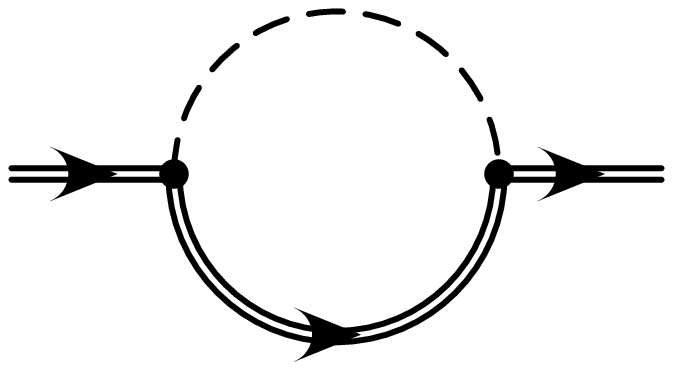, width=1.9cm}} +
\parbox[c]{2cm}{\psfig{figure=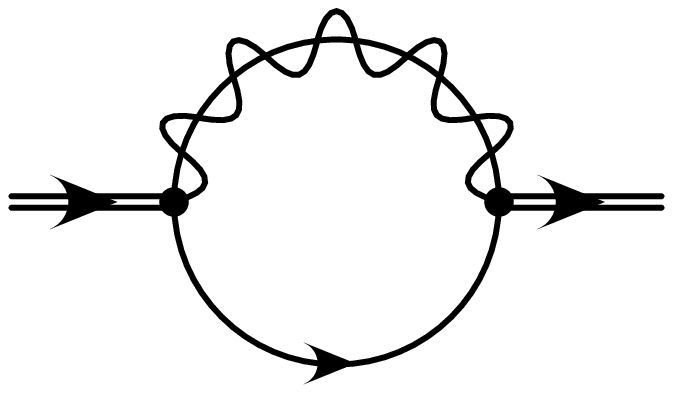, width=1.9cm}} +
\parbox[c]{2cm}{\psfig{figure=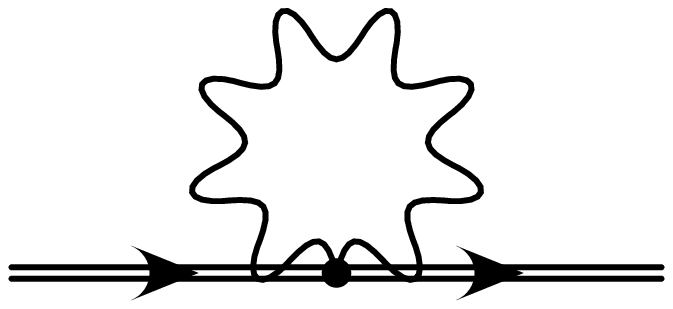, width=1.9cm}} +
\parbox[c]{2cm}{\psfig{figure=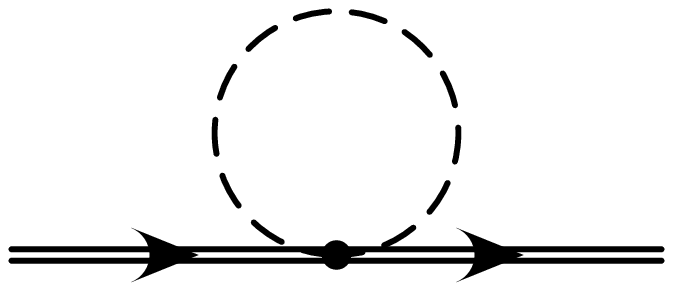, width=1.9cm}} +
\parbox[c]{2cm}{\psfig{figure=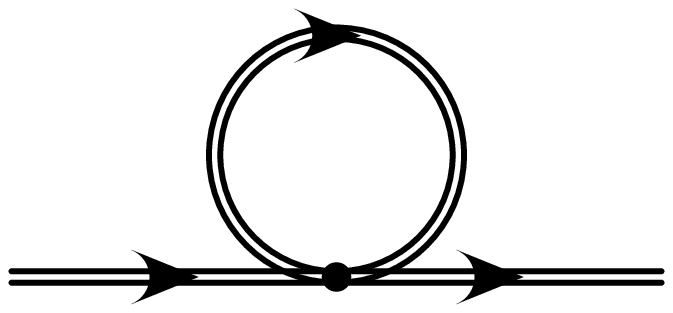, width=1.9cm}}
$\\[2em]
$
\parbox[c]{2cm}{\psfig{figure=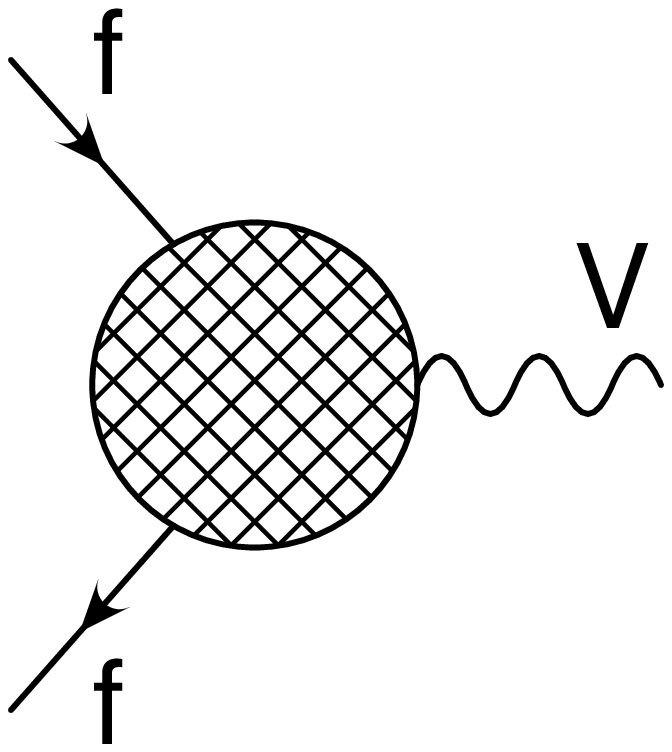, width=2cm}} =
\parbox[c]{2cm}{\psfig{figure=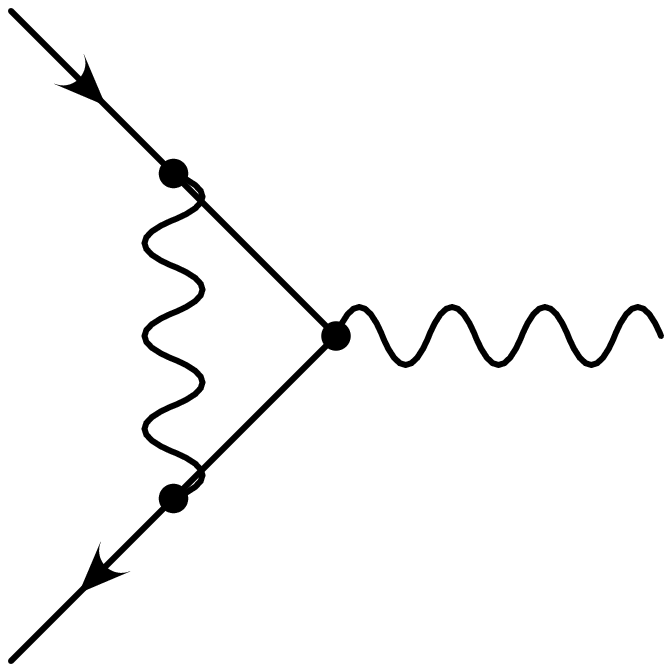, width=2cm}} +
\parbox[c]{2cm}{\psfig{figure=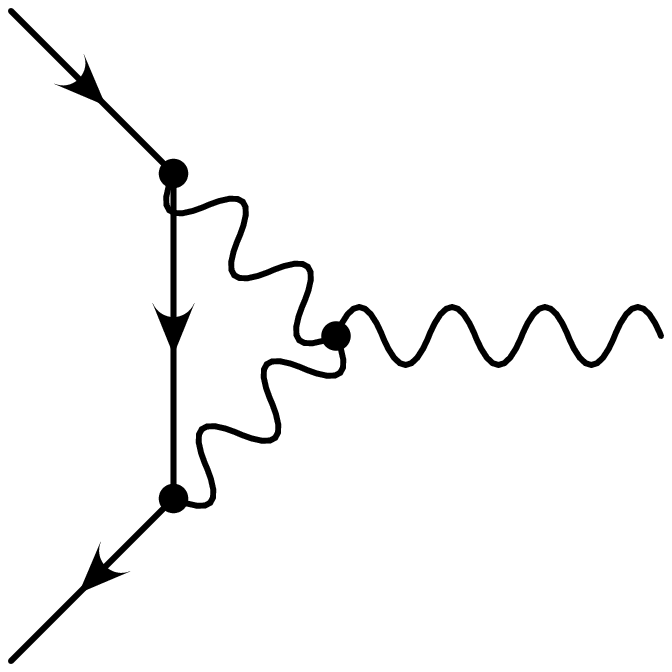, width=2cm}} +
\parbox[c]{2cm}{\psfig{figure=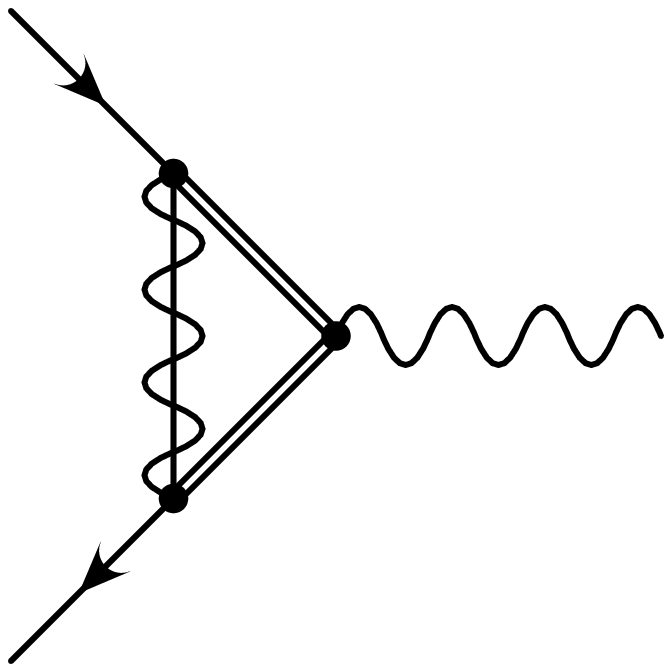, width=2cm}} +
\parbox[c]{2cm}{\psfig{figure=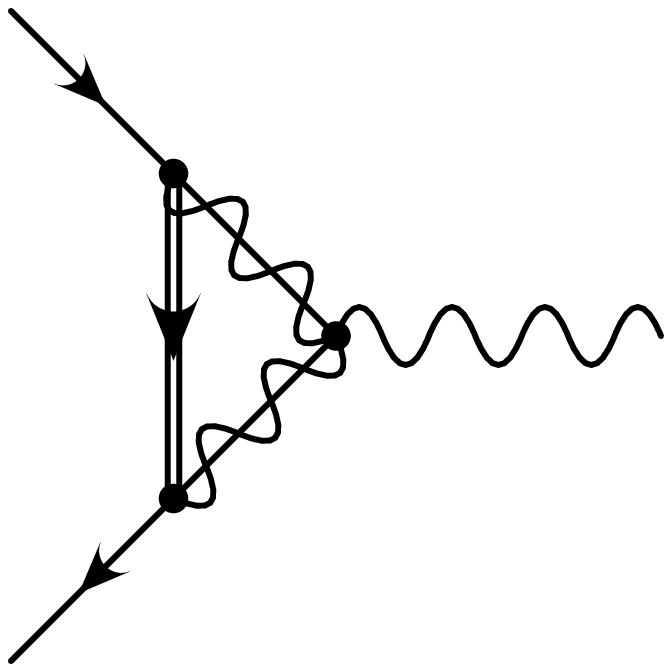, width=2cm}}
\phantom{ + \hspace{2cm}+ \hspace{2cm}\,\anc}
$\\[2em]
$
\parbox[c]{2cm}{\psfig{figure=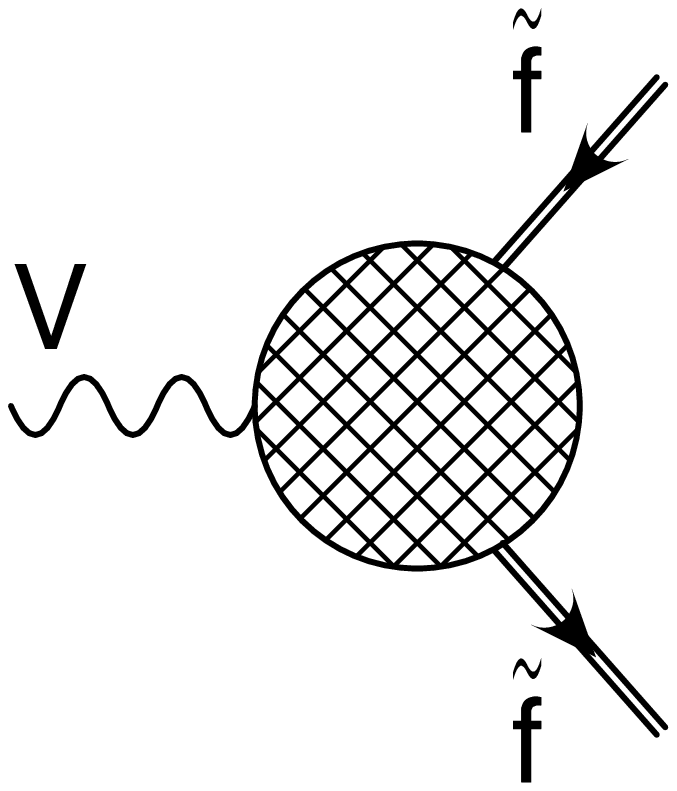, width=2cm}} =
\parbox[c]{2cm}{\psfig{figure=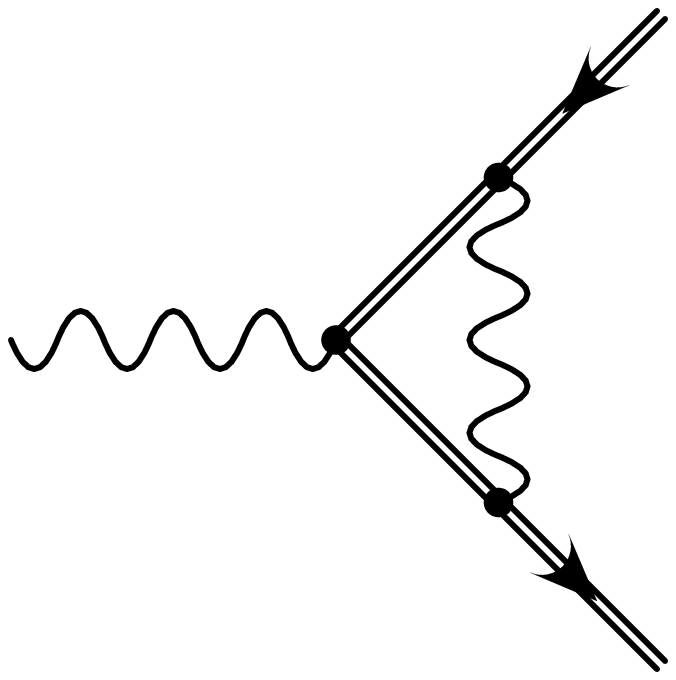, width=2cm}} +
\parbox[c]{2cm}{\psfig{figure=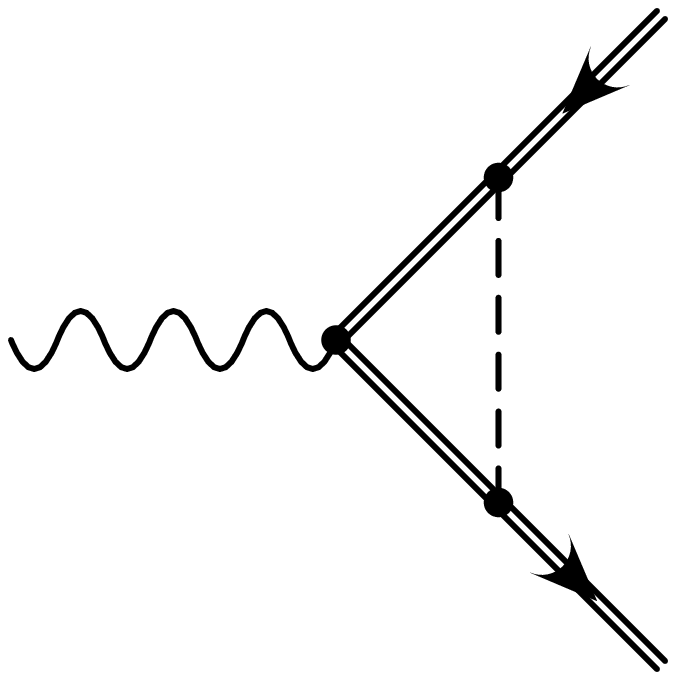, width=2cm}} +
\parbox[c]{2cm}{\psfig{figure=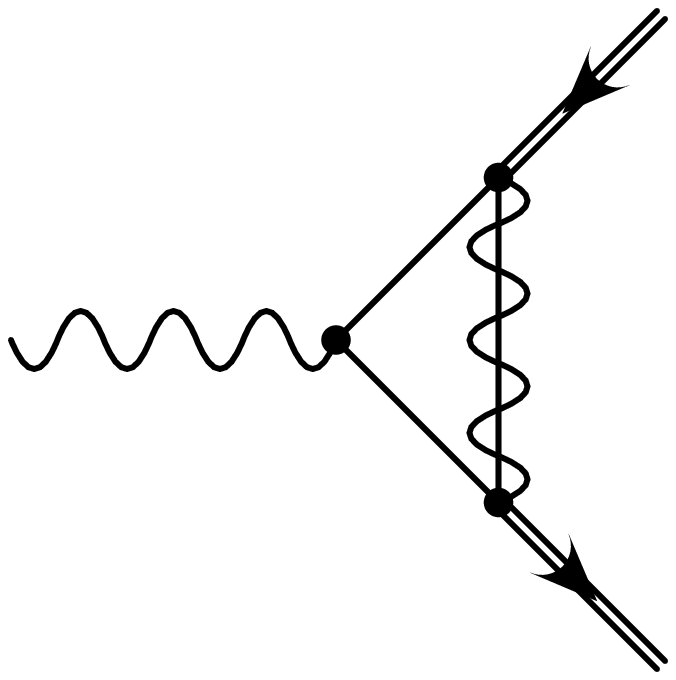, width=2cm}} +
\parbox[c]{2cm}{\psfig{figure=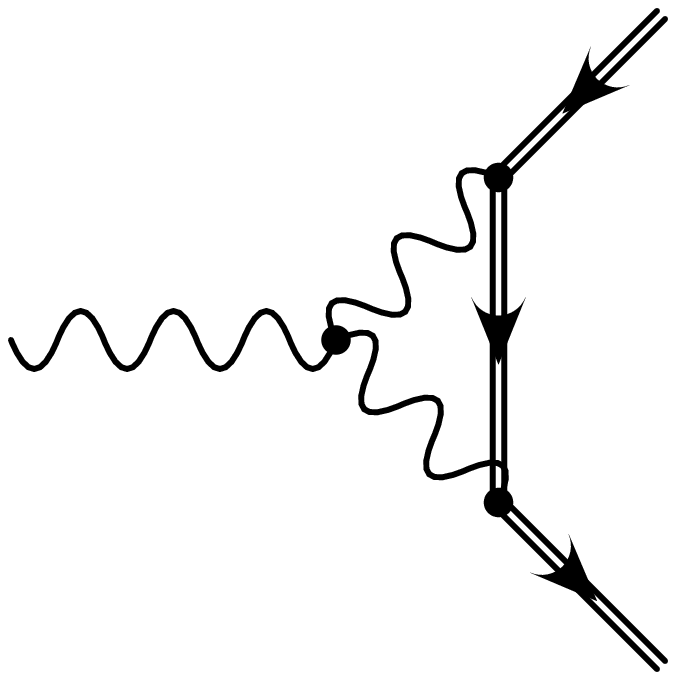, width=2cm}} +
\parbox[c]{2cm}{\psfig{figure=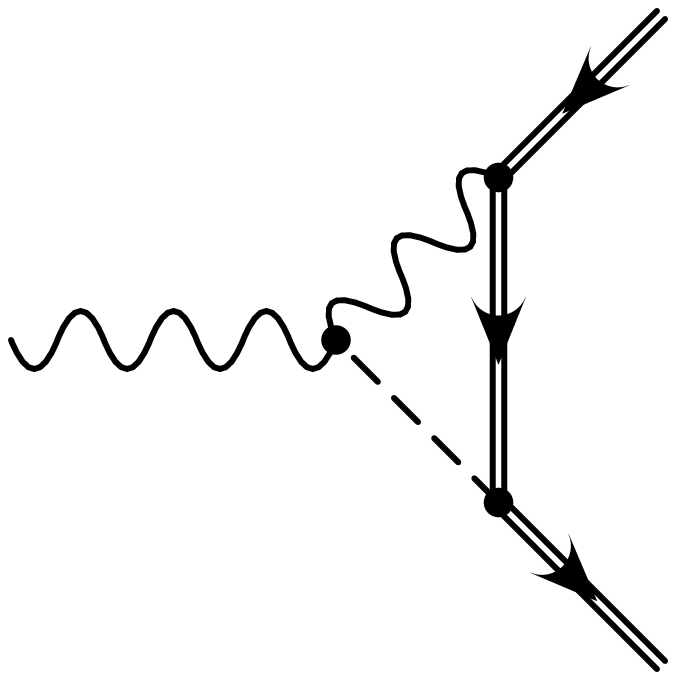, width=2cm}} +
\parbox[c]{2cm}{\psfig{figure=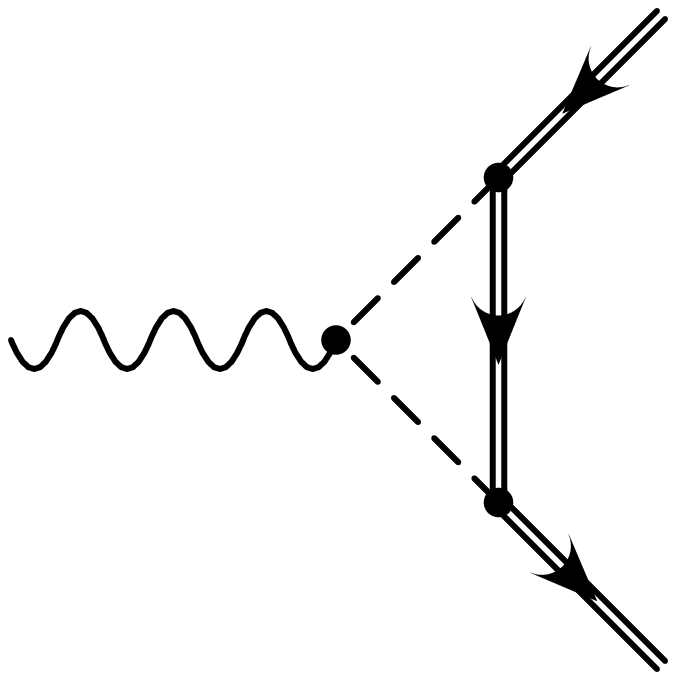, width=2cm}}
$\\[1em]
$\anc\hspace{2cm}\, +
\parbox[c]{2cm}{\psfig{figure=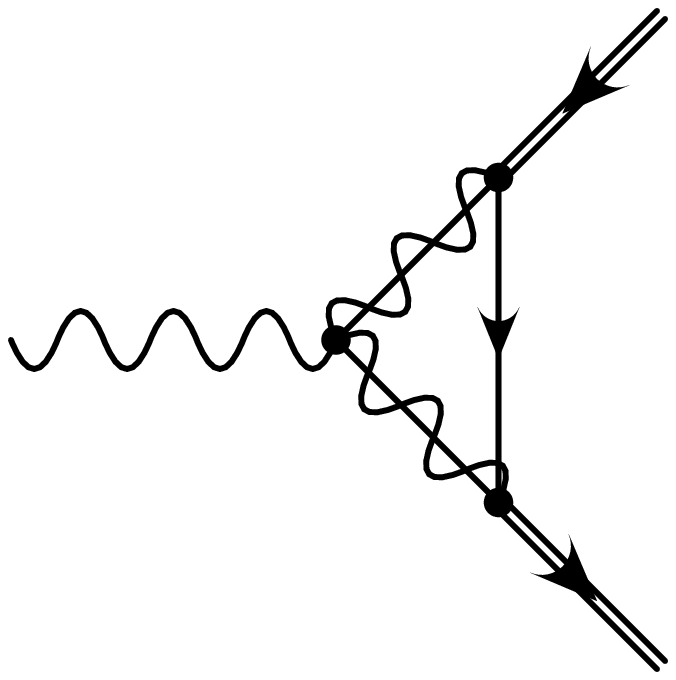, width=2cm}} +
\parbox[c]{2cm}{\psfig{figure=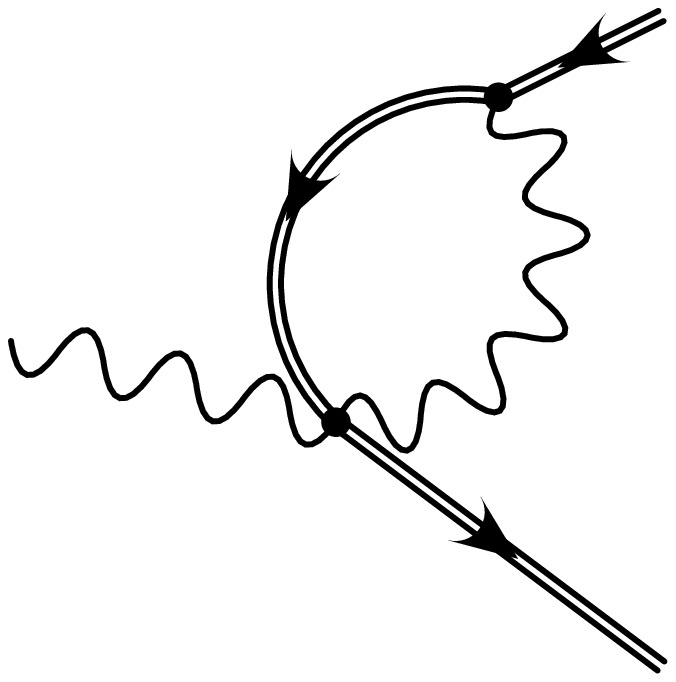, width=2cm}} +
\parbox[c]{2cm}{\psfig{figure=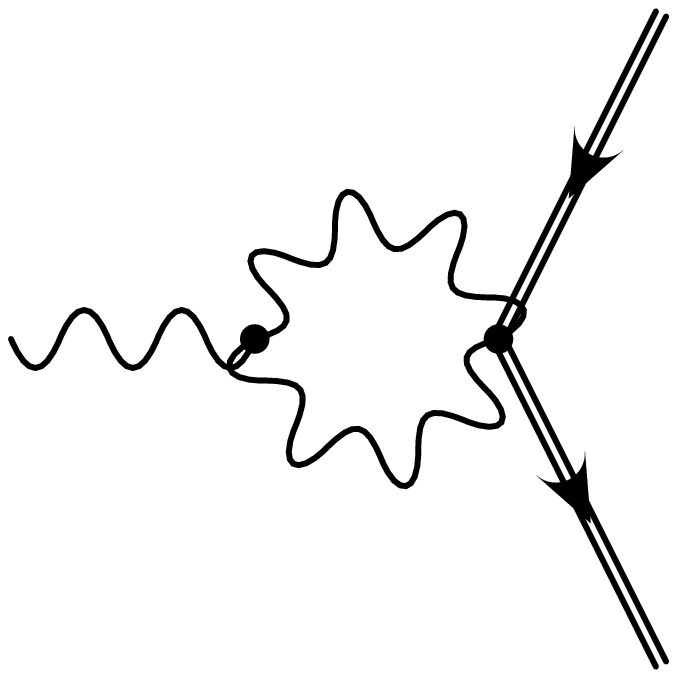, width=2cm}} +
\parbox[c]{2cm}{\psfig{figure=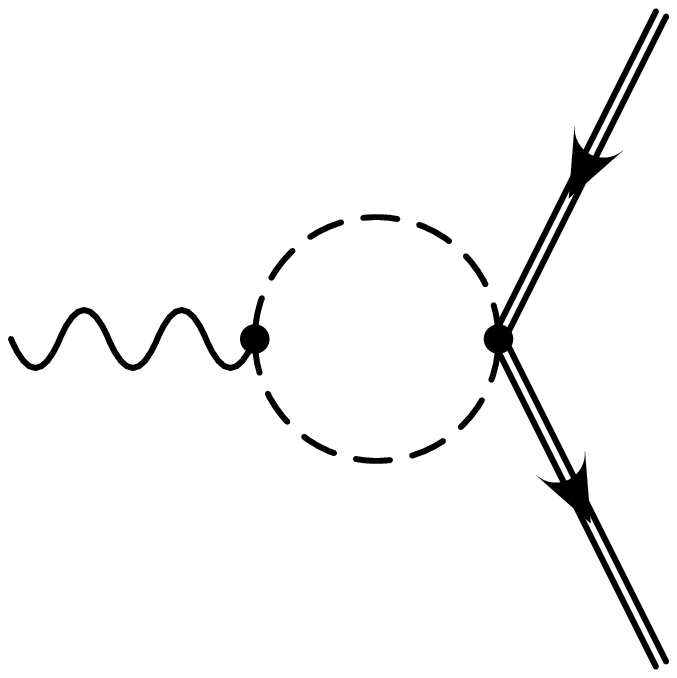, width=2cm}} +
\parbox[c]{2cm}{\psfig{figure=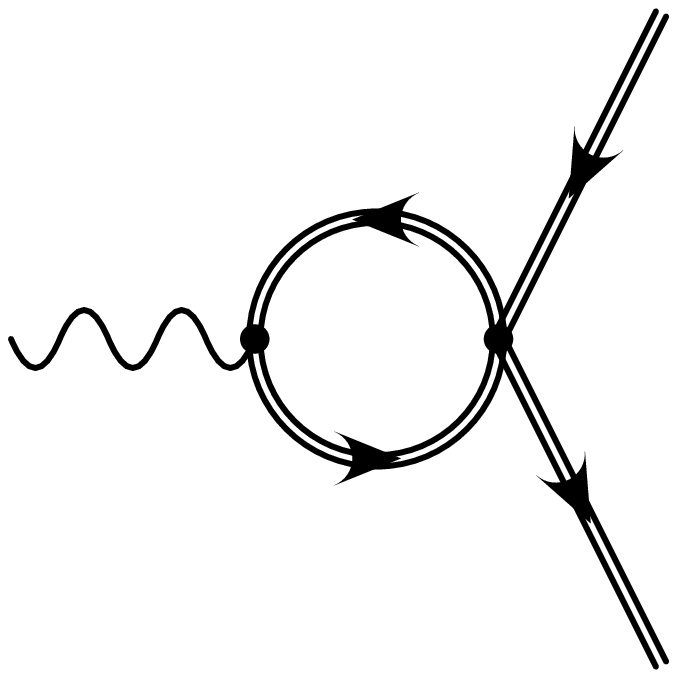, width=2cm}}
\phantom{ + \hspace{2cm}\,\anc}
$\\[1.5em]
$
\parbox[c]{2cm}{\psfig{figure=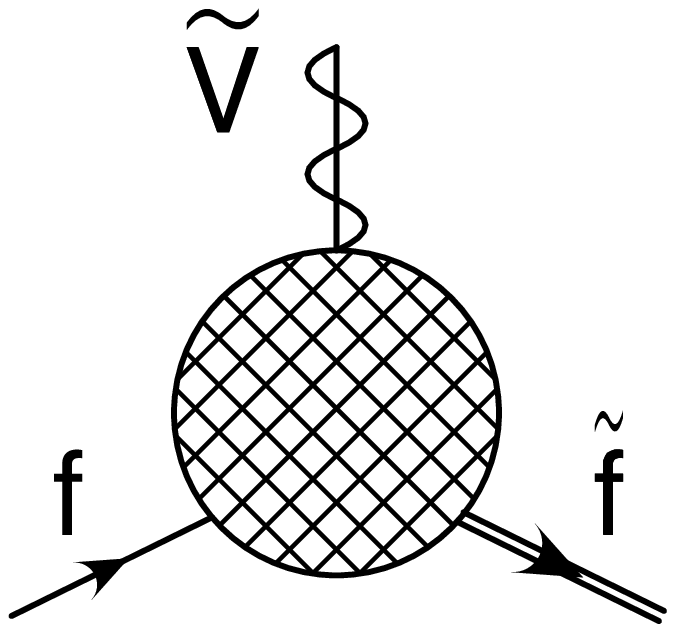, width=2cm}} =
\parbox[c]{2cm}{\psfig{figure=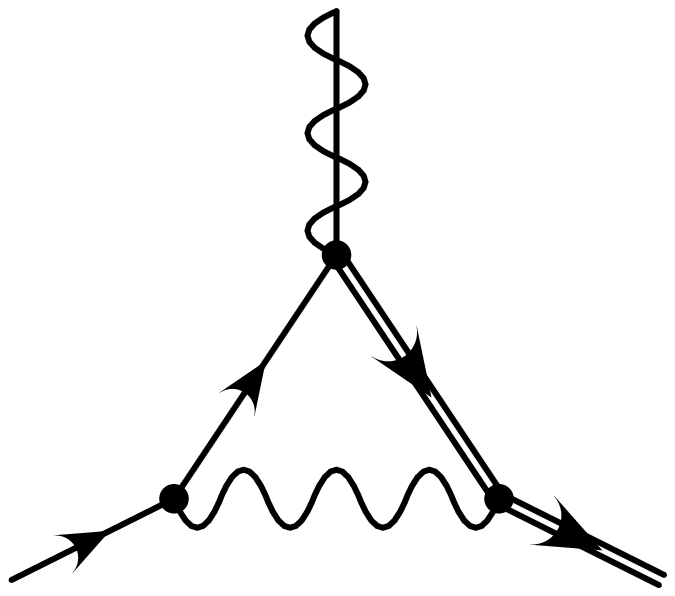, width=2cm}} +
\parbox[c]{2cm}{\psfig{figure=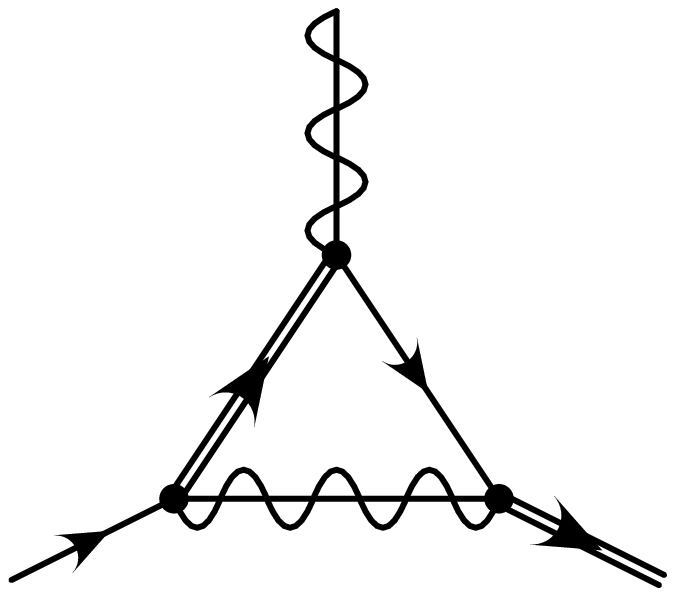, width=2cm}} +
\parbox[c]{2cm}{\psfig{figure=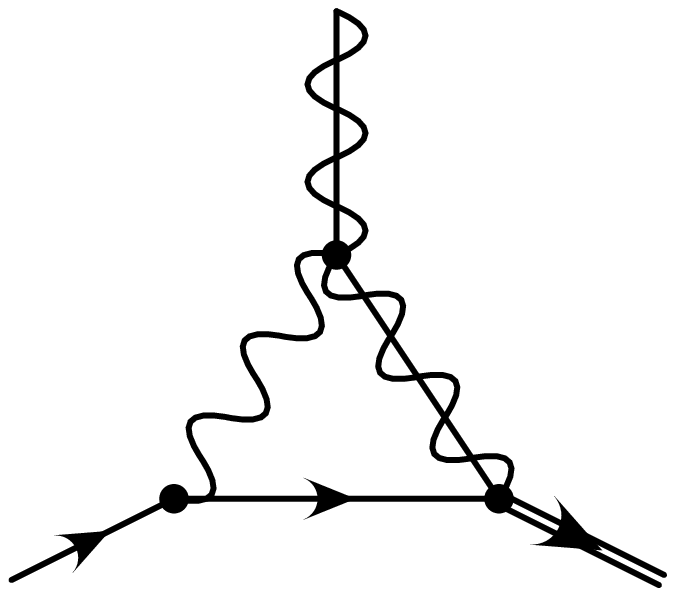, width=2cm}} +
\parbox[c]{2cm}{\psfig{figure=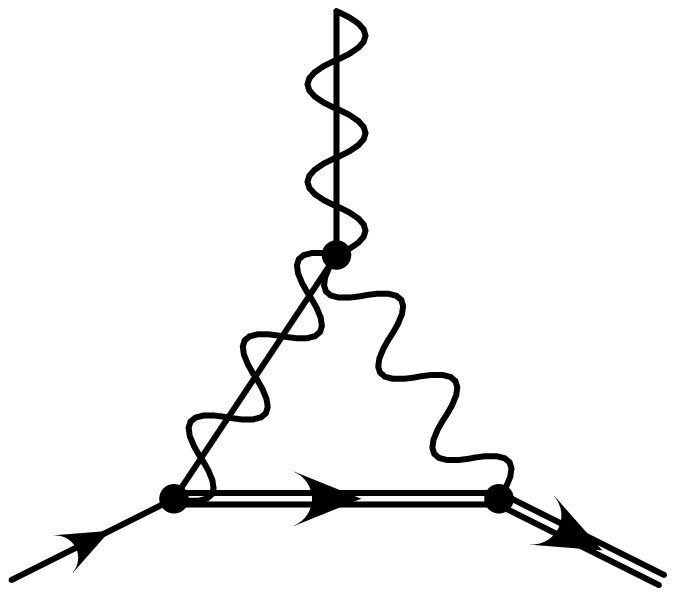, width=2cm}} +
\parbox[c]{2cm}{\psfig{figure=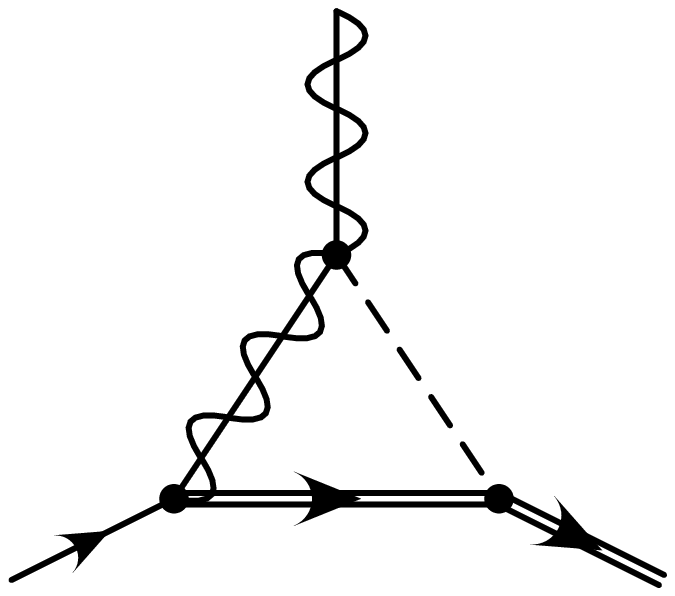, width=2cm}}
\phantom{ + \hspace{2cm}\,\anc}
$\\\anc
\mycaption{Generic sets of Feynman diagrams for the virtual self-energy and vertex
corrections to slepton
production. Solid, dashed and wiggly lines indicate fermions, (Higgs) scalars and vector
bosons, respectively, whereas sfermions and gauginos are denoted by double lines
and wiggly/solid lines.
The selectron-electron-neutralino vertex in the last line only contributes to
selectron production.}
\label{fig:loopdiag1}
\end{figure}
\begin{figure}[tp]
(a) $\quad
\parbox[c]{3cm}{\psfig{figure=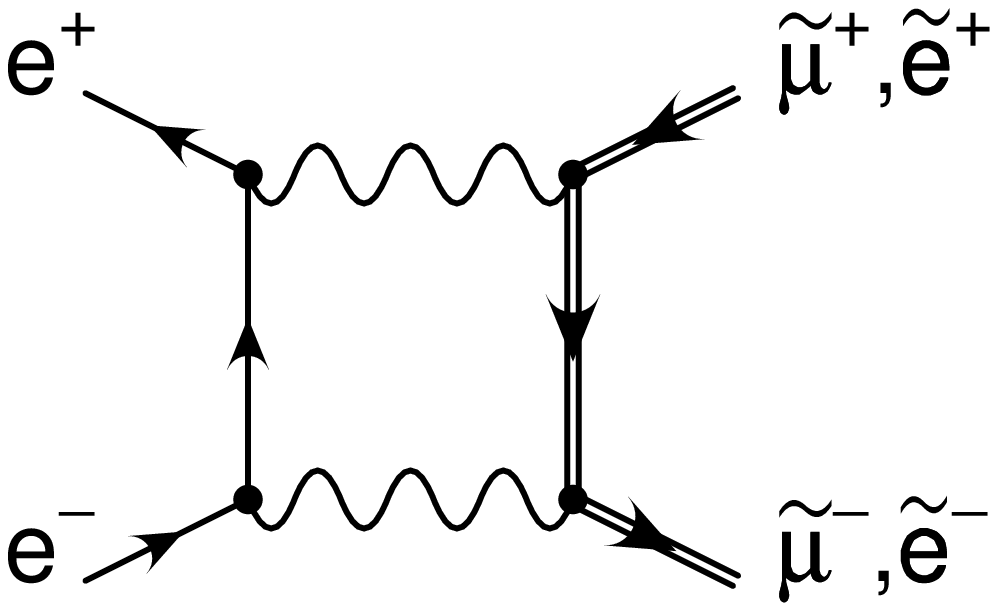, height=1.9cm}} \quad
\parbox[c]{2cm}{\psfig{figure=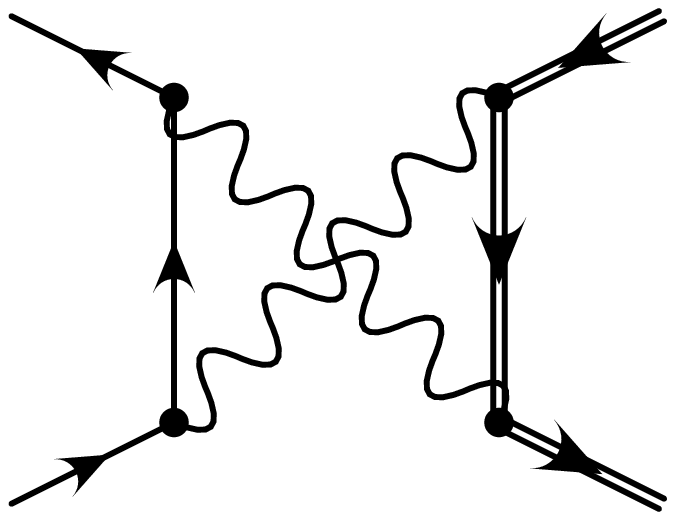, height=1.5cm}} \quad
\parbox[c]{2cm}{\psfig{figure=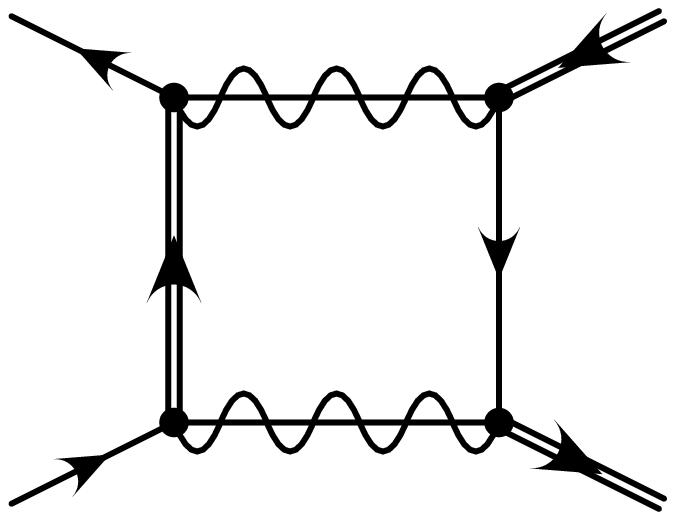, height=1.5cm}} \quad
\parbox[c]{2cm}{\psfig{figure=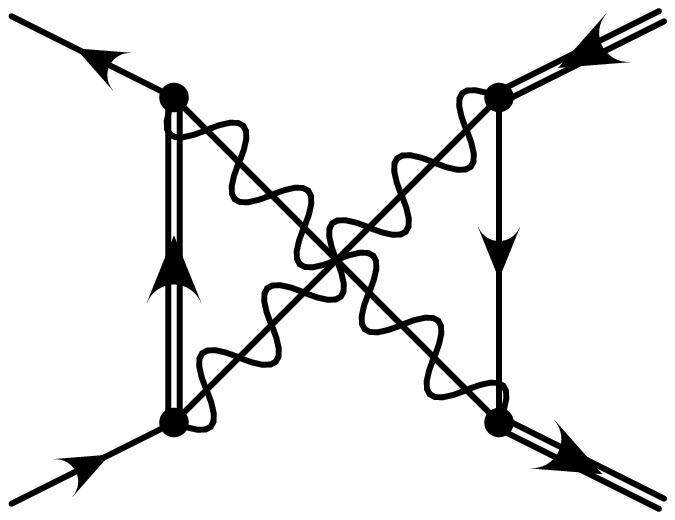, height=1.5cm}} \quad
\parbox[c]{2cm}{\psfig{figure=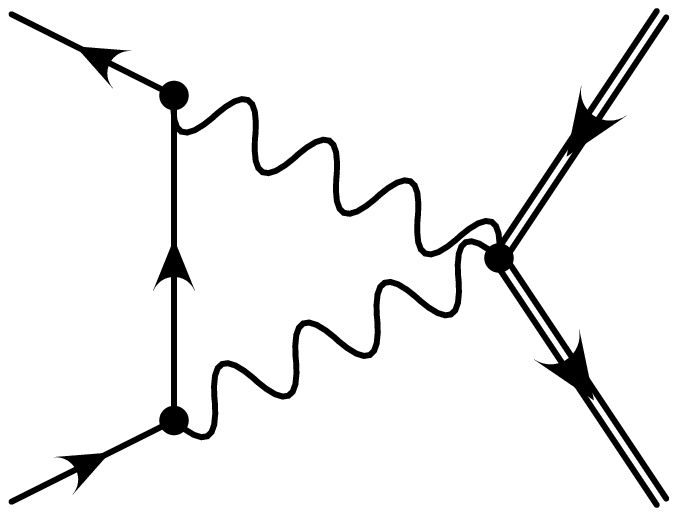, height=1.5cm}} \quad
\parbox[c]{2cm}{\psfig{figure=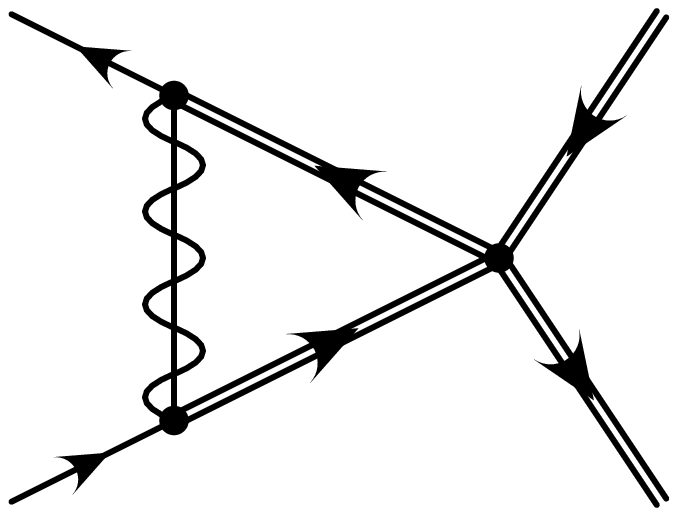, height=1.5cm}}
$\\[1em]
(b) $\quad
\parbox[c]{3cm}{\psfig{figure=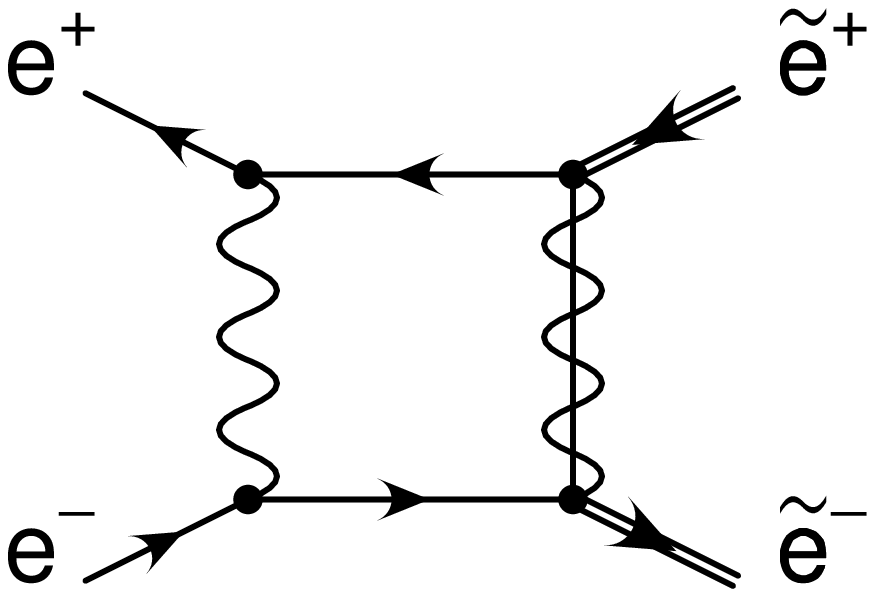, height=1.9cm}} \quad
\parbox[c]{2cm}{\psfig{figure=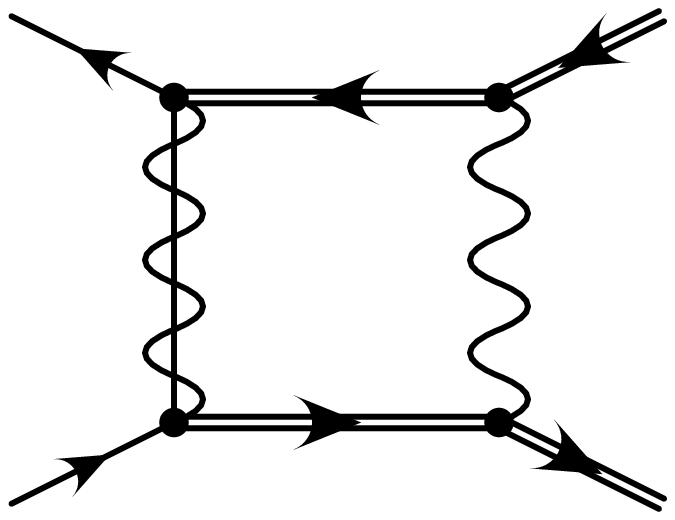, height=1.5cm}} \quad
\parbox[c]{2cm}{\psfig{figure=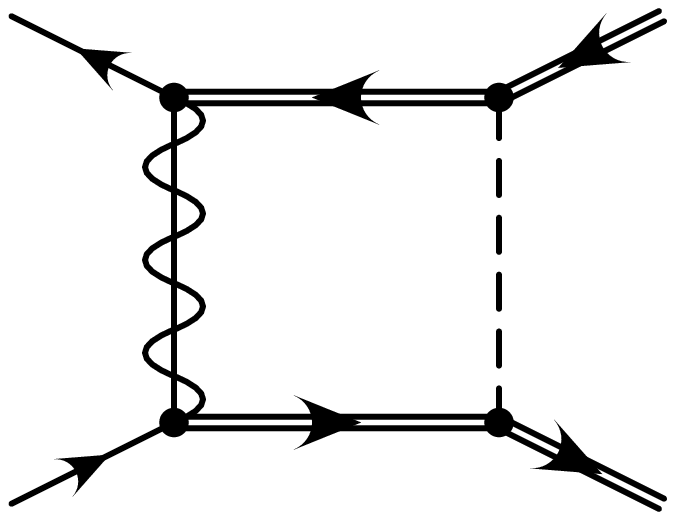, height=1.5cm}} \quad
\parbox[c]{2cm}{\psfig{figure=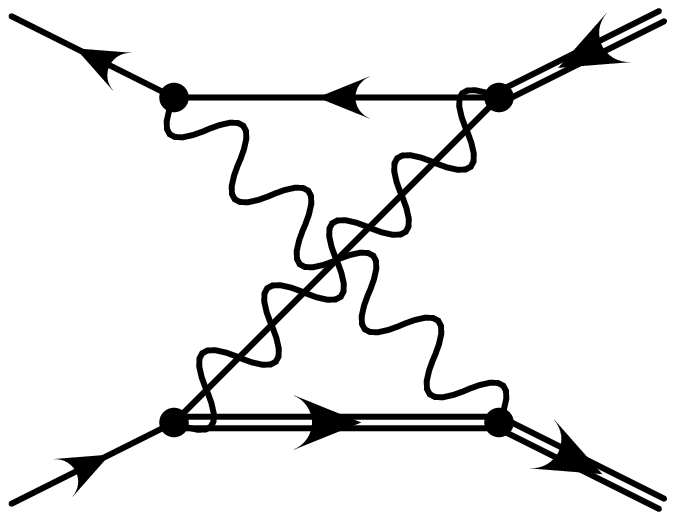, height=1.5cm}} \quad
\parbox[c]{2cm}{\psfig{figure=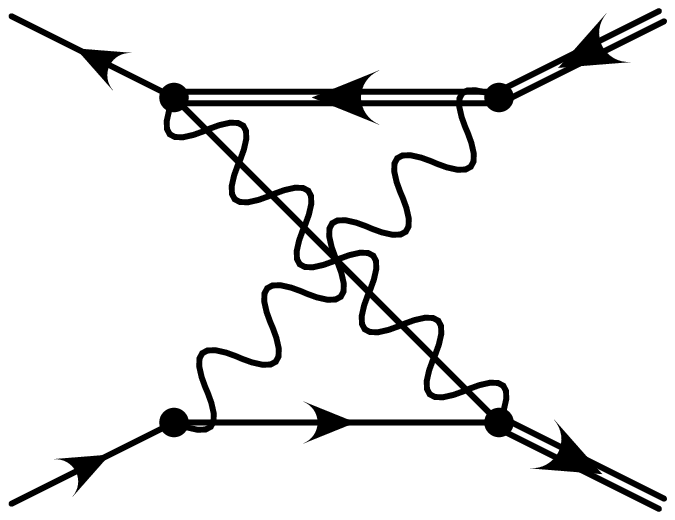, height=1.5cm}}
$
\mycaption{Box-type Feynman diagrams for slepton pair production. The first row
(a) applies both to smuon and selectron production, while the second row (b)
only contributes to selectron production.
}
\label{fig:loopdiag2}
\end{figure}
Characteristic classes of higher-order diagrams for propagators and vertices
are depicted in Fig.~\ref{fig:loopdiag1}. Additional box diagrams, cf.\
Fig.~\ref{fig:loopdiag2}, finally conclude the set of elements contributing to
the 2-2 transitions%
\footnote{Most of the analytical results 
for self-energy operators etc.\ are 
too lengthy to be presented in this report; therefore computer codes for the
calculation of the loop results are made available on the web, cf.\ the
concluding remarks in section~\ref{concl}.}.

\begin{figure}[tp]
\vspace{1ex}
(a)\hspace{10cm}(b)\\[1ex]
$
\raisebox{-1.5mm}{%
\parbox[c]{3.5cm}{\psfig{figure=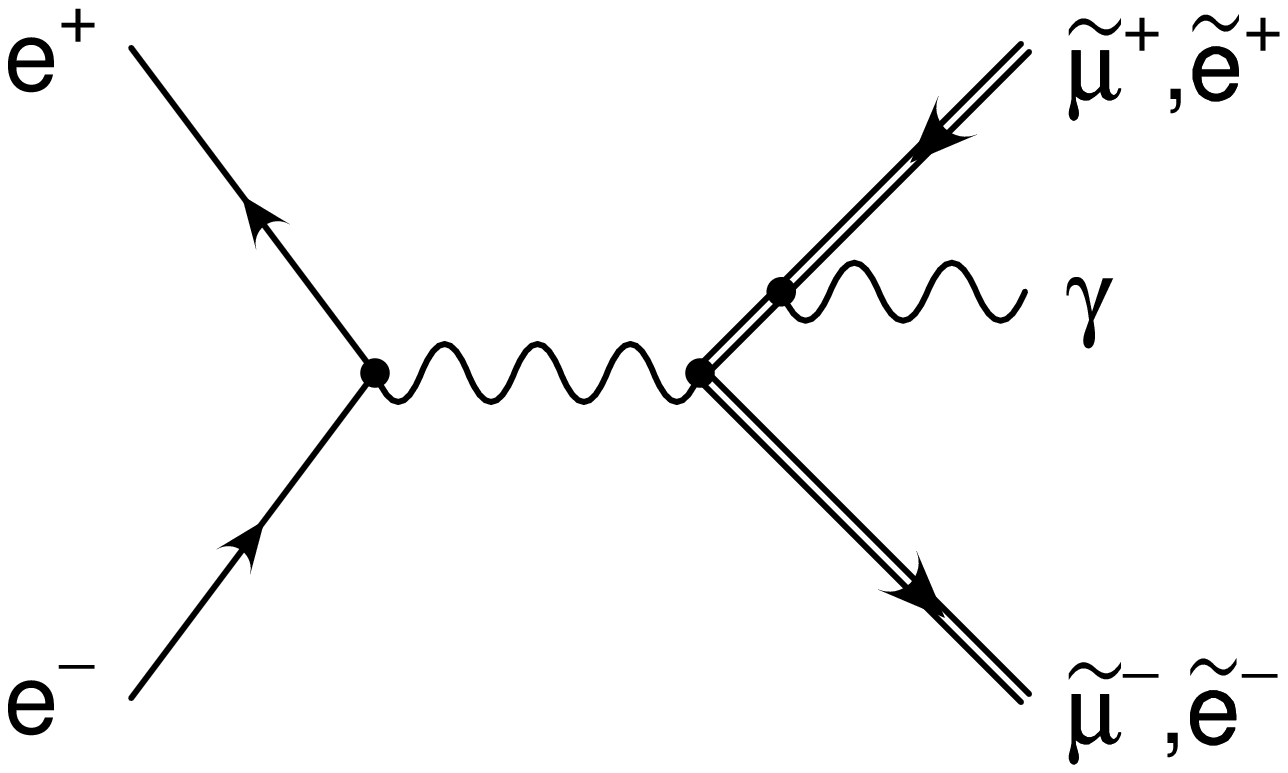, height=2cm}}} \quad
\parbox[c]{2.3cm}{\psfig{figure=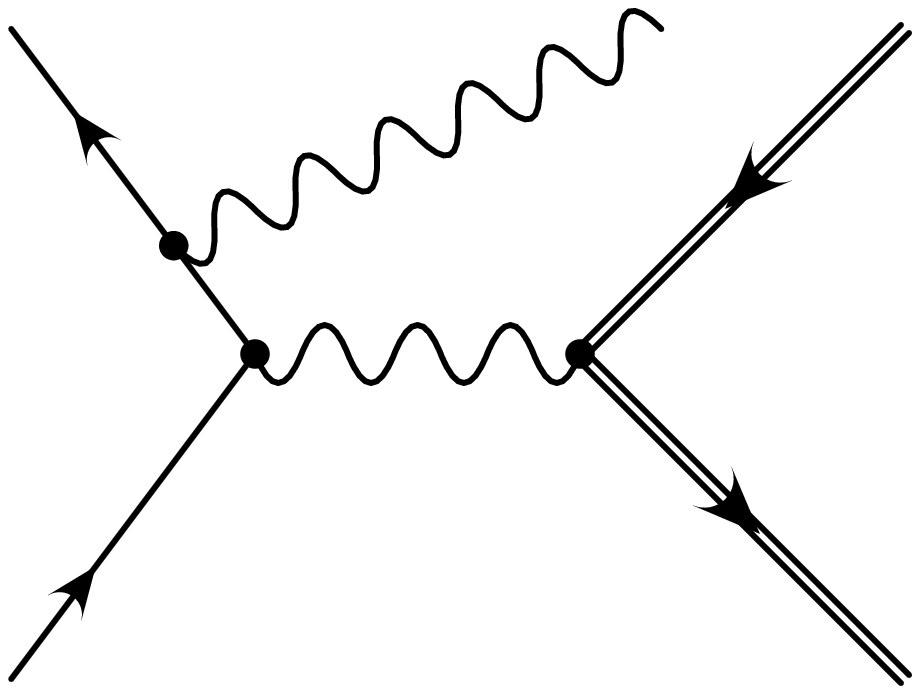, height=1.8cm}} \quad
\parbox[c]{2.3cm}{\psfig{figure=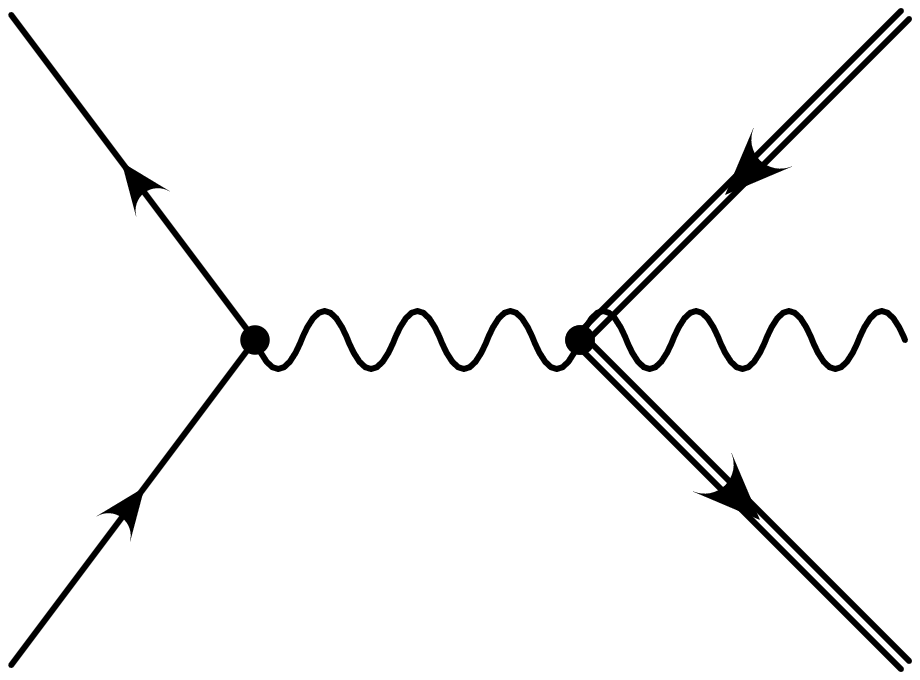, height=1.8cm}}
\qquad\qquad
\raisebox{-1mm}{%
\parbox[c]{3cm}{\psfig{figure=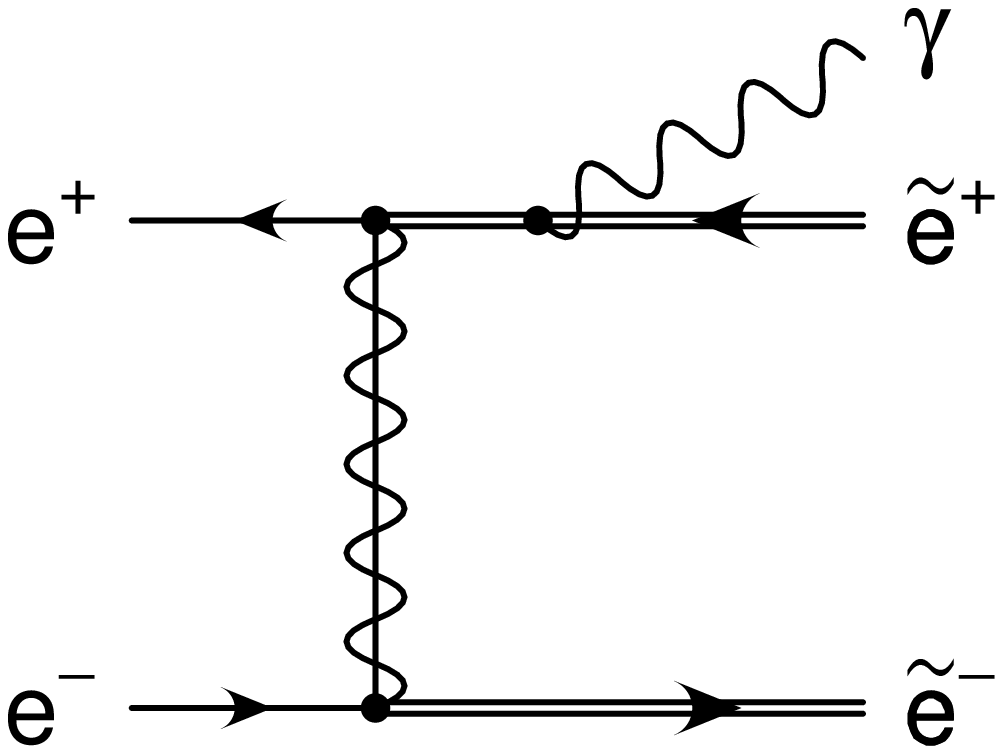, height=2.3cm}}} \quad
\parbox[c]{2.3cm}{\psfig{figure=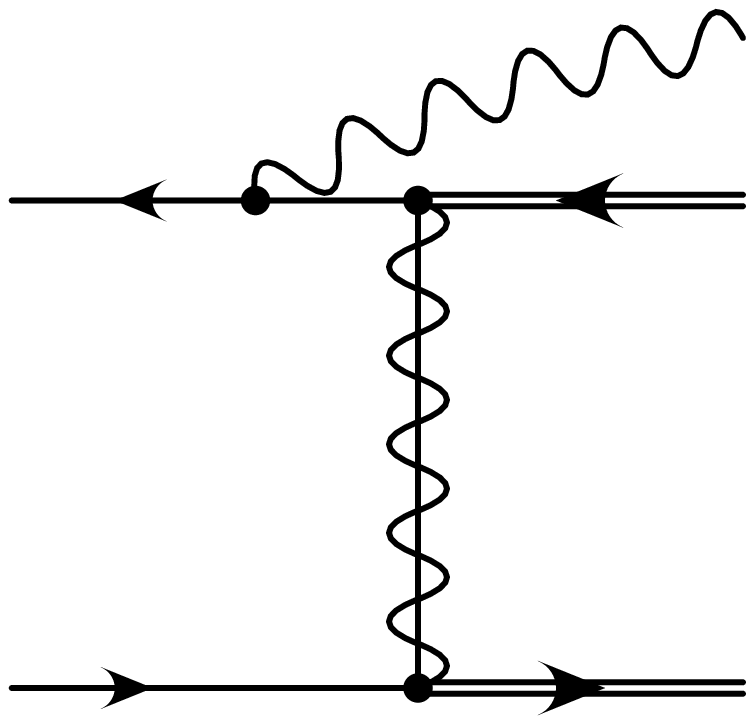, height=2cm}}
$
\mycaption{Feynman diagrams for real photon emission in smuon production (a) and
selectron production (a)+(b).
}
\label{fig:realdiag}
\end{figure}
After carrying out the renormalization program in the ultraviolet sector,
infrared and collinear divergences associated with the massless photon and
lepton fields can be absorbed by adding the real photon emission contributions,
Fig.~\ref{fig:realdiag}. Finite results are automatically guaranteed by
proceeding to experimentally well defined cross-sections, {\it i.e.} the total 
cross-sections in the present analysis.

Due to large number of diagrams involved, the use of computer algebra tools for
the computation is necessary. The generation of diagrams and amplitudes is
performed with the package \textsl{FeynArts} \cite{feynarts}. Throughout the
calculation, the CKM matrix is taken diagonal and mixing between the sfermions
of the first two generations is neglected. For the third generation sfermions,
the mixing between the L- and R-states is consistently taken into account.
A general
covariant $R_\xi$ gauge is used in order to facilitate an additional check of
the result. Using the program \textsl{FeynCalc 2.2} \cite{feyncalc}, the
Lorentz and Dirac algebra is evaluated and the loop integrals are reduced to a
set of fundamental scalar one-loop functions \cite{oneloopint}. 
Since the explicit analytical expressions of the virtual loop contributions 
are generally very lengthy, they have been
implemented into a computer code that calculates the one-loop-corrected
cross-sections, using the package \textsl{LoopTools} \cite{looptools} for the
numerical evaluation of the basic scalar one-loop functions. 

\subsubsection{Gauge sector}

The extension of the Standard Model to a supersymmetric theory in
minimal form (MSSM) does not introduce new couplings in the
gauge/gaugino sector. The gauge sector of the MSSM is therefore
renormalized in parallel to the Standard Model. Just the self-energies
are expanded by the contributions of the supersymmetric fields in a
straightforward way. We briefly summarize the results, adopting the
standard conventions of Ref.~\cite{Denner:93}.

The masses of the W and Z gauge bosons are shifted by
\begin{equation}
\MW^2 \to \MW^2 + \dMW  
\qquad\quad \mbox{and} \qquad\quad
\MZ^2 \to \MZ^2 + \dMZ.
\end{equation}
Imposing the on-shell renormalization conditions defined earlier, the mass
shifts can be expressed in terms of the transverse self-energies $\Sigma_{\rm
T}$,
\begin{equation}
\dMW = \re \, \Sigma_{\rm T}^{\rm WW}(\MW^2)
\qquad\quad \mbox{and} \qquad\quad
\dMZ = \re \, \Sigma_{\rm T}^{\rm ZZ}(\MZ^2),
\end{equation}
with the self-energies for the gauge boson propagation in the Standard Model
expanded by supersymmetric particle contributions:
\begin{equation}
\Sigma_{\rm T}^{\rm V_1V_2}(k^2) = \Sigma_{\rm T}^{\rm V_1V_2}(k^2)\bigr|_{\rm SM} +
\Sigma_{\rm T}^{\rm V_1V_2}(k^2)\bigr|_{\rm SUSY}  \qquad
[{\rm V}_i = {\rm \gamma,Z,W}]. \label{eq:sigmasus}
\end{equation}
The first term includes the usual Standard Model loop contributions as in the
first line of Fig.~\ref{fig:loopdiag1}, while
the second term accounts for the additional loops
involving pairs of supersymmetric fields, gaugino and sfermion fields, as given
in the second line of Fig.~\ref{fig:loopdiag1}.

The SU(2) and U(1) gauge couplings $g$ and $g'$ can be traced back to the
electromagnetic coupling $e$ and the electroweak weak mixing angle $\sw = \sin
\theta_{\rm W}$. $e$ is renormalized as in standard QED apart from the
removal of $\gamma$-$Z$ mixing,
\begin{equation}
e \to (1+ \dZe)\, e 
\qquad \mbox{with} \qquad
\dZe = \h
  \left. \frac{\partial \Sigma^{\gamma\gamma}_{\rm T}(k^2)}{\partial k^2}
  \right|_{k^2 = 0} 
 - \frac{\sw}{\cw} \, \frac{\Sigma^{\gamma Z}_{\rm T}(0)}{\MZ^2}.
\end{equation}
Again the gauge-boson self-energies decompose into Standard Model and
specific supersymmetric contributions as in \eqref{eq:sigmasus}.

Introducing the electroweak mixing angle in on-shell definition through the W
and Z masses as $\sw^2 = 1 - \MW^2/\MZ^2$, the renormalized value is formally
related to the bare value by
\begin{align}
\sw &\to \sw + \dsw & \mbox{with} &&
\frac{\dsw}{\sw} &= \frac{\cw^2}{2\sw^2}
  \left[ \frac{\dMZ}{\MZ^2} - \frac{\dMW}{\MW^2} \right].
\end{align}
Finally, the renormalized left- and right-handed electron fields,
\begin{align}
\eL &\to (1+ \hh \delta Z^{e \rm L}) \, \eL, &
\eR &\to (1+ \hh \delta Z^{e \rm R}) \, \eR,
\end{align}
are related to the electron self-energies by
\begin{align}
\delta Z^{e \rm L} &= -\re \biggl\{
  \Sigma^{e \rm L}(\me^2) + \me^2 \, \frac{\partial}{\partial p^2}
    \Bigl [ \Sigma^{e \rm L}(p^2) + \Sigma^{e \rm R}(p^2) +
           2/\me \, \Sigma^{e \rm S}(p^2) \Bigr ]_{p^2 = \me^2}
  \biggr\}, \label{eq:dZeL} \\
\delta Z^{e \rm R} &= -\re \biggl\{
  \Sigma^{e \rm R}(\me^2) + \me^2 \, \frac{\partial}{\partial p^2}
    \Bigl [ \Sigma^{e \rm L}(p^2) + \Sigma^{e \rm R}(p^2) +
           2/\me \, \Sigma^{e \rm S}(p^2) \Bigr ]_{p^2 = \me^2}
  \biggr\}, \label{eq:dZeR}
\end{align}
with the decomposition
\begin{equation}
\Sigma^e(p) \, = \, \pslash \, \wL \Sigma^{e \rm L}(p^2) \;
  + \pslash \, \wR \Sigma^{e \rm R}(p^2)
  + \Sigma^{e \rm S}(p^2) 
\qquad \mbox{with} \qquad
\omega_{\LL,\RR} = (1 \pm \gamma_5)/2.
   \\
\end{equation}
Apart from the calculation of the (singular) QED corrections, the chiral limit
of vanishing electron mass can
be safely applied, simplifying eqs. \eqref{eq:dZeL}, \eqref{eq:dZeR} to
\begin{align}
\delta Z^{e \rm L}_{\rm weak} &= -\re \,
  \Sigma^{e \rm L}_{\rm weak}(0), &
\delta Z^{e \rm R}_{\rm weak} &= -\re \,
  \Sigma^{e \rm R}_{\rm weak}(0).
\end{align}

\subsubsection{Sfermion sector}

In the limit of vanishing lepton masses in the first and second
generation, the L- and R-selectron and smuon fields do not mix and the
mass matrices are approximately diagonal. The ``chiral'' L and R states
coincide with the mass eigen-states. This remains true in higher orders
as the sfermion mixing is proportional to the associated lepton mass.
The L and R fields may therefore be treated independently so that the
renormalization follows the standard procedure. With
\begin{equation}
\msl{i}^2 \to \msl{i}^2 + \delta \msl{i}^2  \qquad \mbox{and} \qquad
\sll_i \to (1+\hh \dZsl_i) \, \sll_i  \qquad
 [l = e,\mu; i = \LL,\RR],
\end{equation}
we find for the sfermion mass shift
\begin{equation}
\delta \msl{i}^2 = \re \Sigma^\sll_i(\msl{i}^2),
\end{equation}
and for the wave-function renormalization
\begin{equation}
\dZsl_i = -\re \frac{\partial \Sigma^\sll_i(k^2)}{\partial k^2}
  \biggr|_{k^2 = \msl{i}^2}
\end{equation}
Here $\Sigma^\sll_i(k^2)$ denotes the self-energy for the slepton $\sll_i$; $l =
e,\mu$; $i = \LL,\RR$. Since the external fields are superpartners, the slepton
self-energies cannot be separated into a Standard Model and a genuinely
supersymmetric part.

Note that the mass shift and wave-function renormalization for L-sneutrinos
coincide with those of L-selectrons in the chiral limit we consider in this
report.

\subsubsection{Chargino and neutralino sector}

The spectrum of two charginos and four neutralinos in the MSSM is described by
the three mass parameters $\mu$, $M_2$ and $M_1$, see section \ref{nota}. Apart
from other electroweak parameters, the system is also affected by the Higgs
mixing $\tan\beta$. Three chargino/neutralino masses are
sufficient to fix the mass parameters $\mu$, $M_2$ and $M_1$. 
The renormalization of $\tan\beta$ is performed outside the chargino and
neutralino sector, as will be discussed in section~\ref{tanbeta}

The other three masses and the mixing parameters are then uniquely determined
once the parameters in the loop corrections are known \cite{chaneuren,chaneuv}.
Following \cite{chaneuren}, the renormalization is performed in the current
eigen-basis.

{\bf 1.)} Starting from the \underline{chargino} Lagrangian
\begin{equation}
{\cal L}_{\rm ch} = i \bigl[ {\psi^-}^\top \sigma^\mu \partial_\mu \,
 \overline{\psi^-} + \overline{\psi^+}^\top \bar{\sigma}^\mu \partial_\mu \,
 {\psi^+} \bigr]
- \bigl[ {\psi^-}^\top X \, {\psi^+} + \overline{\psi^+}^\top X^\dagger \,
 \overline{\psi^-} \, \bigr],
\end{equation}
with the current fields
\begin{equation}
\psi^+ \equiv \begin{pmatrix} \psi^+_1 \\ \psi^+_2  \end{pmatrix} =
  \begin{pmatrix} \widetilde{W}^+ \\ \widetilde{H}_{\rm u}^+ \end{pmatrix},
\;\;\;\;\;
\psi^- \equiv \begin{pmatrix} \psi^-_1 \\ \psi^-_2  \end{pmatrix} =
  \begin{pmatrix} \widetilde{W}^- \\ \widetilde{H}_{\rm d}^- \end{pmatrix},
\end{equation}
the mass matrix $X$ is renormalized by
\begin{equation}
X \to X + \delta X
\quad \mbox{with} \qquad
\delta X
= \begin{pmatrix} \delta M_2 & \sqrt{2} \; \delta \left( \MW \sin\beta \right) \\
  \sqrt{2} \; \delta \left( \MW \cos\beta \right) & \delta \mu \end{pmatrix},
\label{eq:Mchact}
\end{equation}
and the current fields are replaced by the normalized mass eigen-fields
$\chi^\pm$,
\begin{equation}
\psi^+ \to
  V^\dagger \left( 1 + \hh \dzL \right) \chi^+, \qquad
\psi^- \to
  U^\dagger \left( 1 + \hh \dzR \right) \chi^-.
\end{equation}
Besides the renormalization of the new parameters $\mu$ and $M_2$, $\delta X$
includes the renormalization of $\tan\beta$ and the W mass discussed earlier.
The [infinite] multiplicative renormalization of the wave functions is
absorbed in the matrices $\dzL,\dzR$, and so is the [finite] renormalization
of the matrices $V,U$ rotating the current to the mass fields. The renormalized
chargino Lagrangian and the associated counter terms may be written in
the form
\begin{align}
 {\cal L}_{\rm ch} &\to {\cal L}_{\rm ch} + \delta {\cal L}_{\rm ch}, \nonumber
\\
 {\cal L}_{\rm ch} &= \begin{pmatrix}
   \overline{\cha^+_1},\!\!\!& \overline{\cha^+_2} \end{pmatrix}
 \left[ i\dslash - U^* X V^\dagger \, \wL - V X^\dagger U^\top \wR \right]
  \begin{pmatrix} \cha^+_1 \\ \cha^+_2 \end{pmatrix}, \label{eq:l0ren} \\
 \delta {\cal L}_{\rm ch} &= \begin{pmatrix}
   \overline{\cha^+_1},\!\!\!& \overline{\cha^+_2} \end{pmatrix}
   \biggl [ i\frac{\dslash}{2} \bigl({\dzL}^\dagger + \dzL\bigr) \, \wL +
          i\frac{\dslash}{2} \bigl({\dzR}^* + {\dzR}^\top\bigr) \, \wR
  \nonumber \\
&\qquad -
    \bigl( \hh {\dzR}^\top U^* X V^\dagger +
     \hh U^* X V^\dagger \dzL + U^* \delta X V^\dagger \bigr) \, \wL
    \\
&\qquad -
    \bigl( \hh {\dzL}^\dagger V X^\dagger U^\top +
     \hh V X^\dagger U^\top {\dzR}^* + V \delta X^\dagger U^\top \bigr) \, \wR
    \biggr ] \begin{pmatrix} \cha^+_1 \\ \cha^+_2 \end{pmatrix}. \nonumber
\end{align}
The physical $\chi^\pm$ masses can be introduced in \eqref{eq:l0ren} after
diagonalizing this part of the Lagrangian by rotation through $U,V$. The
counterterms $\delta\mu$ and $\delta M_2$ can thereby be adjusted such that 
the propagator matrix develops poles at the on-shell chargino masses
$\mcha{1,2}$. In addition, the $\tilde{Z}^{\LL,\RR}$ factors can be uniquely
fixed by requiring that the propagator matrix is diagonal and that the pole
residues are normalized to unity for on-shell momenta.

{\bf 2.)} The analogous program can be carried out 
in the \underline{neutralino} system, though the
doubling of degrees of freedom renders the analysis more cumbersome. The
bilinear part of the neutralino Lagrangian in the current eigen-basis is given by
\begin{equation}
{\cal L}_{\rm n} = \frac{i}{2} \bigl [ {\psi^0}^\top \sigma^\mu \partial_\mu \,
 \overline{\psi^0} + \overline{\psi^0}^\top \bar{\sigma}^\mu \partial_\mu \,
 {\psi^0} \bigr ]
- \h \bigl [ {\psi^0}^\top Y \, {\psi^0} + \overline{\psi^0}^\top Y^\dagger \,
 \overline{\psi^0} \, \bigr ],
\end{equation}
with $\psi^0$ and $Y$ given in \eqref{eq:neuLagr} and \eqref{eq:Mneu},
respectively. In this representation the renormalization of the mass matrix $Y$
is defined as
\begin{align}
Y &\to Y + \delta Y, \nonumber
\intertext{with}
\delta Y &= \begin{pmatrix} 
\delta M_1 & 0 & -\delta (\MZ\,\sw\,\cbt) & \delta (\MZ\,\sw\,\sbt) \\
0 & \delta M_2 & \delta (\MZ\,\cw\,\cbt) & -\delta (\MZ\,\cw\,\sbt) \\
-\delta (\MZ\,\sw\,\cbt) & \delta (\MZ\,\cw\,\cbt) & 0 & -\delta \mu \\
\delta (\MZ\,\sw\,\sbt) & -\delta (\MZ\,\cw\,\sbt) & -\delta \mu & 0 \end{pmatrix},
\label{eq:Mneuct}
\end{align}
while the renormalization and the rotation from current fields $\psi^0$ to mass
eigen-spinors $\chi^0$ can be performed as
\begin{equation}
\psi^0 \to
N^\dagger \left( 1 + \hh \dzn \right) \chi^0.
\end{equation}
The matrix $\dzn$ absorbs the multiplicative renormalization of the
current fields as well as the renormalization of the rotation matrix.
Thus the renormalized neutralino Lagrangian can be cast into the form
\begin{align}
 {\cal L}_{\rm n} &\to {\cal L}_{\rm n} + \delta {\cal L}_{\rm n}, \nonumber \\
 {\cal L}_{\rm n} &= \h \begin{pmatrix}
   \overline{\neu_1},\!\!\!& \overline{\neu_2} \end{pmatrix}
 \left[ i\dslash - N^* Y N^\dagger \, \wL - N Y^\dagger N^\top \wR \right]
  \begin{pmatrix} \neu_1 \\ \neu_2 \end{pmatrix},
\intertext{together with the counter terms}
 \delta {\cal L}_{\rm n} &= \h \begin{pmatrix}
   \overline{\neu_1},\!\!\!& \overline{\neu_2} \end{pmatrix}
   \biggl [ i\frac{\dslash}{2} \bigl({\dzn}^\dagger + \dzn\bigr) \, \wL +
          i\frac{\dslash}{2} \bigl({\dzn}^* + {\dzn}^\top\bigr) \, \wR
  \nonumber \\
&\qquad -
    \bigl ( \hh {\dzn}^\top N^* Y N^\dagger +
     \hh N^* Y N^\dagger \dzn + N^* \delta Y N^\dagger \bigr ) \, \wL
    \\
&\qquad -
    \bigl ( \hh {\dzn}^\dagger N Y^\dagger N^\top +
     \hh N Y^\dagger N^\top {\dzn}^* + N \delta Y^\dagger N^\top \bigr ) \, \wR
    \biggr ] \begin{pmatrix} \neu_1 \\ \neu_2 \end{pmatrix}. \nonumber
\end{align}
where the matrix $N$ rotates the neutralino mass matrix into diagonal form
according to \eqref{eq:neumix}. The mass $\mneu{1}$ of the lightest neutralino
$\neu_1$, that will be under excellent experimental control, may be chosen to
define the remaining U(1) gaugino mass parameter $M_1$. The masses of the
heavier neutralinos are thereafter fixed uniquely by the Higgs/higgsino and
gaugino parameters $\mu$ and $M_{2,1}$. Again, the elements of the $\tilde{Z}^0$
wave-function renormalization matrix can be
adjusted such that the elements of the neutralino propagator matrix are diagonal
with unit residues of the mass poles for on-shell momenta.

In the case of CP conservation,
the renormalization of the Higgs/higgsino parameter $\mu$ and the gaugino
parameters $M_1$ and $M_2$ may be cast in the following form
\begin{align}
\delta M_2 &= \!\! \begin{aligned}[t]
  \Bigl [ &\hh (\mcha{2} \mu - \mcha{1} M_2)
  \; \re \! \bigl\{ \mcha{1} \seCL_{11}(\mcha{1}^2) +
  \mcha{1} \seCR_{11}(\mcha{1}^2) + 2\, \seCSL_{11}(\mcha{1}^2) \bigr\} \\
+ &\hh (\mcha{1} \mu - \mcha{2} M_2) \;
  \re \! \bigl\{ \mcha{2} \seCL_{22}(\mcha{2}^2) +
  \mcha{2} \seCR_{22}(\mcha{2}^2) + 2\, \seCSL_{22}(\mcha{2}^2) \bigr\} \\
+ & M_2 \; \dMW + \mu \; \delta \bigl ( \MW^2 \sin 2 \beta \bigr )  \Bigr]
\, /\, (\mu^2-M_2^2) ,
\end{aligned} \label{eq:M2-ct} 
\displaybreak[1] \\
\delta \mu &= \!\! \begin{aligned}[t]
  \Bigl [ &\hh (\mcha{2} M_2 - \mcha{1} \mu)
  \; \re \! \bigl\{ \mcha{1} \seCL_{11}(\mcha{1}^2) +
  \mcha{1} \seCR_{11}(\mcha{1}^2) + 2\, \seCSL_{11}(\mcha{1}^2) \bigr\} \\
+ &\hh (\mcha{1} M_2 - \mcha{2} \mu) \;
  \re \! \bigl\{ \mcha{2} \seCL_{22}(\mcha{2}^2) +
  \mcha{2} \seCR_{22}(\mcha{2}^2) + 2\, \seCSL_{22}(\mcha{2}^2) \bigr\} \\
+ & \mu \; \dMW + M_2 \; \delta \bigl ( \MW^2 \sin 2 \beta \bigr )  \Bigr]
\, /\, (M_2^2-\mu^2),
\end{aligned} \label{eq:mu-ct}
\displaybreak[1] \\
\delta M_1 &= \begin{aligned}[t]
  \frac{1}{N_{11}^2} \Bigl [ &\re \! \bigl\{
  \mneu{1} \seNL_{11}(\mneu{1}^2) + \seNSL_{11}(\mneu{1}^2) \bigr \}
  - N_{12}^2 \, \delta M_2 + 2 N_{13} N_{14} \, \delta\mu \\
+ \; &2 N_{11} \bigl [ N_{13} \; \delta \bigl (\MZ \sw \cos\beta \bigr )
                - N_{14} \; \delta \bigl (\MZ \sw \sin\beta \bigr ) \bigr ] \\
+ \; &2 N_{12} \bigl [ N_{13} \; \delta \bigl (\MZ \cw \cos\beta \bigr )
                - N_{14} \; \delta \bigl (\MZ \cw \sin\beta \bigr ) \bigr ]
\Bigr ], \end{aligned} \label{eq:M1-ct}
\end{align}
which is in agreement with \cite{chaneuren}. Here the following decomposition of
the chargino/neutralino self-energies has been used:
\begin{equation}
\Sigma^{\chi}_{ij}(p)  = \pslash \, \wL \Sigma^{\chi\PL}_{ij}(p^2) \;
  + \pslash \, \wR \Sigma^{\chi\PR}_{ij}(p^2)
  + \wL \Sigma^{\chi\PSL}_{ij}(p^2) + \wR \Sigma^{\chi\PSR}_{ij}(p^2).
\end{equation}
The combinations of self-energies in \eqref{eq:M2-ct}--\eqref{eq:M1-ct}
are equivalent to the counterterms of the on-shell chargino and neutralino
masses,
\begin{align}
\delta \mcha{k} &= \h \,
  \re \! \bigl\{ \mcha{k} \seCL_{kk}(\mcha{k}^2) +
  \mcha{k} \seCR_{kk}(\mcha{k}^2) + 2\, \seCSL_{kk}(\mcha{k}^2) \bigr\}
  \qquad [k=1,2],\\
\delta \mneu{1} &= \phantom{\h\,} \re \! \bigl\{
  \mneu{1} \seNL_{11}(\mneu{1}^2) + \seNSL_{11}(\mneu{1}^2) \bigr \}.
\end{align}
Once the two $\tilde{\chi} ^\pm _{1,2}$ chargino masses
 and the $\neu_1$ mass are fixed, the remaining
heavier neutralino masses are shifted by finite amounts relative to the Born
terms [which, by definition, are the eigenvalues of the renormalized mass matrix
\cite{chaneuren,majsusy}],
\begin{equation}
\mneu{k} - \mneu{k}^{\rm Born} = -\re \! \bigl\{
  \mneu{k} \seNL_{kk}(\mneu{k}^2) + \seNSL_{kk}(\mneu{k}^2) \bigr \}
+ (N^* \delta Y N^\dagger)_{kk} \qquad [k = 2,3,4].
\end{equation}
The cancellation of the divergences between the neutralino self-energies and
the mass matrix counterterm $\delta Y$ in this expression
is a non-trivial check of the method.

\subsubsection{Higgs mixing \boldmath $\tan\beta$}
\label{tanbeta}

At tree level, the Higgs mixing parameter $\tan\beta$ is determined by three
soft SUSY breaking parameters, 
the diagonal mass parameters $m_1^2$ and $m_2^2$ as well as the $H_{\rm
u}$--$H_{\rm d}$ mixing term $m_3^2$,
that is connected with the soft parameter $B\mu$.
They define the bilinear part of the scalar potential of the two Higgs doublets,
\begin{equation}
H_{\rm u} =
\begin{pmatrix}
  \phi^+_{\rm u} \\
  v_{\rm u} + \tfrac{1}{\sqrt{2}} (\phi_{\rm u} + i \rho_{\rm u})
\end{pmatrix}  
\qquad \mbox{and} \qquad
H_{\rm d} =
\begin{pmatrix}
  v_{\rm d} + \tfrac{1}{\sqrt{2}} (\phi_{\rm d} + i \rho_{\rm d}) \\
  - \phi^-_{\rm d}
\end{pmatrix}, \label{eq:higgsdoubl}
\end{equation}
\begin{equation}
\begin{gathered}
V_{\rm bilin} = 
 m_1^2 \left( \hh\phi_{\rm d}^2 +\hh\rho_{\rm d}^2 +|\phi^-_{\rm d}|^2 \right)
 + m_2^2 \left( \hh\phi_{\rm u}^2 +\hh\rho_{\rm u}^2 +|\phi^+_{\rm u}|^2 \right)
\\
+ m_3^2 \left( \phi_{\rm u} \phi_{\rm d} + \rho_{\rm u} \rho_{\rm d} +
                \phi^+_{\rm u} \phi^-_{\rm d} +
                {\phi^+_{\rm u}}^* {\phi^-_{\rm d}}^* \right).
\end{gathered}
\label{eq:vhiggs}
\end{equation}
The three soft SUSY breaking parameters can be reexpressed in terms of the
vacuum expectation values $v_{\rm u}$ and $v_{\rm d}$, and the pseudoscalar mass
\MA.
The tree-level relation $\tan\beta = v_{\rm u}/v_{\rm d}$ is modified by
higher-order corrections to the Higgs potential and needs to be renormalized.

Various renormalization prescriptions have been proposed in the literature,
which however do not lead to satisfactory solutions on all accounts
\cite{stoecki:02}. 
Renormalization prescriptions that are derived from the Higgs potential 
introduce dangerously large corrections to $\tan\beta$, which effectively
invalidate the convergence of the perturbation series. Other methods impose the
renormalization condition that the Goldstone bosons to not mix with the physical
Higgs bosons for on-shell momenta \cite{unmix}, but the value of $\tan\beta$
in these schemes depends on the gauge choice.

Alternatively, one may define $\tan\beta$ in higher orders by relating it to a
specific physical process. 
However, any definition of $\tan\beta$ through a physical process is
afflicted with technical difficulties that are introduced by the particular
process. Moreover, the value of $\tan\beta$ may be extracted from 
different observables, for example from Higgs decays
\cite{tb_higgsdec,stoecki:02},
Higgs production processes \cite{Gunion:2002ip}, or
from the cross-sections for mixed chargino pair
production for moderate values of $\tan\beta \lesim 10$ \cite{ckmz}, while for
large $\tan\beta$ the  $\tau$ polarization in $\tilde\tau$ decays provides an
attractive opportunity \cite{taupol}. Any such approach does not lead
to a unique and universal choice for the renormalization of $\tan\beta$.

In the following analysis we adopt the most convenient solution by just
subtracting the divergent part in the \drbar scheme from the unrenormalized
parameter to define the renormalized parameter. Without loss of generality,
the universal \drbar counterterm for
$\tan\beta$ may be extracted from the mixing self-energy of the $Z$-boson and the
pseudo-scalar Higgs boson $A^0$, with the natural choice 
for the renormalization scale
being the $A^0$-mass $\MA$:
\begin{equation}
\tan\beta \to \tan\beta + \delta \tan\beta
\qquad \mbox{with} \qquad
\delta \tan\beta = - \frac{1}{2 \cos^2\beta\,\MZ} \,\im\, \Sigma_{A^0Z}(\MA^2)
  \big|_{\rm div},
\end{equation}
with the subscript ``div'' indicating that only the divergent part of the
self-energy is retained.

Though being
process-independent, this definition is not perfect either as the value
depends on the chosen gauge. [By accident it remains independent on the
gauge fixing parameter $\xi$ in the $R_\xi$ gauge to one-loop order]. Of course,
the predictions for physical observables remain gauge independent as the gauge
dependence of $\tan\beta|_\drbar$ is neutralized by equivalent terms
in the amplitudes themselves.

\subsection{Effective Yukawa Couplings}

In general, quantum corrections are reduced with increasing mass of the virtual
particles inside the loops, as generally expected by the uncertainty principle
and formalized by the decoupling theorem \cite{Appelquist:75}. However, in
theories with broken symmetries, these corrections may grow to 
large values if the mass splitting in the particle multiplets is large. High
mass scales in theories with broken symmetries can thus manifest themselves in
the radiative corrections to precision observables at much lower energies. A
classical example for this phenomenon is provided by the mass splitting in the
top-bottom iso-doublet of the Standard Model \cite{oblique} which strongly
affects the $\rho$ parameter in the ratio of $W$- and $Z$-boson masses.

In theories with broken supersymmetry such a phenomenon arises if the splitting
between the masses of SM particles and some of the SUSY partners becomes very
large [such scenarios are realized, for instance, in focus point theories
\cite{focusp}],
leading to superoblique corrections that grow logarithmically with the mass splitting
\cite{eYuk, superoblique}. The
splitting affects in particular the relation between the gauge couplings and
the associated Yukawa couplings. In parallel to the bare couplings, the
renormalized Yukawa couplings can naturally be defined to be the same as the
gauge couplings at the renormalization point  if supersymmetry is broken
softly. Accordingly, in the on-shell renormalization scheme the renormalized
Yukawa couplings are defined to be equal to the renormalized on-shell gauge
couplings, so that the renormalized Lagrangian manifestly reflects the
supersymmetry [{\it modulo} the soft-breaking terms].
Nevertheless, the loops will modify \emph{in toto} the two types
of vertices associated with the two couplings differently in physical
amplitudes evaluated near the light SUSY scale, or equivalently the electroweak
scale.
In a more intuitive language, the running of the two couplings 
with energy from
the high SUSY mass scale down to the low SUSY mass scale (or electroweak scale)
is different \cite{eYuk,superoblique}.

\begin{figure}[tb]
\centering{
\begin{tabular}{c@{\hspace{2cm}}c}
\epsfig{file=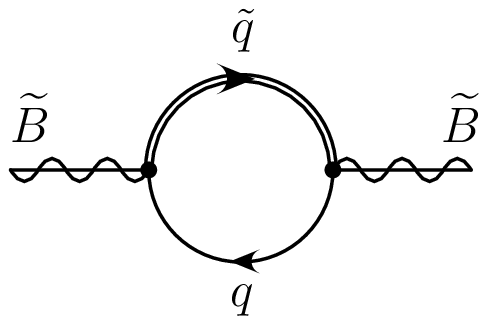,width=5cm} &
\epsfig{file=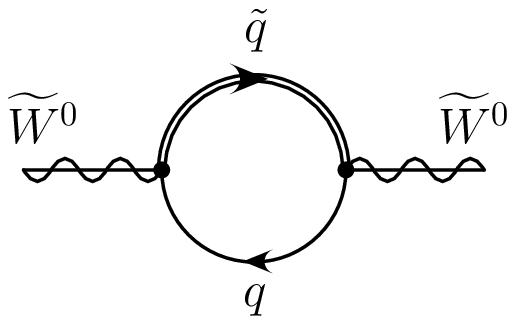,width=5cm} \\[-1em]
 ~~~(a) & ~~~(b)
\end{tabular}
}
\mycaption{Quark-squark loop corrections (a) to the U(1) and (b) to the SU(2) 
neutral gaugino self-energies.}
\label{fig:neuprop}
\end{figure}

These effects are rooted in the self-energies of the gaugino lines, shown in
Fig.~\ref{fig:neuprop}~(a) and (b) for bino and wino lines, respectively. In
parallel to the iso-multiplets of the Standard Model, the non-decoupling
corrections arise from large mass splittings within the supersymmetric particle
spectrum, for example for exceedingly high squark masses. The wave-function
renormalization associated with these diagrams can be projected onto effective
U(1) and SU(2) Yukawa couplings, $\hat{g}'_{\rm eff}$ and $\hat{g}_{\rm eff}$,
defined near the light SUSY/electroweak scale $\mwe$; in the same way as the
wave-function renormalization $Z^\gamma$ of the photon defines the effective
electromagnetic coupling $e_{\rm eff}$ = $\sqrt{Z^\gamma} \, e$.

The self-energy correction derived from Fig.~\ref{fig:neuprop}~(a)
for the U(1) bino propagator is given, in leading logarithmic order 
of the ratio between the large SUSY and the electroweak scale, by
\begin{equation}
\seN_{\rm \tilde{B},log}(p) = - \! \pslash \, \frac{g'^2}{16\pi^2} \,
\frac{11}{2} \, \ln \MsQ^2/\mwe^2.
\end{equation}
This may be reinterpreted as a shift of the effective U(1) Yukawa coupling,
\begin{equation}
\frac{\hat{g}'^2_{\rm eff}}{\hat{g}'^2_0} = 1 + \frac{g'^2}{16\pi^2} \,
\frac{11}{2} \, \ln \MsQ^2/\mwe^2  \label{eq:hshift}
\end{equation}
in leading logarithmic approximation, in relation to the bare Yukawa/gauge
couplings $\hat{g}'_0 = g'_0$. The effective U(1) gauge coupling $g'^2_{\rm
eff,log}$ can be introduced in parallel. It is identical to the renormalized
gauge coupling $g'$ in the on-shell scheme,
\begin{equation}
\frac{g'^2_{\rm eff}}{g'^2_0} 
= 1 + \frac{g'^2}{16\pi^2} \,
\frac{11}{6} \, \ln \MsQ^2/\mwe^2. \label{eq:gshift}
\end{equation}
The ratio of U(1) Yukawa to gauge coupling is therefore given effectively
by
\begin{equation}
\frac{\hat{g}'^2_{\rm eff}}{g'^2_{\rm eff}}
= 1 + \frac{g'^2}{16\pi^2} \, \frac{11}{3} \, \ln \MsQ^2/\mwe^2. \label{eq:hgdiff}
\end{equation}
Thus, even in spite of the fundamental identity of the renormalized Yukawa
and gauge couplings in supersymmetric theories in soft SUSY breaking
scenarios, we find nevertheless logarithmically enhanced departures
from universality in effective couplings at the electroweak scale
if the mass splitting in the
supersymmetric multiplets is large.

The SU(2) Yukawa and gauge couplings can be treated
analogously. From the wino self-energy in Fig.~\ref{fig:neuprop}~(b) the effective
SU(2) Yukawa coupling $\hat{g}^2_{\rm eff}$ can be defined. Together with
the on-shell renormalization of the SU(2) gauge coupling $g$, this leads to
\begin{equation}
\frac{\hat{g}^2_{\rm eff}}{g^2_{\rm eff}}
= 1 + \frac{g^2}{16\pi^2} \, 3 \, \ln \MsQ^2/\mwe^2
\end{equation}
for the ratio of the two effective Yukawa and gauge couplings in leading
logarithmic order.

The logarithmic growth of the one-loop corrections will later be analyzed
numerically when the production of selectron pairs, involving the Yukawa
couplings in t-channel neutralino exchange amplitudes, will be compared
with the production of smuon pairs in detail.

\subsection{Anomalous Thresholds}

\begin{figure}[tb]
\vspace{1ex}
\begin{tabular}{@{}p{6cm}p{10cm}@{}}
\raisebox{-2.2cm}{\psfig{file=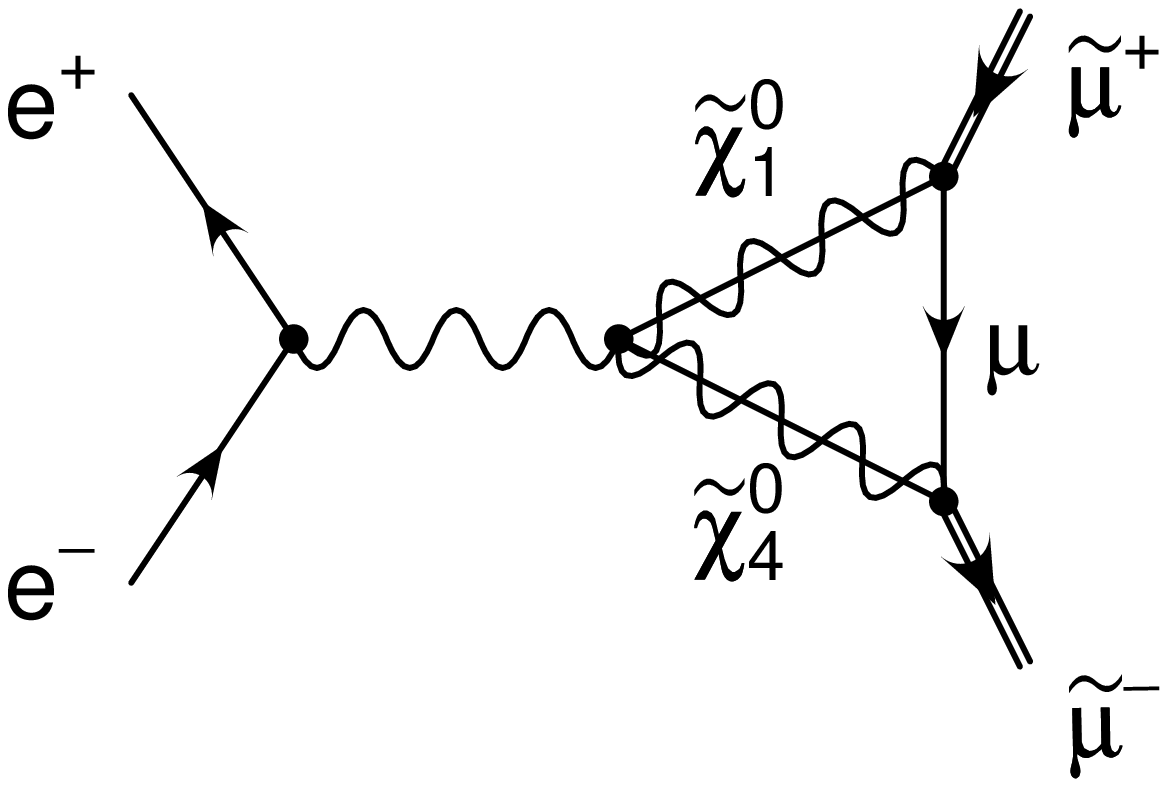,width=6cm}}
&
\mycaption{Vertex graph for the process $e^+e^- \to \smu^+ \smu^-$ that
bears an anomalous threshold for the mass hierarchy $\mneu{1} <
\msmu{} < \mneu{4}$.
}
\label{fig:anthr_diag}
\end{tabular}
\end{figure}
The rich pattern of different masses in supersymmetric theories gives rise
to anomalous threshold singularities \cite{anom,Liu:xy} in vertex and box graphs 
[which play no role
in general in the Standard Model\footnote{As an exceptional case,
anomalous thresholds do occur in $W^+W^-$ scattering in
the Standard Model [comment by T. Hahn].}]. While the vertex graph for 
$e^+e^- \to \smu^+ \smu^-$
in Fig.~\ref{fig:anthr_diag} generates a normal threshold singularity when the
energy $\sqrt{s}$ passes the threshold for $\neu_1 \neu_4$ production, an
additional anomalous singularity occurs for a special set of mass values
$\mneu{1} < \msmu{} < \mneu{4}$
at the kinematical point
\begin{equation}
s = s_{\rm a} \equiv 
\frac{\msmu{}^2 (\mneu{4}^2-\mneu{1}^2)^2}{(\msmu{}^2-\mneu{1}^2)
  (\mneu{4}^2-\msmu{}^2)}. \label{eq:athr}
\end{equation}
Here, as before, the muon mass has been neglected.
The singularity can be traced back to a zero value of the denominator function $D$ in the
vertex amplitude $I$,
in a configuration where all intermediate particles in the loop become
on-shell at the same time,
\begin{equation}
I = \int {\rm d}^4 q \; \frac{f(q, p_{\smu{}^+}, p_{\smu{}^-})}{D},
\qquad
D = q^2 \, \bigl [(q+p_{\smu{}^+})^2 - \mneu{1}^2 \bigr ] \,
  \bigl [(q+p_{\smu{}^+}-p_{\smu{}^-})^2 - \mneu{4}^2 \bigr ],
\end{equation}
where $f$ is a polynomial function.
Even though the singularity is mild
and can be integrated over, it leaves its trace in a discontinuity of the
cross-section.

\begin{figure}[p]
\hspace{-6mm}
\epsfig{file=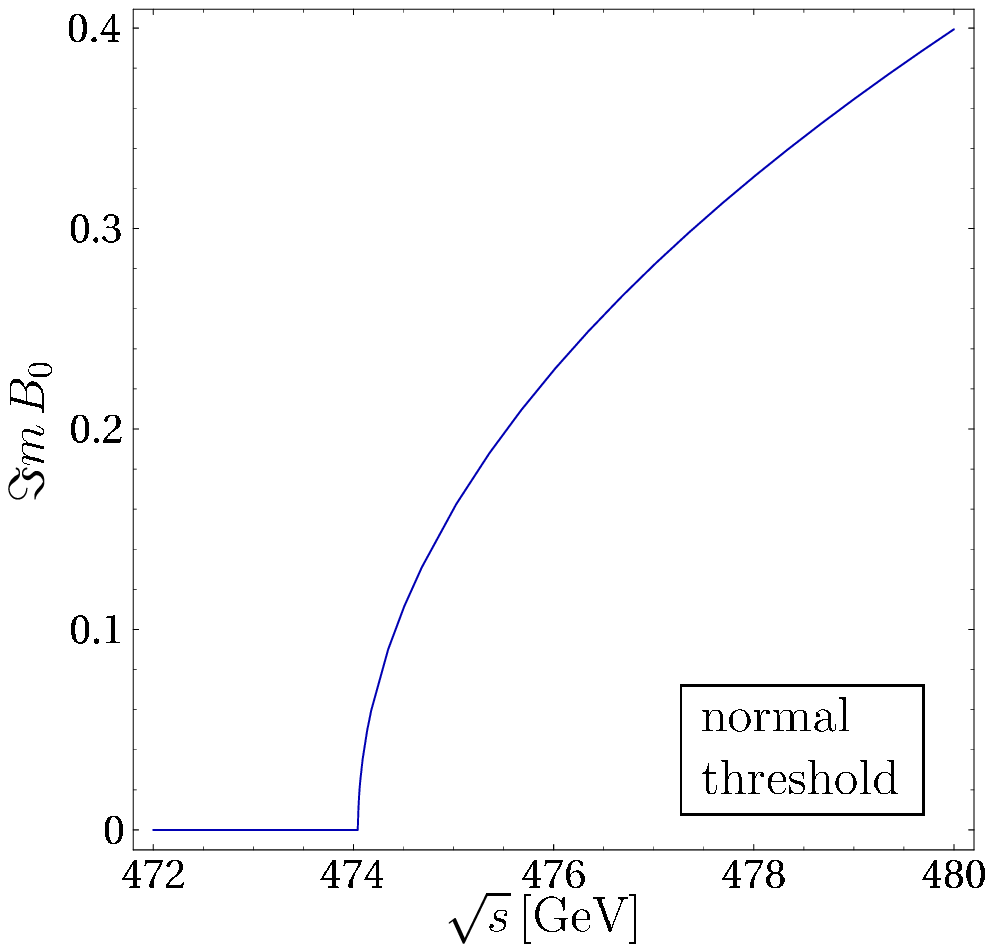,height=7.5cm}
\hspace{-4mm}
\epsfig{file=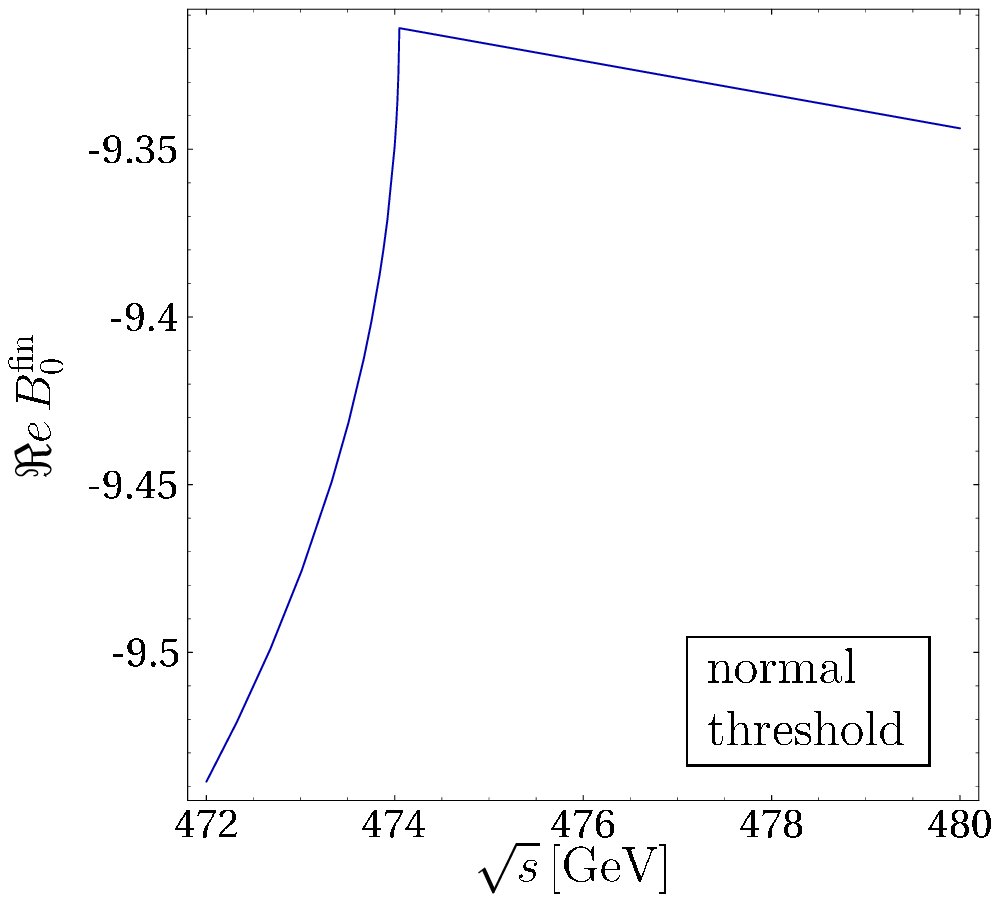,height=7.5cm}
\mycaption{Effect of the normal threshold in the finite part of the 2-point
function $B_0(s,\mneu{1}^2,\mneu{4}^2)$ at the branch cut $\sqrt{s_{\rm n}} =
\mneu{1}+\mneu{4} = 474.05$ GeV [in the SPS1a scenario].}
\label{fig:nthr}
\end{figure}
\begin{figure}[p]
\hspace{-6mm}
\epsfig{file=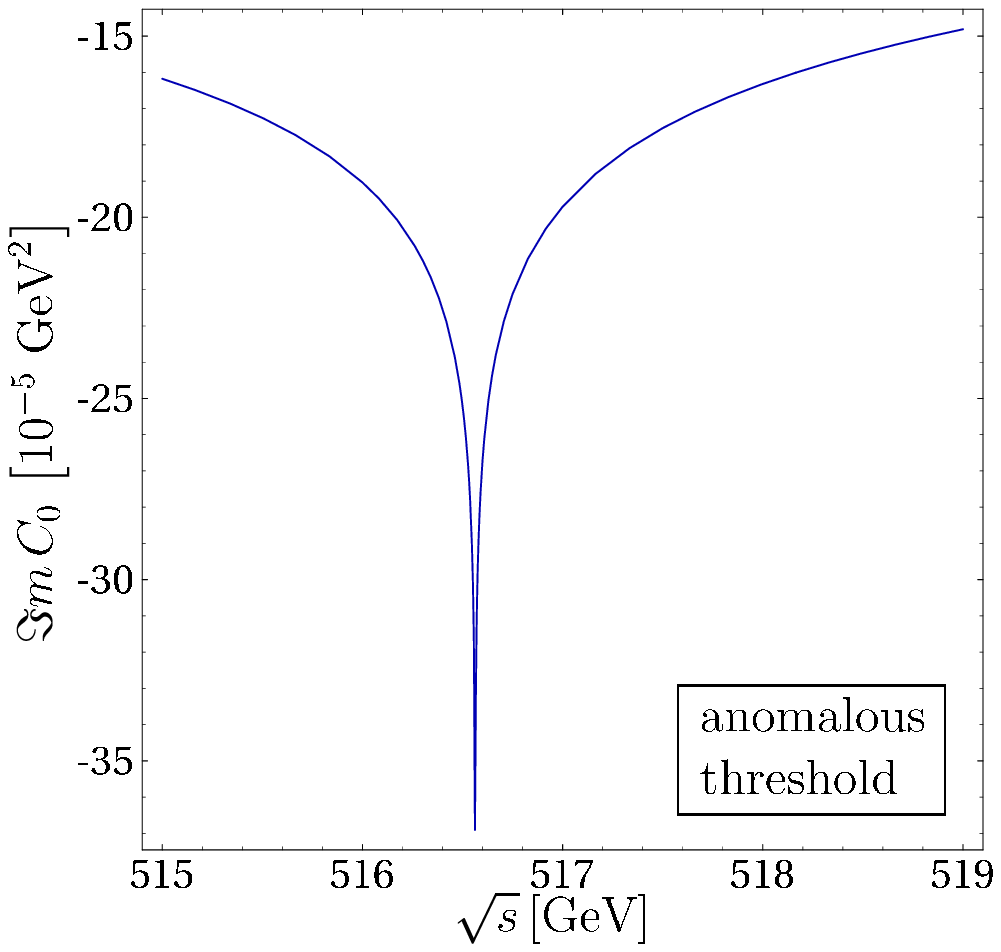,height=7.5cm}
\hspace{-3mm}
\epsfig{file=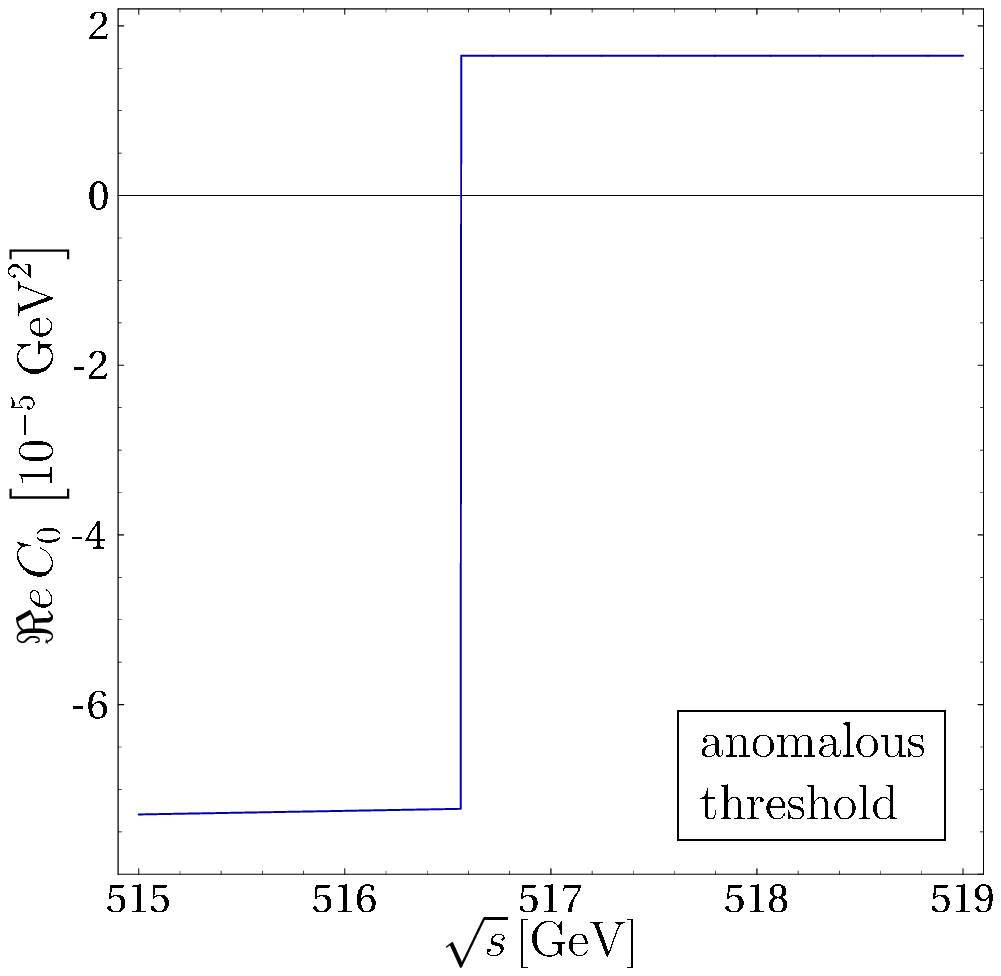,height=7.5cm}
\mycaption{Effect of the anomalous threshold defined in eq.~\eqref{eq:athr}
in the 3-point
function $C_0(s,\msmu{}^2,\msmu{}^2,\mneu{1}^2,\mneu{4}^2,0)$ at 
$\sqrt{s_{\rm a}} = 516.56$ GeV [in the SPS1a scenario].}
\label{fig:athr}
\end{figure}
The two different types of threshold singularities for the vertex introduced
above are exemplified in Fig.~\ref{fig:nthr} and Fig.~\ref{fig:athr}. The
normal threshold singularity at $\sqrt{s_{\rm n}}$ in Fig.~\ref{fig:nthr} is
generated by the non-zero onset of the imaginary part of the 2-point function
$B_0(s,\mneu{1}^2,\mneu{4}^2)$, corresponding to a kink in the real part. In
contrast, the anomalous threshold at $\sqrt{s_{\rm a}}$ in Fig.~\ref{fig:athr}
is generated by the $C_0(s,\msmu{}^2,\msmu{}^2,\mneu{1}^2,\mneu{4}^2,0)$
function, characterized by a kink in the imaginary part which leads, if
integrated out in a dispersion relation, to a step in the real part
\cite{anom}.

\subsection{Production Cross-Sections}

Mediated by the pure s-channel $\gamma/Z$ exchange mechanisms,
the production of smuon pairs is the most basic process
of supersymmetric theories at $e^+e^-$ colliders. The results
for the production cross-section will therefore be presented
for this process first. Subsequently we expand the analysis
to selectron pair production in $e^+e^-$ and $e^-e^-$ collisions which involve
also t-channel neutralino exchange mechanisms. They will serve
as an excellent instrument to measure the electron-selectron-gaugino
SU(2) and U(1) Yukawa couplings, which will be analyzed at the end of this section.

\subsubsection{Smuon production}

After the renormalized transition amplitude for the process $e^+e^- \to \smu^+
\smu^-$ is constructed following the way outlined in the last section, the
experimental parameters must be defined properly. We use the on-shell
definition for all masses, while the electromagnetic coupling $\alpha$ will be
evaluated at the scale of the center-of-mass energy $Q = \sqrt{s}$, so that the
large logarithmic corrections $\propto \log s/\mf^2$ from light fermion loops
in the running of $\alpha(Q^2)$ are absorbed into this definition.

The resulting amplitude is UV finite but still infrared divergent.
This divergence is removed by adding the contributions from photon
radiation in the initial and final states to the cross-section.
The virtual and real QED corrections form a gauge invariant subset separate from
the other virtual loop corrections.

\emph{Initial-state QED corrections:} Adding to the loop-corrected
cross-section the contribution of soft photon radiation from the
initial lepton lines, the ensuing cross-section factorizes into
the Born cross-section and a radiation coefficient that depends
on the cut-off $\Delta E$ of the soft photon energy (defined in the cms
frame),
\begin{equation}
{\rm d} \sigma_{\rm ISR}^{\rm virt+soft} = {\rm d} \sigma_{\rm Born} \,
\frac{\alpha}{\pi} \left [ \log \frac{(\Delta E)^2}{s}
\left( \log \frac{s}{\me^2} -1 \right) + \frac{3}{2} \, \log \frac{s}{\me^2} -2
+\frac{\pi^2}{3} \right ].
\end{equation}
This $\Delta E$ dependence is removed if the radiation of hard photons
is added to the cross-sections. We are still left, however, with the
logarithmic enhancement of the cross-section from collinear radiation
$\propto \log s/\me^2$. In leading logarithmic order the photon radiation
effectively just reduces the cms energy available for the final-state
particles \cite{Chen:1974wv}. This is described by the convolution
\begin{equation}
{\rm d} \sigma_{\rm LL}(s) = \int_{4 \msmu{}^2/s}^1
  {\rm d}z \; \Gamma_{ee}^{\rm LL}(2\alpha,z,s) \, \sigma_{\rm Born}(zs)
\label{eq:collfac}
\end{equation}
of the Born cross-section with the radiator function \cite{LL1}
\begin{equation}
\Gamma_{ee}^{\rm LL}(\alpha,z,Q^2) = \delta(1-z)
+\frac{\alpha}{2\pi} \,
\log\frac{Q^2}{\me^2} \; \frac{1+z^2}{1-z} \Big|_{z \leq 1-\epsilon}
\label{eq:collrad}
\end{equation}
(that can easily be generalized to higher orders \cite{LLexp}).
The variable $z$
denotes the energy fraction left to the electron/positron parton
after the radiation of the collinear photon. The additional non-collinear
photon radiation is treated numerically by applying Monte Carlo
integration techniques, with fast convergence after the leading
logarithmic order is subtracted analytically, as outlined above.

\emph{Final-state QED corrections:} After adding up the vertex correction
plus the final-state photon radiation, the cross-sections for soft photons
factorizes again in the Born cross-section and a radiation function
that depends on the photon cut-off energy $\Delta E$,
\begin{equation}
\begin{aligned}
{\rm d} \sigma_{\rm FSR}^{\rm virt+soft} = {\rm d} \sigma_{\rm Born} \,
\frac{\alpha}{\pi} \left \{ \log \frac{4(\Delta E)^2}{\msmu{}^2}  \right.
\!&\left [ -1 + \frac{1+\beta^2}{2\beta} \log\frac{1+\beta}{1-\beta}
\right ]  -2
 + \frac{1}{\beta} \log\frac{1+\beta}{1-\beta} \\
 + \frac{1+\beta^2}{\beta}
 & \left [ \log\frac{1+\beta}{1-\beta} \left( 1- \h \log\frac{4\beta^2}{1-\beta^2}
 \right) \right. \\
 & \qquad \left. \left. + \frac{\pi^2}{3} +
 {\rm Li}_2 \frac{1-\beta}{1+\beta} - {\rm Li}_2 \frac{2\beta}{1+\beta} \right]
 \right \}.
\end{aligned}
\end{equation}
Since the smuon mass is large, the velocity $\beta = (1-4\msmu{}^2/s)^{1/2}$
of the smuons
in the final state stays sufficiently away from unity not to generate
collinear singularities. The $\Delta E$ dependence is neutralized when
the hard photon contributions are added
and integrated out (numerically) to calculate
the total cross-section.

The initial and final-state QED corrections do not interfere in the
total cross-section as a consequence of CP invariance. 
However, the amplitudes do in general interfere in the
calculation of final-state distributions.

\begin{figure}[p]
(a)\\[-1em]
\anc\hfill
\epsfig{file=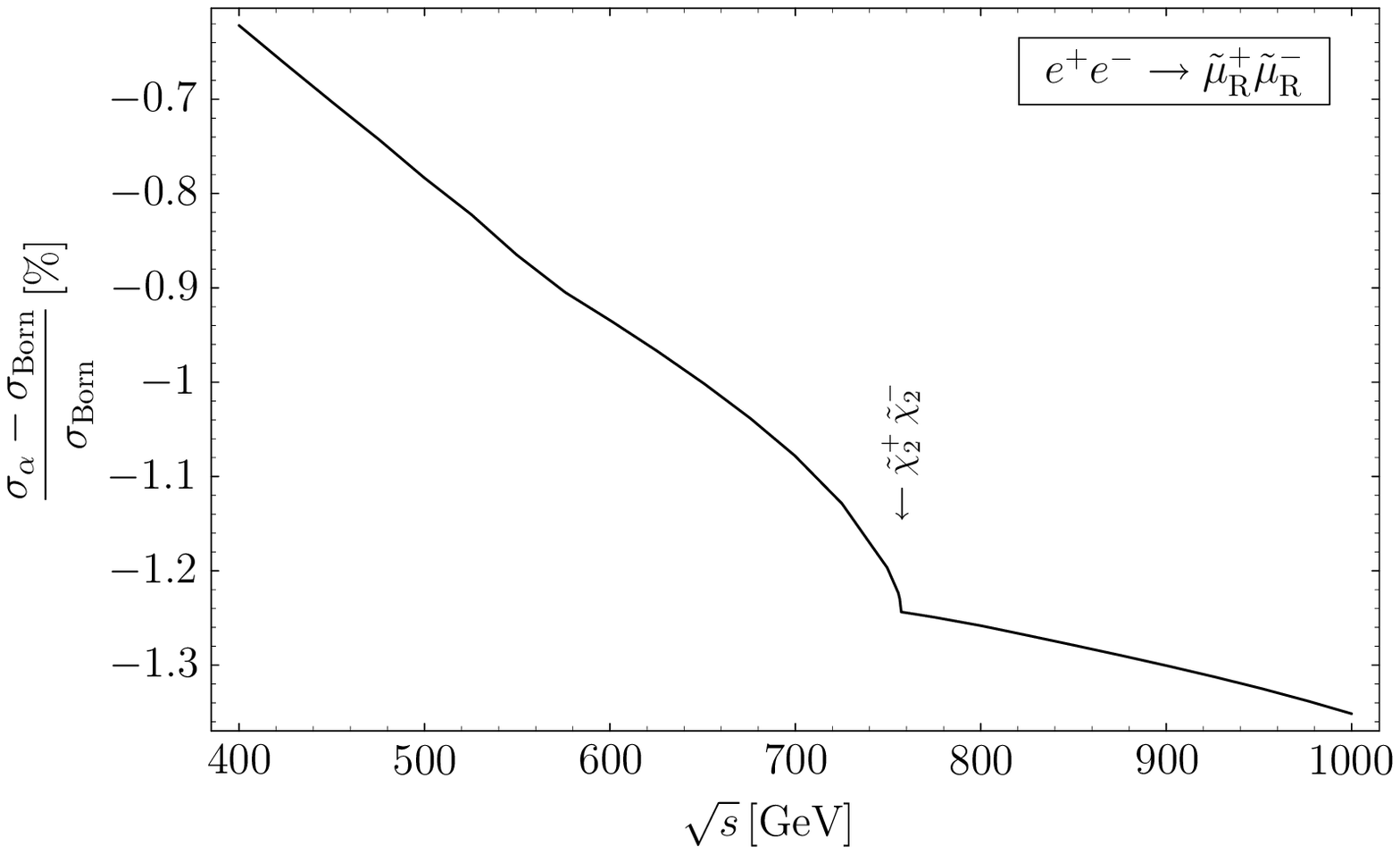,height=3.6in}\\[1em]
(b)\\[-1em]
\anc\hfill
\epsfig{file=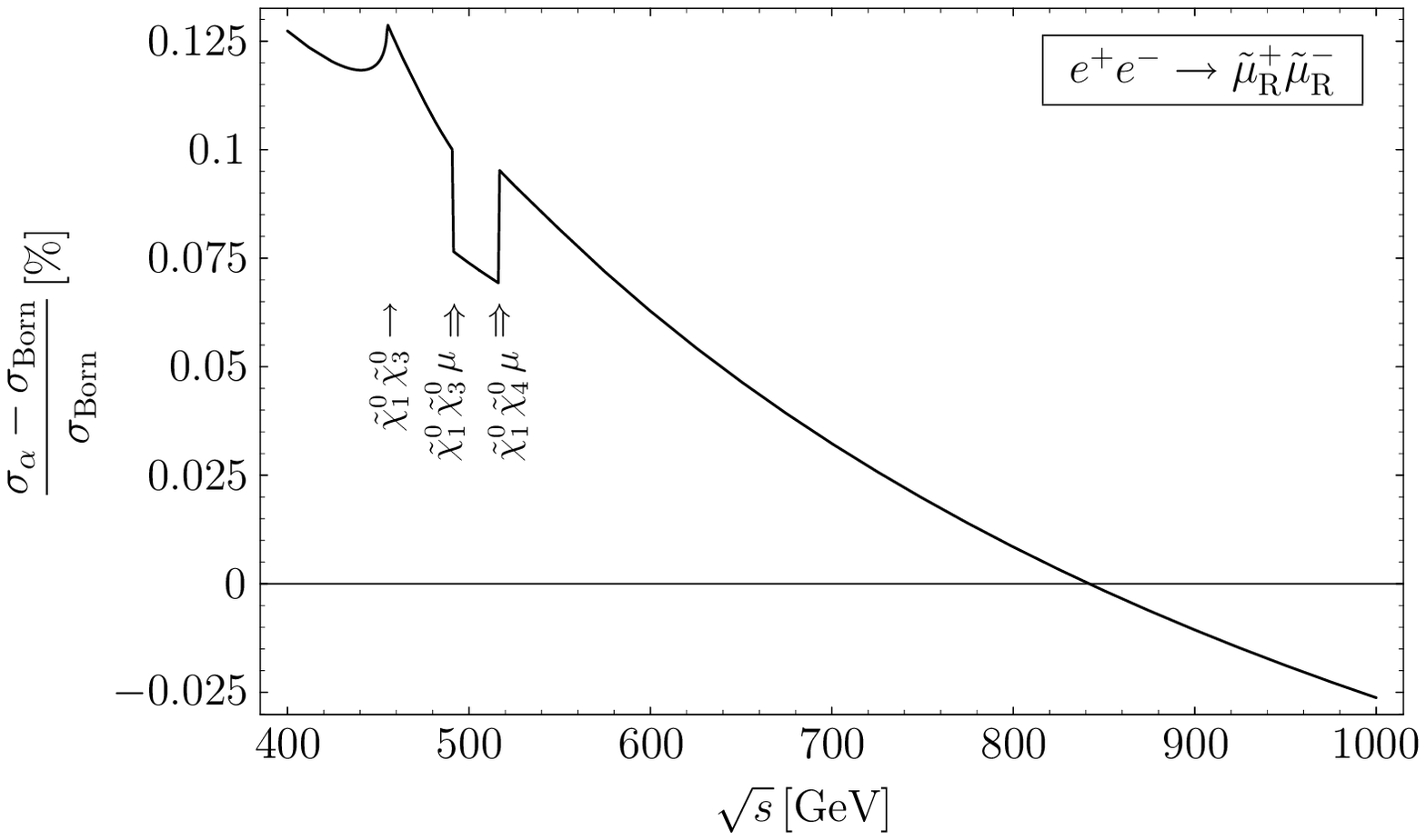,height=3.64in}%
\mycaption{Corrections from genuine SUSY loops to the cross-section for $e^+e^- \to
\smuR^+\smuR^-$ relative to the Born cross-section. The corrections are separated
into vertex and self-energy corrections (a) and box corrections (b).
The kinks are generated by normal (single arrows) and anomalous thresholds
(double arrows).
The values are given for the SPS1a scenario.}
\label{fig:smu_susy}
\end{figure}

\begin{figure}[p]
{\raggedright (a)\\[-1em]}
\centering{
\epsfig{file=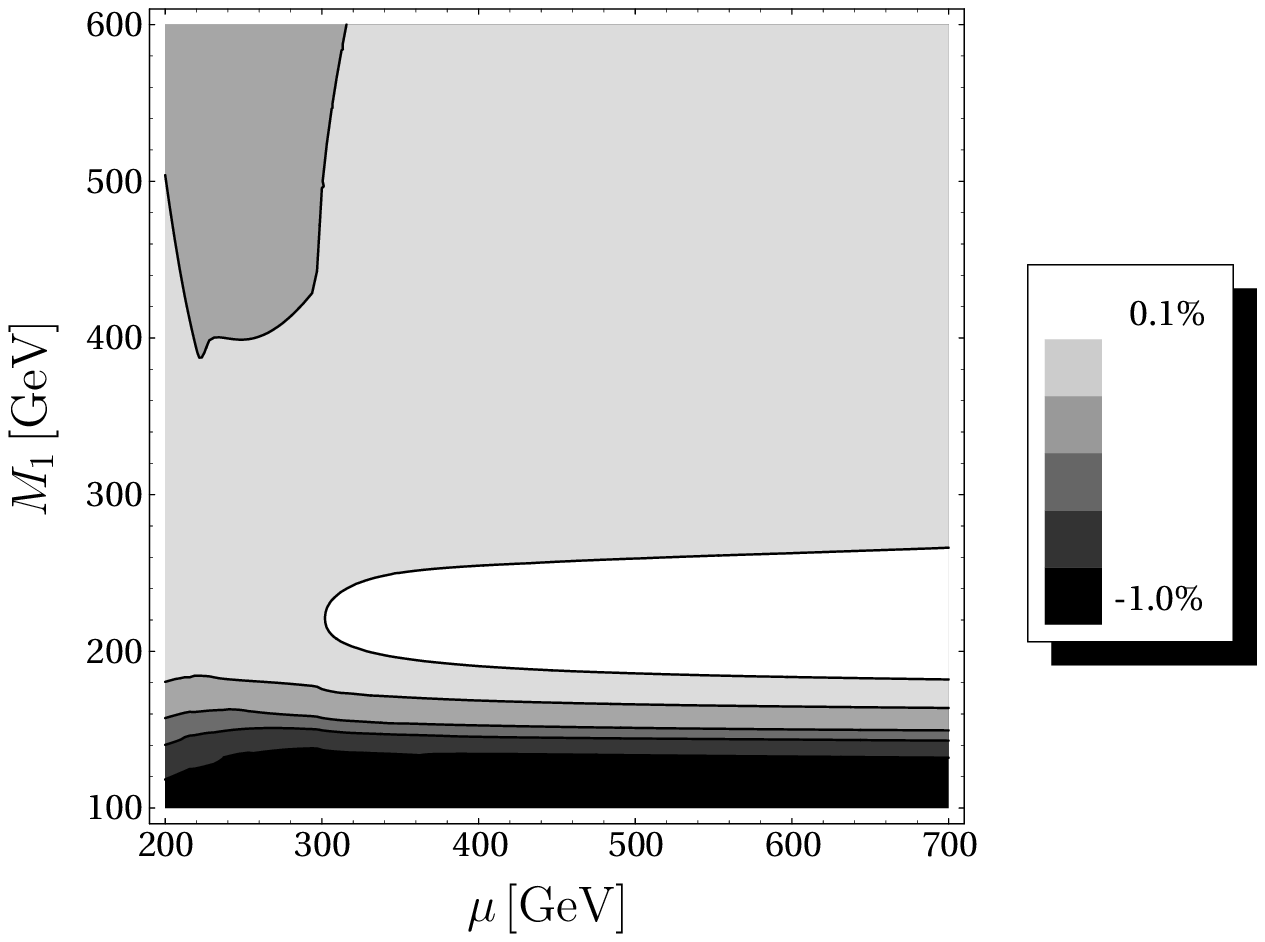,width=5in}
}\\
{\raggedright (b)\\[-1em]}
\centering{
\epsfig{file=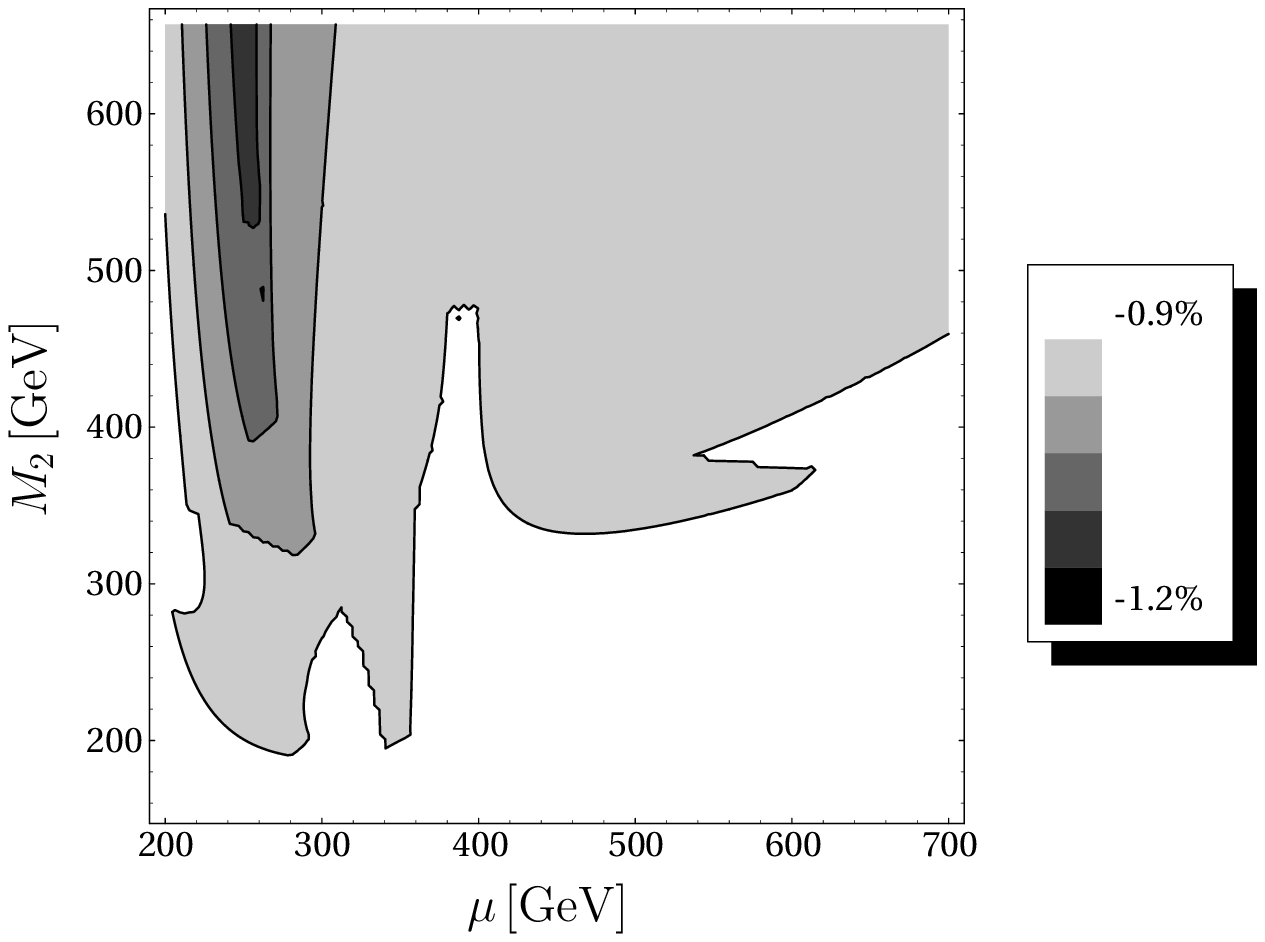,width=5in}
}\\
\vspace{-2ex}
\mycaption{Dependence of the relative one-loop corrections  
$(\sigma_\alpha - \sigma_{\rm Born})/\sigma_{\rm Born}$ to $\smuR$ pair
production on the gaugino parameters $M_1$, $M_2$ and $\mu$ for $\sqrt{s} = 500$ GeV.
The values of the other parameters are taken from the SPS1a scenario.
}
\label{fig:smu_mum2}
\end{figure}

In general it is not possible to divide the virtual loop corrections for slepton
pair production into SM-like corrections and genuine supersymmetric corrections.
However, for the special case of $\smuR$ pair production, a gauge-invariant
and UV-finite subset of SM-like loop contributions can be defined, including all
diagrams where a SM fermion, gauge boson or the lightest Higgs boson is attached
to the tree-level graphs, and taking the mixing angle $\alpha$ of the CP-even
Higgs bosons to be $\alpha = \beta - \pi/2$.
The remaining loop contributions can then be interpreted as the 
genuine virtual SUSY corrections. For this contribution
we find relative corrections of the order
of 1\%, as demonstrated in Fig.~\ref{fig:smu_susy},
which nicely illustrates the onset of normal and anomalous thresholds
in vertex and box diagrams with rising energy. The variation of the
corrections across the $[M_1,\mu]$ and $[M_2,\mu]$ planes has been studied
in Fig.~\ref{fig:smu_mum2}~(a,b).

\begin{figure}[tb]
\centering{
\epsfig{file=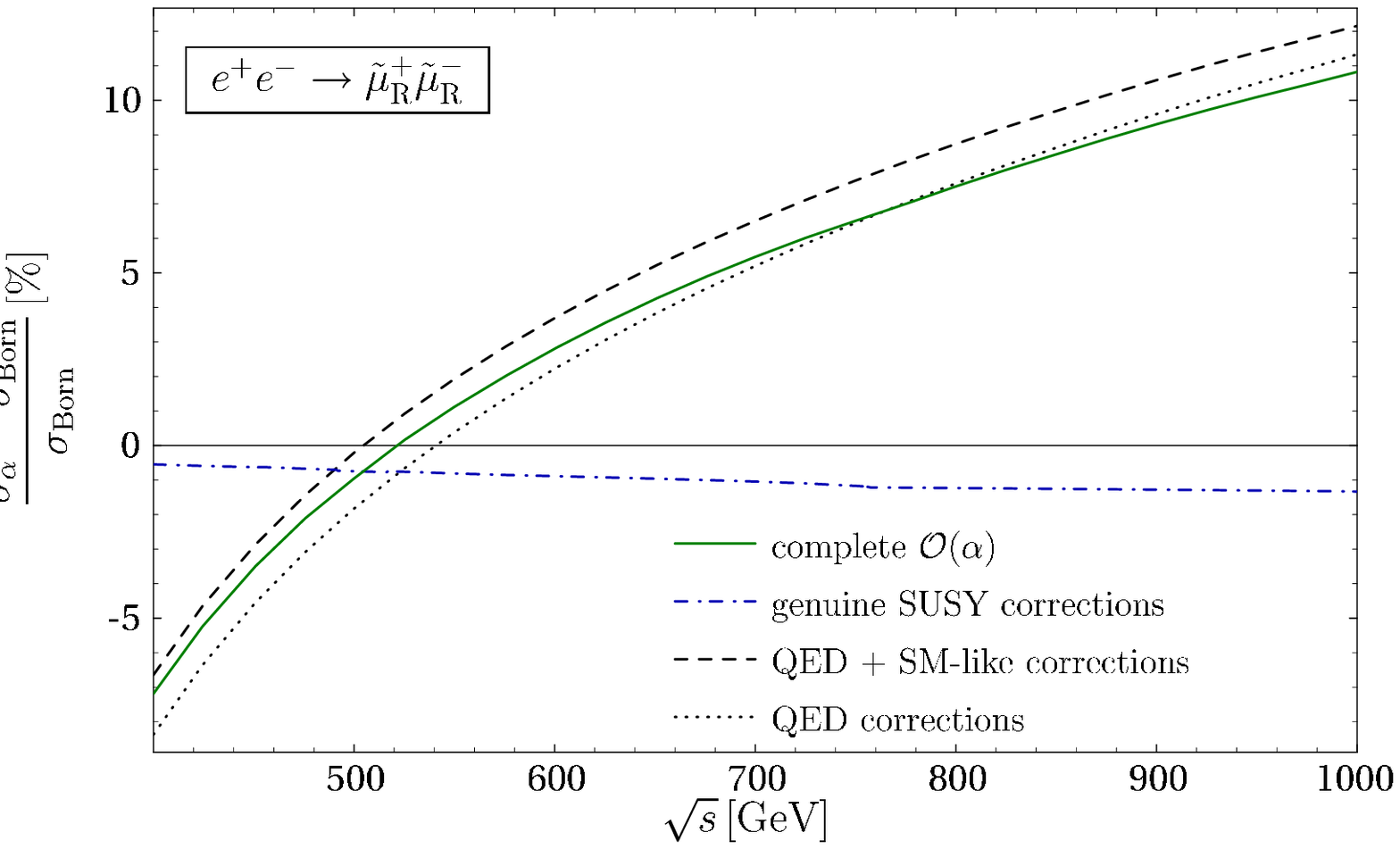,height=3.6in}%
}
\mycaption{Electroweak corrections to the cross-section for $e^+e^- \to \smuR^+
\smuR^-$, relative to the improved Born
cross-section. Besides the full $\OO(\alpha)$ result, contributions from
different subsets of diagrams are shown. Input parameters taken from 
the SPS1a scenario.}
\label{fig:smu_total}
\end{figure}
All crucial elements have now been collected to present the overall
correction of the total cross-section for $\smuR$ pair production. The
parameters of the Snowmass reference point SPS1a have been adopted again
to illustrate the final results. As a function of energy the correction
normalized to the Born cross-section, defined for the running electromagnetic
coupling, is displayed in Fig.~\ref{fig:smu_total}.
Moreover, the total correction is
broken down to initial-state plus final-state QED corrections,
the total SM corrections including just lines of SM particles in the
virtual corrections, and the genuine SUSY corrections introduced above.
The real photon radiation is treated in a fully inclusive way, {\it
i.e.} both soft and hard photon emission are included.
The main contribution to the corrections can trivially be traced back to the
universal factorizable QED terms which are logarithmically enhanced.
However, after these dominant effects
are subtracted, the remaining QED, weak-loop and genuine SUSY contributions
still amount to a level of 5\%.

Thus precision measurements of the total cross-sections for slepton pair
production, that may reach a level of a few per-mille, require the
one-loop radiative corrections to be included properly. In this way
a satisfactory understanding of the expected experimental results will
be achieved.

\subsubsection{Selectron production}

In comparison to smuon pair production, the loop calculations to
selectron pair production are significantly more complex, the reason
being twofold. An extra technical challenge is introduced by the additional
t-channel neutralino exchange mechanisms. These mechanisms also give rise
to delicate problems for gauge invariant subdivisions of diagram classes
and the subsequent renormalization procedures. The origin of the problems
is the continuous flow of charges from the initial to the final states while,
at the same time, the Fermi/Bose character of the charge line changes in
the electron-selectron-neutralino Yukawa vertex---a SUSY vertex \emph{sui
generis}. Related problems were encountered first in $WW$ pair production
via t-channel neutrino exchange as opposed to muon pair production in $e^+e^-$
collisions.

A transparent example is provided by the process $\eR^-\eR^- \to  \seR^-\seR^-$
which is built up solely by t-channel neutralino exchanges:

\begin{enumerate}

\item Closed loops of leptons and sleptons implanted in the virtual
neutralino lines form a gauge invariant subset of diagrams, and so do
loops of quarks and squarks.

\item The diagrams involving massive gauge bosons, Higgs bosons, gauginos
and higgsinos however cannot be separated from the QED loops in a
gauge-invariant manner anymore. This follows from a simple argument.
The set of photonic corrections to the electron-selectron-neutralino
Yukawa vertex is not UV finite, but only so after being supplemented by
the corresponding virtual photino diagram. Since the photino is not a mass
eigenstate, this amplitude is closely linked to the remaining degrees of
freedom in the electroweak sector. Thus only the total set of gauge boson/Higgs
and gaugino/higgsino electroweak diagrams is gauge invariant. 

Nevertheless, as expected on general grounds,
soft real-photon radiation regularizes the
infrared divergences generated by the virtual photon diagrams, and the
selectron pair cross-section for soft photon radiation factorizes again
into the Born term times a radiator function. In leading logarithmic order of
the soft-photon energy $\Delta E$,
\begin{equation}
\begin{aligned}
\anc\hspace{-1.3ex}
{\rm d} \sigma_{\rm ISR+FSR}^{\rm virt+soft,log} = {\rm d} \sigma_{\rm Born} \,
\frac{\alpha}{\pi} \log \frac{(\Delta E)^2}{s}
\left\{ \log \frac{s}{\me^2} \right.
&+ 2 \, \log 4 \biggl [ 1 + \log
  \frac{\me^2 \mse{}^2}{(\mse{}^2 - t)(\mse{}^2 - u)} \biggr ] \\
&+ \left. \frac{1+\beta^2}{2\beta} \log \frac{1+\beta}{1-\beta} \right\},
\end{aligned}
\end{equation}
where $t$ and $u$ are the invariant momentum transfers in the t-/u-channel.

In the same way, in leading logarithmic order,
collinear photon radiation from the initial beam line
can be cast in the convoluted form of eqs.~\eqref{eq:collfac}, \eqref{eq:collrad}.
Note however that virtual initial and final state radiation cannot (even not
theoretically) be disentangled anymore.

\end{enumerate}

The influence on parameters of the Higgs (and higgsino) sector is rather mild
since the couplings of these fields to electron-type lines is negligible.
Effects of Higgs bosons on the self-energies of the $Z$ boson and the neutralinos
may naively be expected non-negligible. However, the leading effects of the
Higgs boson spectrum on these parameters are only proportional to the logarithm
of the mass ratios of two Higgs bosons, {\it e.g.} $\propto \log \MA/\MH$. As a
result, these contributions are naturally suppressed.

\begin{figure}[p]
(a)\\[-1em]
\anc\hfill
\epsfig{file=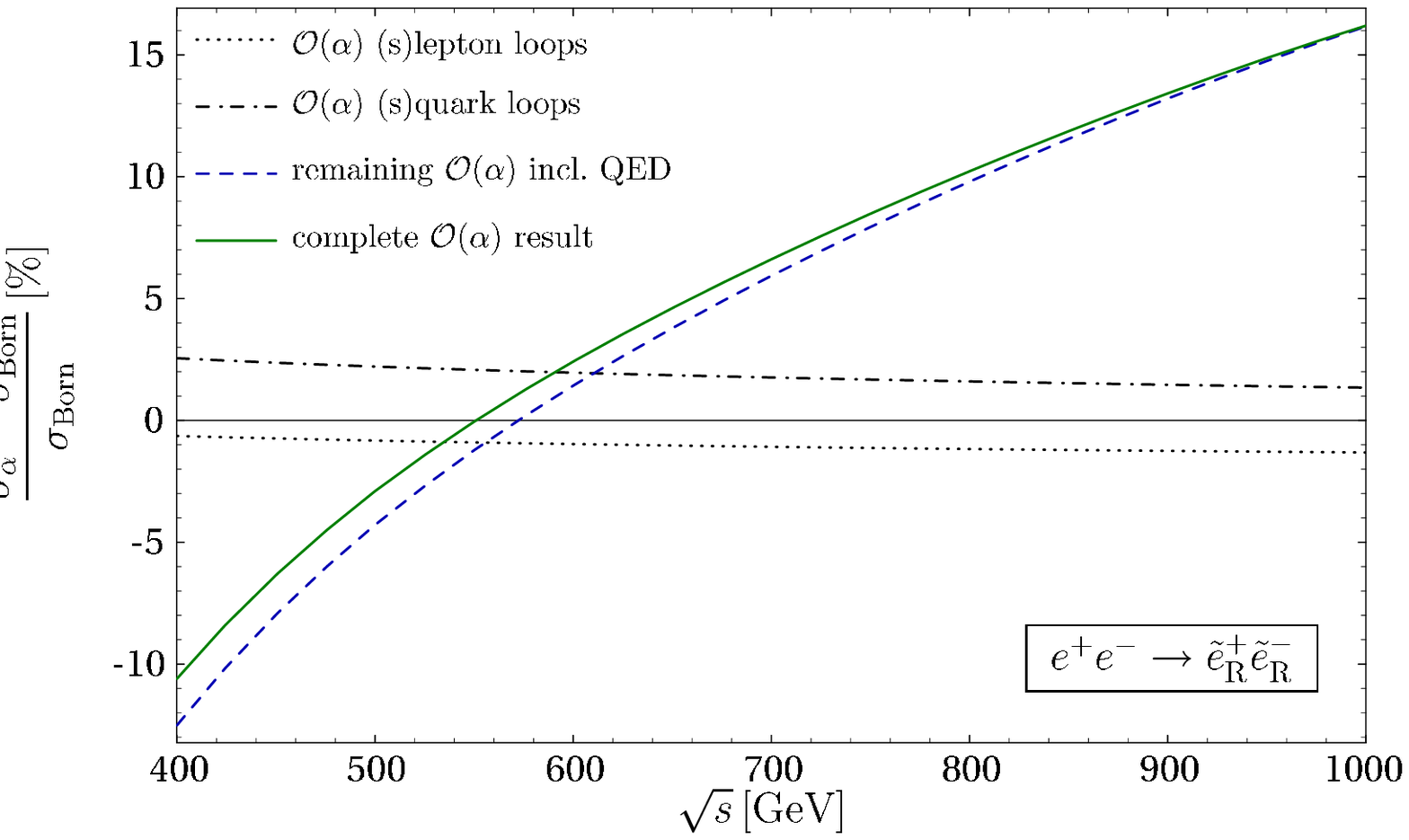,height=3.6in}\\[1em]
(b)\\[-1em]
\anc\hfill
\epsfig{file=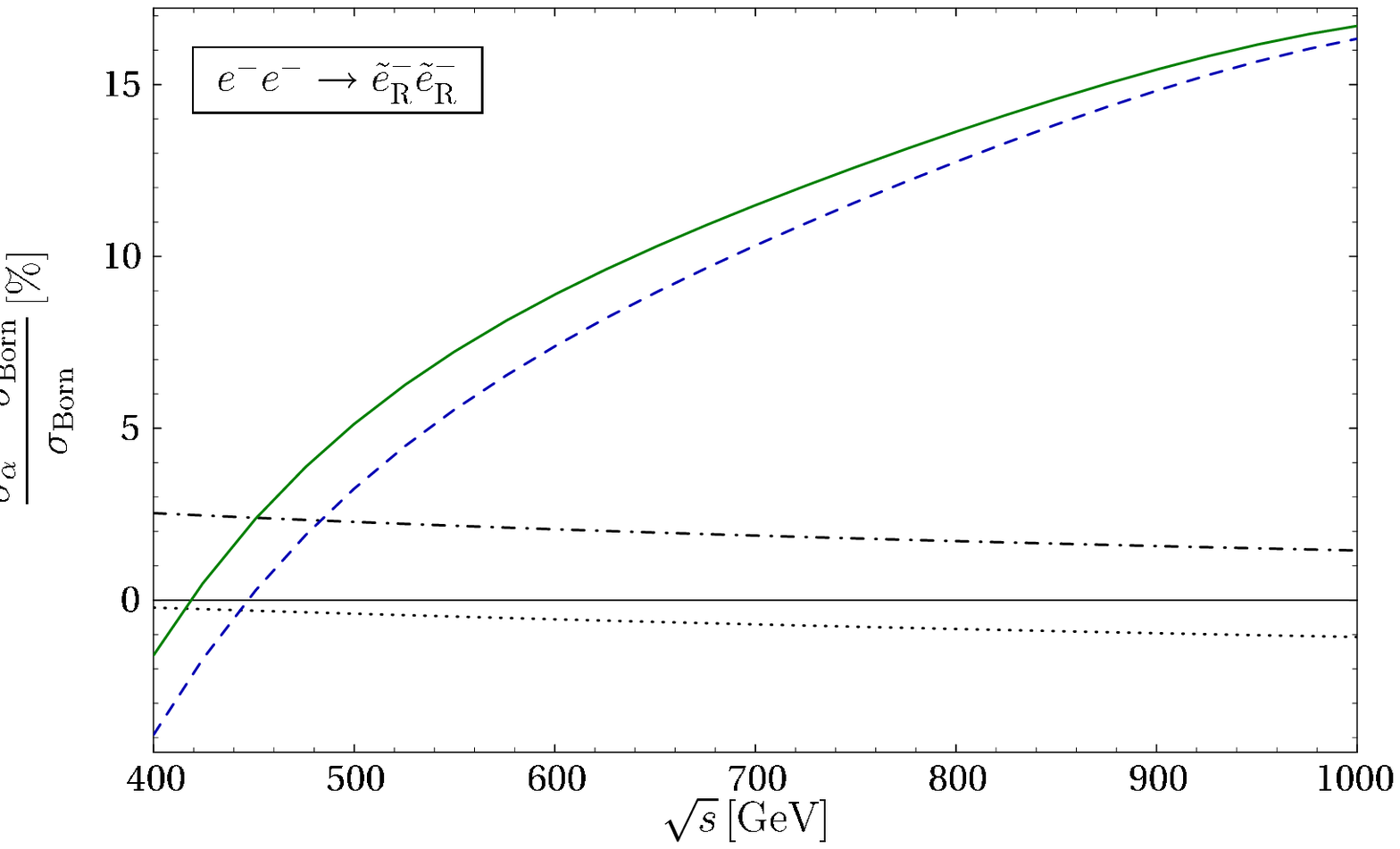,height=3.61in}
\mycaption{Electroweak corrections to the cross-sections (a) 
for $e^+e^- \to \seR^+
\seR^-$ and (b) for $e^-e^- \to \seR^- \seR^-$, relative 
to the improved Born
cross-section. Besides the full $\OO(\alpha)$ result, contributions from
different subsets of diagrams are shown. Input parameters taken from 
the SPS1a scenario.}
\label{fig:sel_total}
\end{figure}
The corrections to the RG improved Born cross-sections for selectron $\seR$
pairs in $e^+e^-$ and $e^-e^-$ collisions are depicted in
Fig.~\ref{fig:sel_total}~(a) and (b). In addition to the full corrections, the
results are broken down to the individual contributions from closed loops of
leptons/sleptons, quarks/squarks and the remaining corrections involving gauge
bosons, Higgs bosons, gauginos and higgsinos, as well as the QED corrections.

As before, we shall study the influence of the corrections induced through
the supersymmetry sector at some detail. Even though the genuine SUSY loops
are intimately correlated with the Standard Model loops, the higher-order
effects vary widely over the supersymmetry parameter space, measuring the
influence of the SUSY degrees of freedom beyond the trivial effects due to
the masses and couplings of the selectrons produced in the final state.

\begin{figure}[p]
{\raggedright (a)\\[-1em]}
\centering{
\epsfig{file=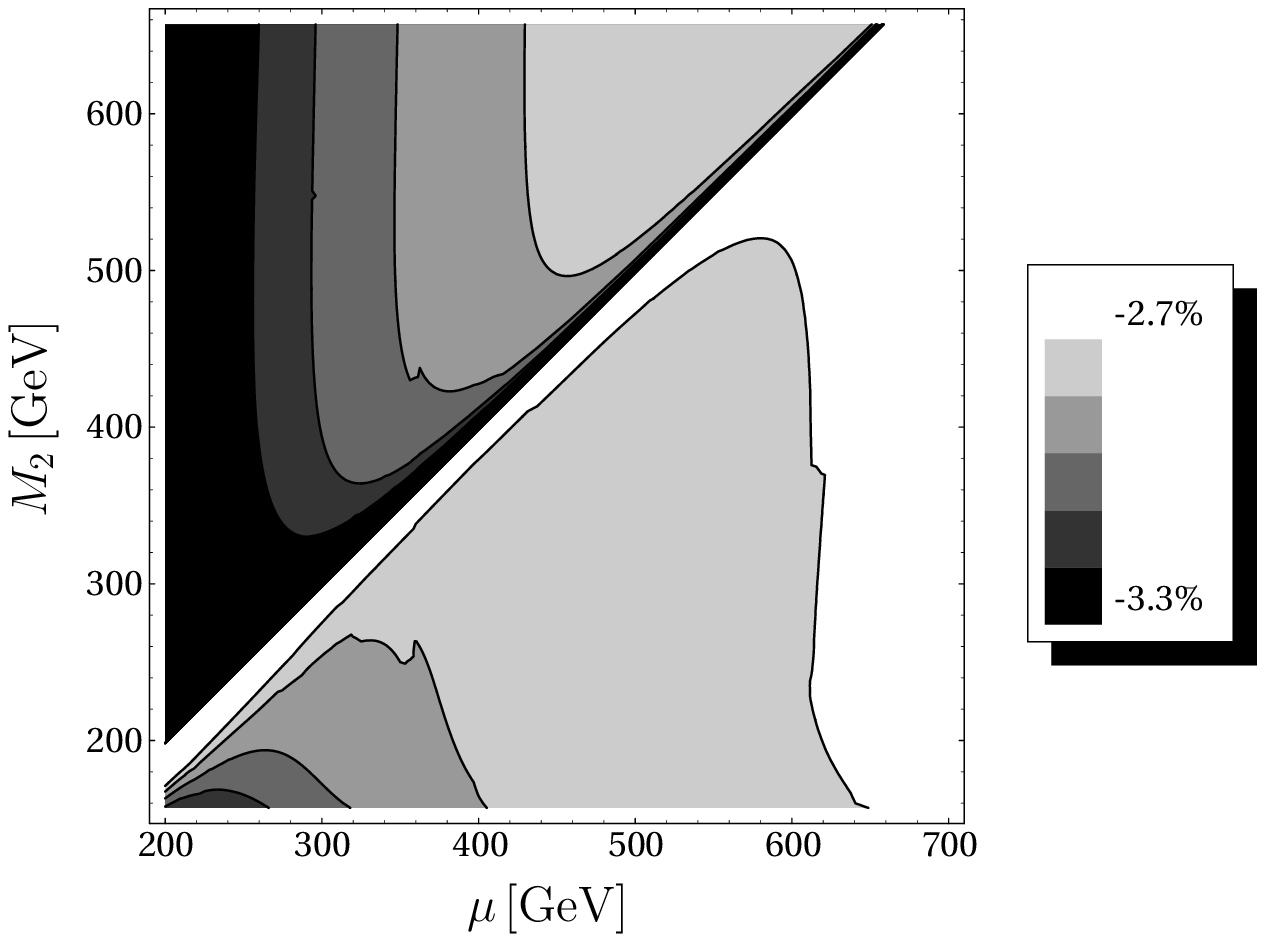,width=5in}
}\\
{\raggedright (b)\\[-1em]}
\centering{
\epsfig{file=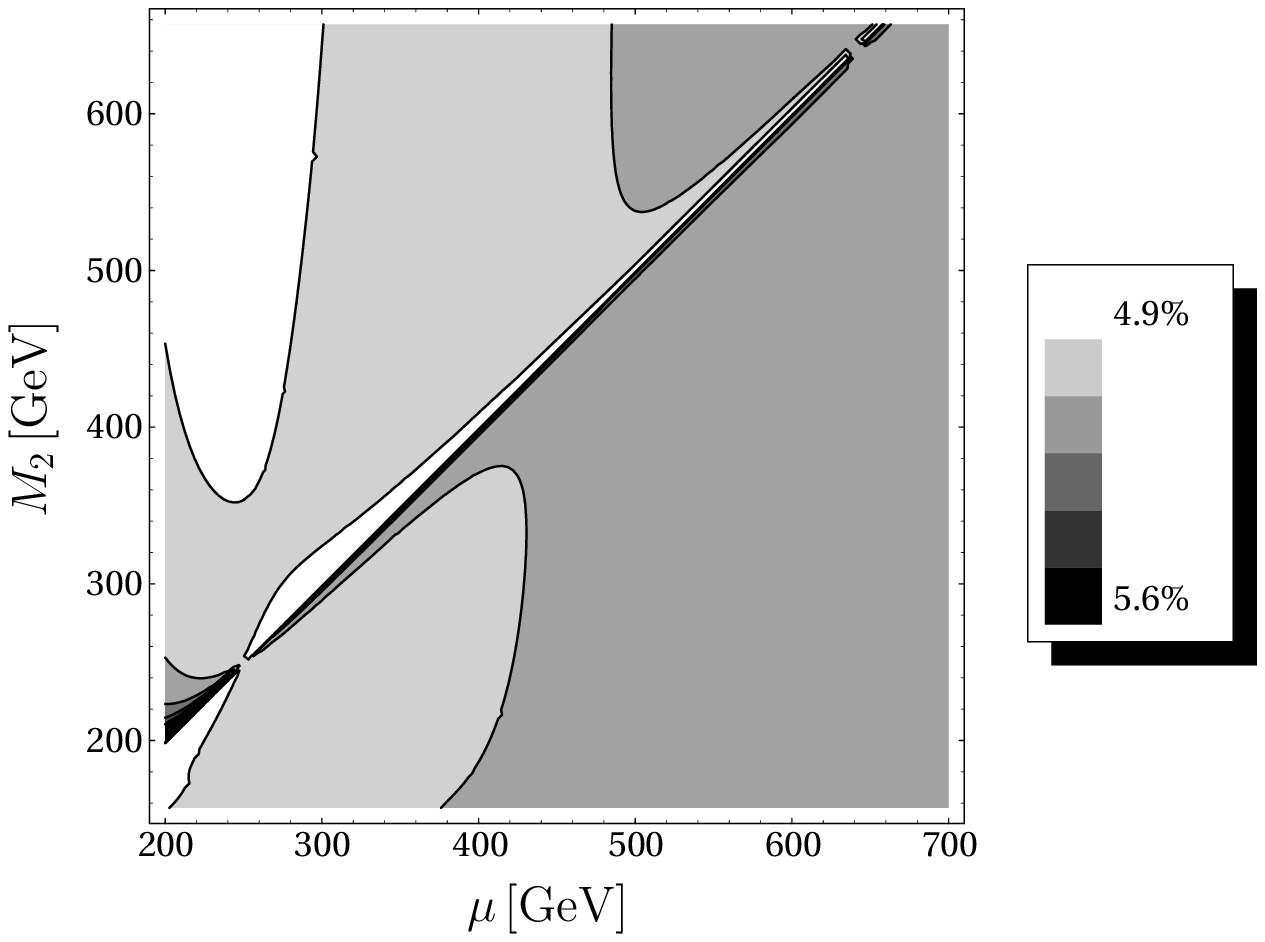,width=5in}
}\\
\vspace{-2ex}
\mycaption{Dependence of the relative one-loop corrections 
$(\sigma_\alpha - \sigma_{\rm Born})/\sigma_{\rm Born}$ to $\seR^+\seR^-$ production (a) and
$\seR^-\seR^-$ production (b) on the gaugino parameters $M_2$ and $\mu$
for $\sqrt{s} = 500$ GeV.
The values of the other parameters are taken from the SPS1a scenario.
}
\label{fig:mum2}
\end{figure}
A significant influence on the one-loop corrections arises from the electroweak
gaugino sector, characterized by the parameters $M_1$, $M_2$ and $\mu$. As an
example, the dependence of the one-loop corrections relative to the Born
cross-section on $M_2$ and $\mu$ is shown in Figs.~\ref{fig:mum2}~(a) and
(b). The effects
are maximal for small $\mu$ due to the higgsino loops affecting the $W$ and
$Z$ self-energies. [The rapid changes along the diagonal $M_2 = \mu$ are a
consequence of the level crossings between the $\neu_i$ states which induces
drastic changes in the couplings to the electroweak gauge bosons.]

\begin{figure}[p]
(a)\\[-1em]
\anc\hfill
\epsfig{file=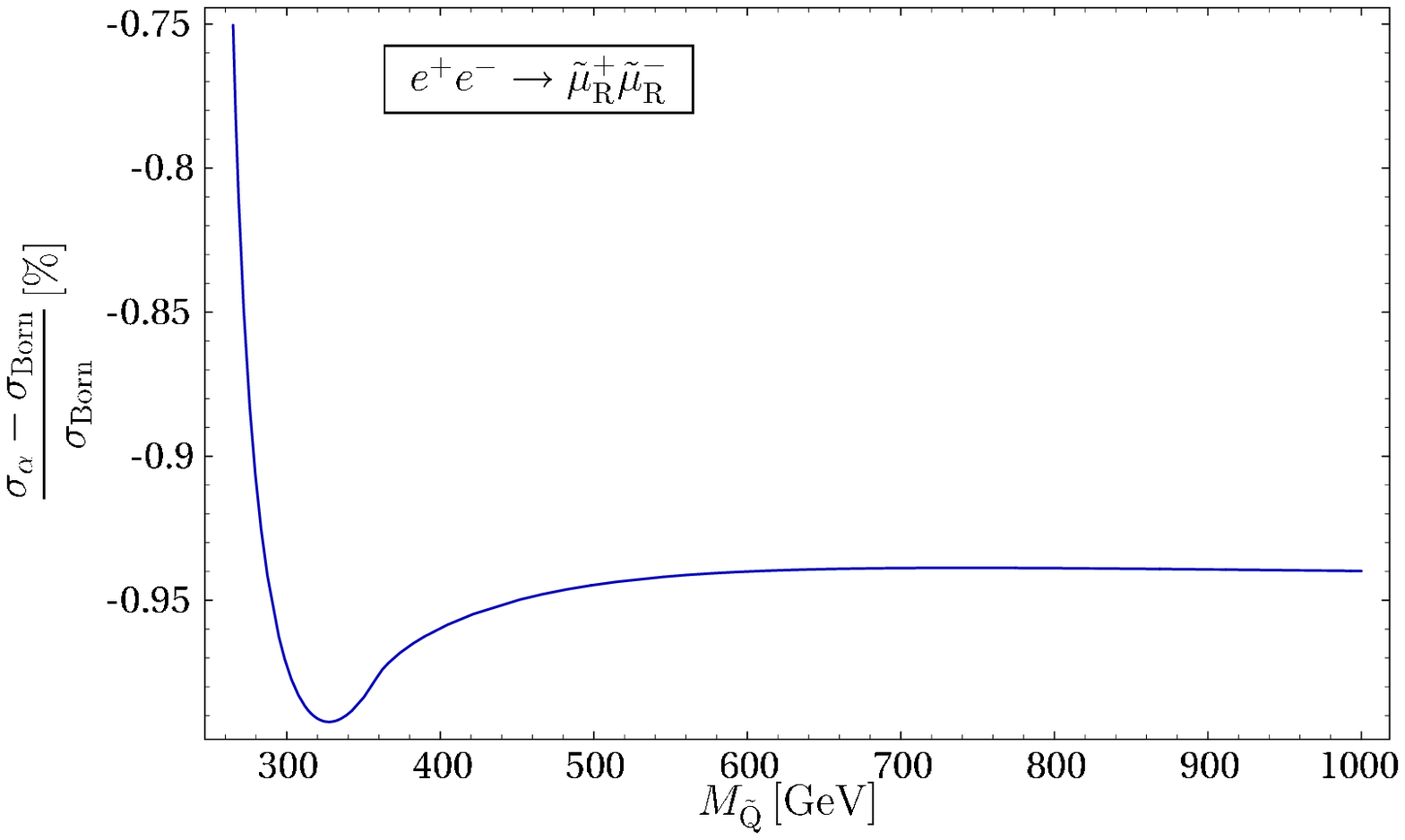,height=3.6in}\\[1em]
(b)\\[-1em]
\anc\hfill
\epsfig{file=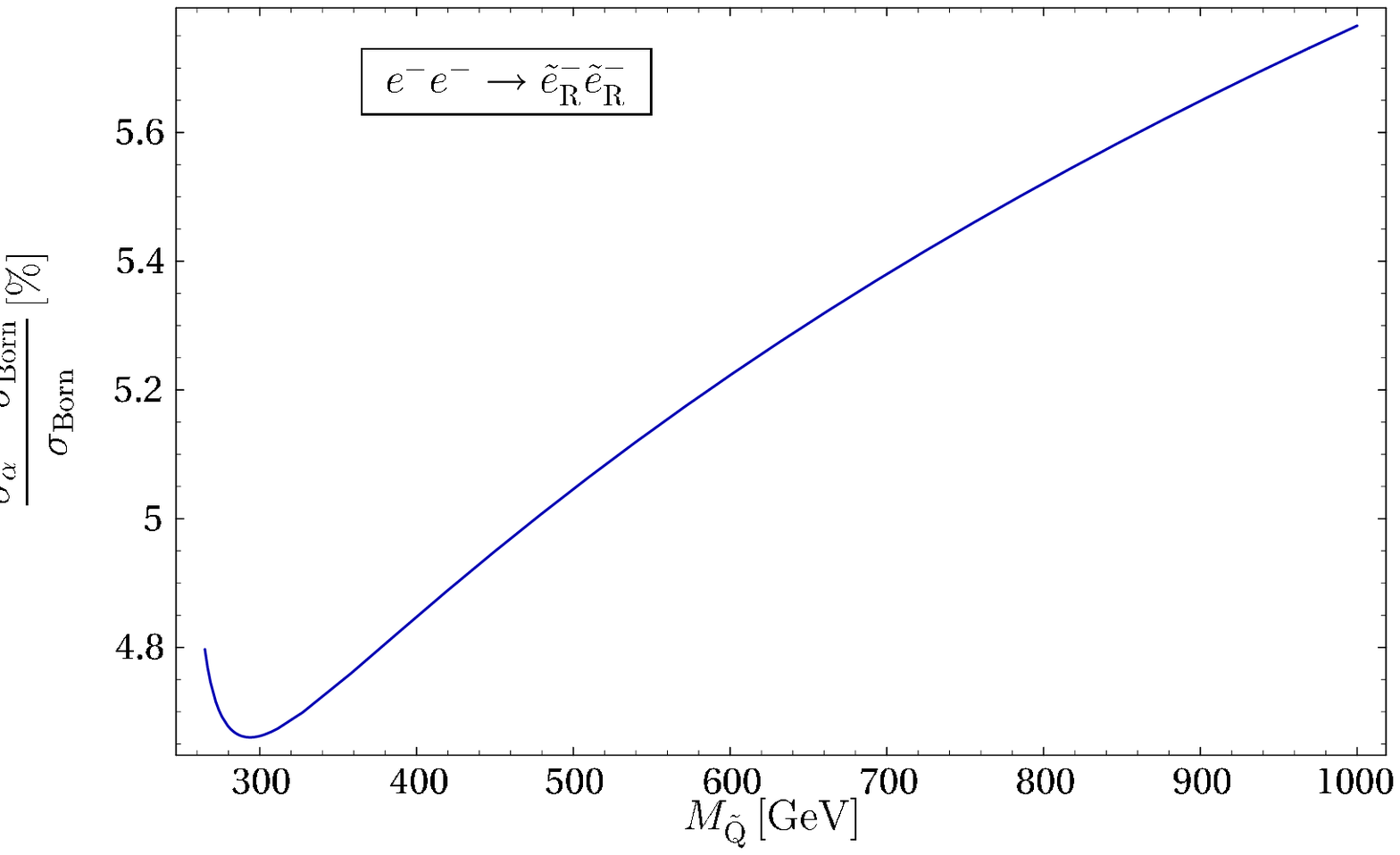,height=3.6in}
\mycaption{Dependence of the one-loop corrections on the soft-breaking
squark mass parameter $M_{\rm\tilde{Q}}$ [assumed to be universal for all
squarks] for (a) $\smuR^+
\smuR^-$ production and (b) $\seR^- \seR^-$ production.
The values of the other parameters are taken from the SPS1a scenario, and
the cms energy is set to $\sqrt{s} = 500$ GeV.
}
\label{fig:sel_msq}
\end{figure}

As outlined earlier, big mass differences
between the SUSY sfermions and the corresponding SM fermions generate large
effective Yukawa couplings and thus large
superoblique corrections to selectron production. This can nicely be
illustrated by comparing the squark loop effect on the smuon pair cross-sections
with the selectron pair cross-section mediated by t-channel neutralino
diagrams. Beyond the low-mass threshold region, the squark contributions are
rising linearly in the logarithm of the squark masses for selectron production
while approaching a plateau for smuon production, where gaugino/higgsino lines
are absent at the Born level, cf.\ Figs.~\ref{fig:sel_msq}~(a) and (b).

\subsubsection{The identity of SUSY Yukawa and SM gauge couplings}

Selectron-pair production in $e^+e^-$ collisions and, particularly, in $e^-e^-$
collisions are excellent channels for testing the identity of the SU(2)
and U(1) SUSY Yukawa couplings and the corresponding SM gauge couplings.
Selectron production in $e^-e^-$ collisions is mediated solely by neutralinos,
and the same is true for initial-state leptons of equal helicities in $e^+e^-$
channels, as evident from Table~\ref{tab:process}.

Asymptotically, the t-channel exchange is the dominant mechanism. As a
consequence of unitarity, the logarithmically leading part of the cross-section
for high energies is independent of the neutralino mixing parameters after all
neutralino exchanges in the t-channel are added up:
\begin{align}
\sigma[e^+ \, e^- \to \seR^+ \, \seR^-] &\limit{s \to \infty}
\frac{\hat{g}'^4}{16\pi} \, \frac{\log s}{s} + \OO \left( \frac{1}{s}
  \right), \label{eq:eRasymp}
\displaybreak[2] \\
\sigma[e^+ \, e^- \to \seL^+ \, \seL^-] &\limit{s \to \infty}
\frac{(\hat{g}^2 + \hat{g}'^2)^2}{16\pi} \, \frac{\log s}{s} +
  \OO \left( \frac{1}{s} \right), \label{eq:eLasymp}
\end{align}
where $\hat{g}$ and $\hat{g}'$ are the SU(2) and U(1) Yukawa couplings,
respectively.
However, very large energies indeed would be needed before the asymptotic
behavior is reached in practice.

In this subsection we shall study the sensitivity of selectron production to
the Yukawa electron-selectron-gaugino couplings and the errors expected
experimentally. We assume that the masses and mixing parameters of the
neutralinos have been pre-determined in chargino/neutralino pair-production,
and we properly take into account the expected errors in this sector. It is
important to note that the mixing parameters affecting the selectron production
cross-sections depend only on the gaugino/higgsino mass parameters $M_1$, $M_2$
and $\mu$. These parameters can be determined in the MSSM from the precision
measurements of three chargino and neutralino masses, for example,
$\cha_1^\pm$, $\cha_2^\pm$ and $\neu_1$, but independently of the
chargino/neutralino cross-sections [which would re-introduce the Yukawa
couplings otherwise].

To derive the size of the errors on the Yukawa couplings, we perform
this study in the one-loop approximation. Proper errors will be associated
to all terms in the improved Born approximation. However, the higher-order
corrections can be calculated for the ideal values of the parameters in the
loops since their errors would affect
the errors in the Yukawa couplings only to second order which is consistently
neglected. Iterative procedures may later be employed for the next-level
improvements.

The selectron production cross-sections are computed using the beam
polarizations and cuts introduced in Section~\ref{fsana}%
, and with beamstrahlung and ISR switched on.
Besides enhancing the statistics, the polarization of both the $e^-$ and
$e^+$ beams is essential for identifying the chiral quantum numbers of the
selectrons \cite{gudichep}.
Since only total
cross-section measurements will be considered here, the one-loop corrections
can be included by means of a simple K-factor. Besides the neutralino
parameters, the cross-sections crucially depend on the selectron masses, which
can be extracted from threshold scans, cf.\ Tab.~\ref{tab:massres}. For the
neutralino/chargino masses the following errors are assumed, 
$\delta \mneu{1} = 50 \mev$, $\mneu{2} = 80 \mev$,
$\delta \mcha{2} = 3000 \mev$, which
are based on a coherent analysis of LHC and LC mass measurements
\cite{lhclc}.
From these masses, the gaugino and higgsino mass parameters can be derived,
in models with two Higgs/higgsino doublets, to which we restrict our general
analysis. These parameters determine the elements of the
chargino and neutralino mixing matrices
including estimates of their errors.
An error of 1\% is assigned to the
polarization degree of the incoming electron/positron beams.

As discussed in Section~\ref{threshold}, the L- and R-selectron states can be
discriminated by considering the decay of the selectron $\seL$ into the
neutralino $\neu_2$, followed by the decay chain $\neu_2 \to \tau^+\tau^- \,
\neu_1$ and leading to the final states listed in Tab.~\ref{tab:massres}. It is
assumed that each tau pair can be identified with an efficiency of
$\epsilon_\tau = 80$\%. In addition a global acceptance factor of
$\epsilon_{\rm det} = 50$\% is assigned for potential detector effects, that
are not simulated in this study.

\begin{figure}[t]
\centering{
\begin{tabular}{l@{\hspace{1cm}}l}
 \underline{(a) $e^+e^-$, $\sqrt{s} = 500 \gev$, $L = 500$ fb$^{-1}$} &
 \underline{(b) $e^-e^-$, $\sqrt{s} = 500 \gev$, $L = 50$ fb$^{-1}$}
 \\[2ex]
\hspace{-1cm}
\epsfig{file=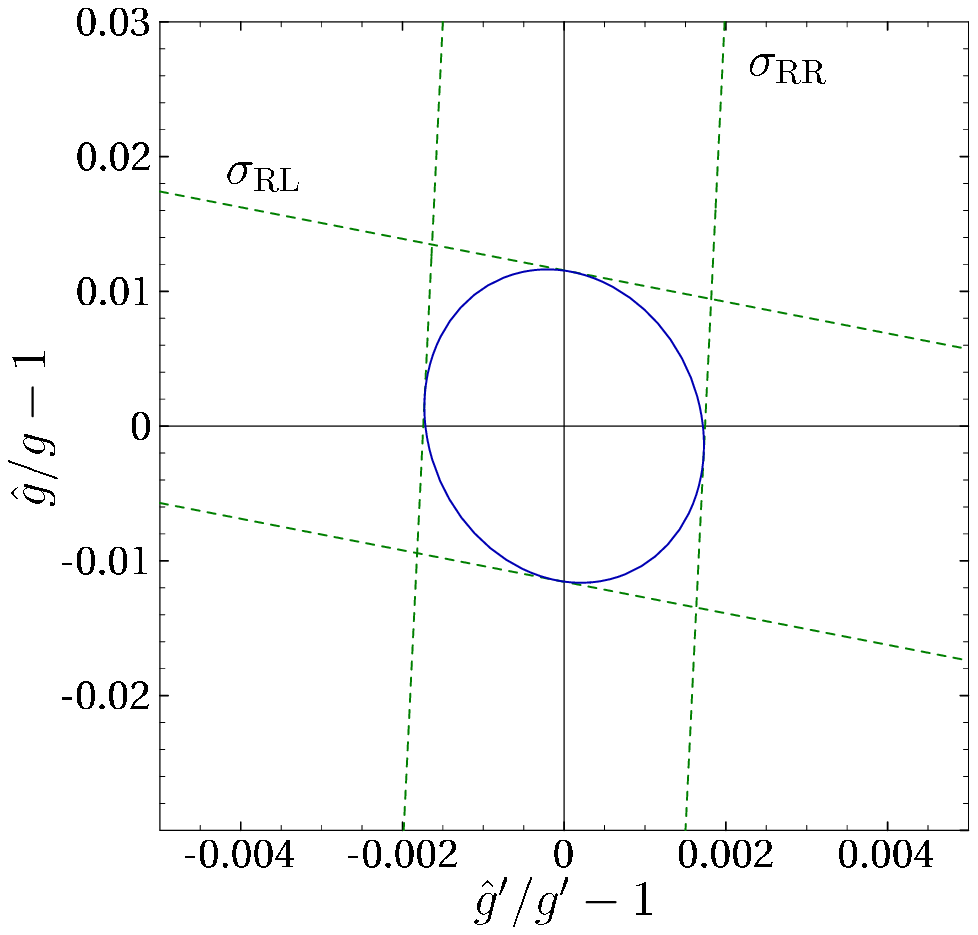,width=8.2cm} &
\hspace{-1cm}
\epsfig{file=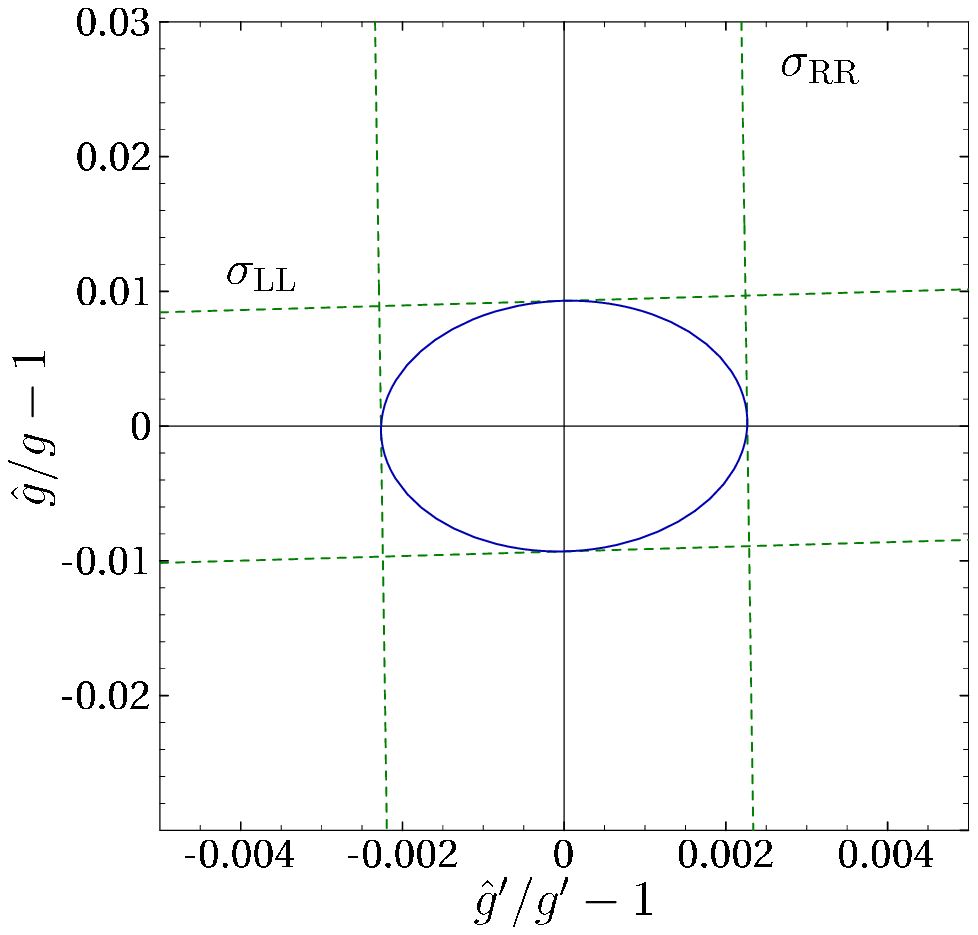,width=8.2cm}
\end{tabular}
}
\mycaption{1$\sigma$ bounds on the determination of the supersymmetric U(1) and
SU(2) Yukawa
couplings $\hat{g}'$ and $\hat{g}$ from selectron cross-section measurements.
The two plots compare the information obtained from
the cross-sections $\sigma_{\rm RR} =
\sigma[e^+ \, e^- \to \seR^+ \, \seR^-]$ and $\sigma_{\rm RL} =
\sigma[e^+ \, e^- \to \seR^\pm \, \seL^\mp]$ in the $e^+e^-$ mode (a)
as well as
$\sigma_{\rm RR} =
\sigma[e^- \, e^- \to \seR^- \, \seR^-]$ and $\sigma_{\rm LL} =
\sigma[e^- \, e^- \to \seL^- \, \seL^-]$ in the $e^-e^-$ mode (b), respectively.
Parameters taken from the SPS1a scenario.
}
\label{fig:ggp}
\end{figure}
Taking into account all these statistical errors and systematic uncertainties,
the constraints on the SU(2) and U(1) Yukawa couplings of the MSSM, $\hat{g}$ and
$\hat{g}'$, are presented in Fig.~\ref{fig:ggp}~(a) and (b) for the SPS1a scenario. 
The results are based on 500 fb$^{-1}$ data accumulated in $e^+e^-$ collisions
at 500 GeV, and 50 fb$^{-1}$ in $e^-e^-$ collisions, respectively. From the
overlap regions the following expected 1$\sigma$ errors are obtained:
\begin{align}
&e^+e^- : & \frac{\delta \hat{g}'}{\hat{g}'} &\approx 0.18 \%, &
                \frac{\delta \hat{g}}{\hat{g}} &\approx 1.2 \%, \\[1ex]
&e^-e^- : & \frac{\delta \hat{g}'}{\hat{g}'} &\approx 0.23 \%, &
                \frac{\delta \hat{g}}{\hat{g}} &\approx 0.9 \%.
\end{align}
As evident from these results, the expected sensitivity for the measurement of
the Yukawa couplings in the $e^+e^-$ and $e^-e^-$ modes are similar. 
While $e^-$ polarization is essential for disentangling the SU(2) and
U(1) couplings, the additional $e^+$ polarization reduces the errors by a factor
of about 1.4 for a degree of 50\%.
The result
for the determination of the U(1) bino Yukawa coupling is comparable to
previous studies, Ref.~\cite{susyid,superoblique2}, while being slightly more
precise  than analyses based on the differential selectron cross-section
\cite{eYuk}.

The precision on the Yukawa couplings expected from selectron pair-production
compares favorably with corresponding measurements in chargino/neutralino pair
production, for which the errors are of similar though slightly larger
magnitude \cite{superoblique2,ckmz}.

Thus it has turned out that one of the basic properties of supersymmetric
theories, the identity of Yukawa and gauge couplings, can be tested with
accuracies down to the per-cent and even per-mille level in selectron pair
production at $e^+e^-$ and $e^-e^-$ colliders.


\section{Conclusions}
\label{concl}

We have presented in this report the theoretical basis for high-precision
studies of the supersymmetric partners to muons and electrons, the scalar
smuons and selectrons, at future $e^+e^-$ and $e^-e^-$ linear colliders. The
theoretical material elaborated in the study is complemented by
phenomenological analyses of the masses and couplings of these particles.

\vspace{1ex}
\underline{{\bf Masses:}}
A central target of experiments exploring the properties
of supersymmetric particles is the measurement of their masses. The
experimentally observed particle masses are connected with mass parameters
in the SUSY Lagrangian that encode the breaking of supersymmetry and are
thus directly related to the basic structure of the supersymmetric theory
at the TeV scale. Extrapolating these parameters to high scales will
allow us to reconstruct the fundamental supersymmetric theory.

Precision measurements can be performed in the clean environment
of high-energy lepton colliders operating with polarized beams at high
luminosity.
The masses of scalar sleptons, selectrons in particular, can be determined
with unparalleled precision from threshold scans as the excitation curves
rise steeply with the energy above threshold, either with the third power
in the velocity for smuons or even linearly for selectron channels.

As a consequence, accurate theoretical predictions for the pair production
are required to match the expected experimental accuracies. The two
theoretical key points in this context, non-zero width effects and
Coulombic Sommerfeld rescattering effects, have been elaborated in
detail. Special attention has been paid to preserving gauge invariance in
truncated subsets of the entire ensemble of Feynman diagrams describing
the final states after the resonance decays. The remaining contributions of this
ensemble are taken into account as part of the SUSY backgrounds,
with the additional backgrounds from SM sources added on.

Based on this procedure, a phenomenological analysis of slepton masses in
threshold scans was performed, improving significantly on the theoretical
reliability compared with earlier simulations. While the excitation curves are
characterized by their distinct rise near the thresholds, including
sub-dominant backgrounds reduces, somewhat, the precision expected from
previous studies. Nevertheless, a precision of about a few 100 MeV can be
expected for slepton masses around 200 GeV in general, corresponding to a
relative error at the per-cent to per-mille level. For the R-selectron mass, an
accuracy of even $0.2 \times
10^{-3}$ can be obtained from threshold scans in the $e^-e^-$ mode, benefiting
from the exceptionally sharp rise of the S-wave selectron excitation.

Moreover, the threshold scans of selectron pair production can also be
exploited to extract the decay widths to accuracies between 10 and 20\%.

\vspace{1ex}
\underline{{\bf Yukawa couplings:}}
A key character of supersymmetric theories is
the identity between the Yukawa couplings $\hat{g}(f\tilde{f}\widetilde{V})$
of the fermions, their
superpartners and the gauginos, and the gauge couplings 
$g(ffV)/g(\tilde{f}\tilde{f}V)$ of the
fermions and sfermions to the gauge bosons. The identity of these
couplings is crucial for the natural solution of the fine-tuning problem.
It must hold not only in theories with exact supersymmetry but also in theories
incorporating the breaking of supersymmetry, as to ensure the stable
extrapolation of the system to energies near the Planck scale---one of
the defining \emph{raisons d'\^etre} for supersymmetry.

In passing it may be noticed that the identity of the gauge couplings
themselves in the SM and SUSY sectors can be tested at the per-cent
level. Smuon pair production is particularly suited for extracting the
gauge couplings in the SUSY sector as this process is mediated solely
by s-channel $\gamma$ and $Z$-boson exchanges.

In contrast, the SUSY Yukawa couplings of the electroweak sector  may be probed
in selectron pair production due to the neutralino t-channel exchange
contribution. By carefully analyzing statistical errors and systematic
uncertainties we could demonstrate that these couplings can be extracted from
measurements of the total cross-sections in the high-energy continuum with a
precision of better than the per-cent level. In particular, this slightly
exceeds the accuracy that can be achieved by other methods, for example in the
analysis of chargino/neutralino pair production.

Matching this expected experimental accuracy 
with its theoretical counterpart requires
the calculation of the cross-sections to per-cent accuracy. For this purpose the
complete next-to-leading order one-loop SUSY electroweak radiative corrections
were calculated for the production of on-shell smuon and selectron pairs. The
corrections were found to be sizable, being of the order of 5--10\%, with
genuine SUSY corrections accounting for about 1\% in $\smuR$ pair production,
where these corrections can be defined unambiguously and separated consistently.

In all examples analyzed in this report, the production of scalar electrons in
$e^+e^-$ annihilation has been compared to the corresponding processes 
in $e^-e^-$
scattering. The $e^-e^-$ mode turns out to be particularly favorable for the
measurement of the selectron masses in threshold scans. In addition 
it can provide complementary information on the selectron Yukawa couplings.

The detailed analytical expressions for the slepton pair production
cross-sections are too lengthy to be reported here in writing. The results are
implemented in computer programs that return the 
cross-sections for smuons and selectrons to one-loop order for
whatever set of Lagrangian parameters in the Minimal Supersymmetric Standard
Model MSSM is chosen. The computer codes are available from the web at
\texttt{http://theory.fnal.gov/people/\linebreak[0]afreitas/}. [Technical information on
installing and running the programs are given at the web site.]

The theoretical analysis presented in this report for smuons and selectrons
is one of the cornerstones for precision analyses of the supersymmetry
sector in particle physics. The precision that is expected to be achieved
in future linear collider experiments requires the analysis of many different
channels in parallel---sleptons, charginos/neutralinos and squarks/gluinos.
The parameters of all these states affect mutually the theoretical predictions
at the one-loop level so that all the associated production channels must be
analyzed simultaneously. Such an overall analysis demands complementary
and coherent experimental action at lepton and proton colliders---a
program for which the present analysis is a crucial building block%
\footnote{Such a comprehensive study is presently underway,
Ref.~\cite{spa}.}. This ensures
finally a self-consistent picture of the SUSY sector at the phenomenological
level.

Beyond drawing a high-resolution picture of supersymmetry at low energies,
the precise determination of these parameters provides the base for
exploring the mechanism of supersymmetry breaking and the reconstruction
of the fundamental supersymmetric theory, potentially at scales near the
Planck scale. In short, the high precision analyses provide us with a
telescope for exploring the structure of physics at the scale of ultimate
unification of genuine particle physics with gravity, as expected to be
realized near the Planck scale.

\section*{Appendix}

The theoretical results and the phenomenological analyses presented in
this report, have been based on the specific reference scenario SPS1a for the
MSSM,
defined in the set of the ``Snowmass Points and Slopes'' \cite{sps}.

The SPS1a point is a typical mSUGRA scenario characterized by fairly
light sfermion masses. If realized in nature, a wealth of experimental
information would become available on this supersymmetric theory from
a linear collider operating in the first phase at energies up to about
1 TeV.

The SUSY parameters of SPS1a are defined at the GUT scale for the following
universal values:
\begin{equation}
\begin{aligned}
m_0 &= 100 \gev, &
M_{1/2} &= 250 \gev, &
A_0 &= -100 \gev, &
\tan\beta &= 10, &
\mu &> 0.
\end{aligned} \label{eq:sps1highpar}
\end{equation}
The overall set is completed by the Standard Model parameters specified
at the electroweak scale as
\begin{equation}
\begin{aligned}
\alpha(\MZ) &= 1/127.70, & \qquad \mt &= 175 \gev, \\
\MZ &= 91.1875 \gev,		& \mb &= 4.25 \gev, \\
\MW &= 80.426 \gev,		& m_\tau &= 1.777 \gev.
\end{aligned} \label{eq:SMpar}
\end{equation}
The evolution of the soft SUSY breaking parameters \eqref{eq:sps1highpar} down
to the electroweak scale by means of the program \textsc{Isajet 7.58} leads to 
the weak-scale parameters listed in Ref.~\cite{spsval}. Uncertainties due to
the implementation of the renormalization group evolution are not relevant for
the purpose of the present study, since the MSSM parameters at the weak scale
are taken as the starting point for our analysis. Using the MSSM soft-breaking
parameters from Ref.~\cite{spsval} together with the SM parameters
\eqref{eq:SMpar} the SUSY particles spectrum in Tab.~\ref{tab:spectrum} is
obtained. The widths and branching ratios of the particles involved in the
decay chains of the sleptons can be found in Tab.~\ref{tab:sps1}.
\renewcommand{\arraystretch}{1.2}
\begin{table}[tb]
\begin{center}
\fbox{\parbox{15.275cm}{\centering{SUSY particles and masses}}}\\
\begin{tabular}{|r|l|}
\hline
$\slR = \seR/\smuR$ & $142.72$ \\
$\slL = \seL/\smuL\,$ & $202.32$ \\
$\tilde{\nu}_l = \tilde{\nu}_e\,/\tilde{\nu}_\mu\:$ & $185.99$ \\
\cline{1-2}
$\tilde{\tau}_1\,$ & $132.97$ \\
$\tilde{\tau}_2\,$ & $206.29$ \\
$\tilde{\nu}_\tau\,$ & $185.05$ \\
\hline
\end{tabular}\hspace{1pt}%
\begin{tabular}{|r|l|}
\hline
$\tilde{u}_\PR$ & $520.50$ \\
$\tilde{u}_\PL$ & $537.20$ \\
$\tilde{d}_\PR$ & $520.11$ \\
$\tilde{d}_\PL$ & $543.07$ \\
& \\ & \\
\hline
\end{tabular}\hspace{1pt}%
\begin{tabular}{|c|l|}
\hline
$\tilde{t}_1$ & $375.74$ \\
$\tilde{t}_2$ & $585.15$ \\
$\tilde{b}_1$ & $488.01$ \\
$\tilde{b}_2$ & $528.23$ \\
& \\ & \\
\hline
\end{tabular}\hspace{1pt}%
\begin{tabular}{|c|l|}
\hline
$\neu_1$ & $96.18$ \\
$\neu_2$ & $176.62$ \\
$\neu_3$ & $358.80$ \\
$\neu_4$ & $377.87$ \\
$\cha_1^\pm$ & $176.06$ \\
$\cha_2^\pm$ & $378.51$ \\
\hline
\end{tabular}\hspace{1pt}%
\begin{tabular}{|c|l|}
\hline
$h^0$ & [89.28]\phantom{0} 122.71\\
$H^0$ & [394.07] 393.56 \\
$A^0$ & [393.63] 393.63 \\
& \\ & \\ & \\
\hline
\end{tabular}
\end{center}
\vspace{-1em}
\mycaption{Mass spectrum in GeV of SUSY particles relevant for this study for the
reference scenario SPS1a \cite{sps}. The Higgs masses are given in Born
approximation [square brackets] and radiatively corrected.}
\label{tab:spectrum}
\end{table}

The staus, squarks and Higgs bosons only enter in the loop corrections to
smuon and selectron pair-production. The masses of
the Higgs bosons in the loop contributions must be evaluated at
tree level, in order to ensure the cancellation of gauge-parameter dependence
in the next-to-leading order result. On the other hand, some of the background
processes considered in Section \ref{fsana} involve tree-level Higgs boson
exchanges. In order to get reliable predictions for these processes, 
it is necessary to include the large radiative corrections to
the Higgs masses, which for this purpose were calculated with the program
\textsl{FeynHiggs} \cite{feynhiggs}.

\subsubsection*{Added Note:}

The SPS1a set of parameters generates the following predictions for the
low-energy
precision observables: BR$[b \to s\gamma] = 2.7 \cdot 10^{-4}$ and 
$\Delta (g_\mu-2)/2 = 17 \cdot 10^{-10}$. The amount of cold dark
matter is, with $\Omega_{\chi}h^2 = 0.18$, still compatible with
WMAP data but somewhat on the high side if they are supplemented
by the ACBAR and CBI data. 

Shifting the scalar mass parameter slightly downwards to $m_0 = 70$ GeV,
but not altering any of the other universal parameters in SPS1a, drives
the value for the density of cold dark matter to the central band of
the data, $\Omega_{\chi}h^2 = 0.126$, without violating the bounds on
BR$[b \to s\gamma]$ and $\Delta (g_\mu-2)/2 $ \cite{belyaev}.
Slepton, chargino/neutralino masses and branching ratios
relevant for the present analysis change within so limited a margin
that none of the conclusions in this report is affected to a significant
amount.

\vspace*{2mm}

\section*{Acknowledgements}

We benefited from many helpful remarks by H.~U.~Martyn 
on experimental slepton analyses at $e^+e^-$ linear colliders. 
Special thanks go also to A.~Belyaev for providing us with crucial 
information on the area in mSUGRA parameter space that is compatible
with WMAP data on cold dark matter. We are also very grateful to 
J.~Kalinowski and T.~Tait for carefully reading the manuscript.  


\end{document}